# High-resolution estimates of the foreign-born population and international migration for the United States.

Nicolas A Menzies PhD[1,2]

**Affiliations:** [1] Department of Global Health and Population, Harvard T.H. Chan School of Public Health, Boston, [2] Center for Health Decision Science, Harvard T.H. Chan School of Public Health, Boston.

**Corresponding author:** Nicolas A Menzies, nmenzies@hsph.harvard.edu, (+1) 617 432 0492, Department of Global Health and Population, Harvard T.H. Chan School of Public Health, 665 Huntington Ave, Boston, MA 02115, USA.

**Abstract**

Detailed estimates of migration stocks and flows provides evidence for understanding population dynamics, and the impact of economic and political changes that influence migration. Using data from the 2000 decennial census and 2001-2016 American Community Survey (ACS), this study derives highly-disaggregated estimates of the foreign-born population residing in the United States for the period 2000-2018, and annual foreign-born entries to the ACS population as a measure of immigration volume. These estimates are derived from an evidence synthesis combining pooled survey data with auxiliary data on potential biases in raw survey estimates and other trends affecting the foreign-born population. For an individual population stratum (defined by current age, entry year, country of origin, and calendar year) direct estimates using survey data can have substantial sampling uncertainty. By imposing logical and probabilistic constraints, data are pooled across survey years to produce more precise estimates. Corrections are implemented for respondent misreporting of demographic information, and undercount of the foreign-born population in the ACS. This paper describes the statistical approach used to model population change, demonstrates the validity of the approach via in- and out-of-sample predictive performance, provides the population estimates, and highlights potential applications.

## Introduction

In 2016, 44 million foreign-born individuals resided in the United States (U.S. Census Bureau 2017), and these individuals play important roles across multiple domains of economic and civic life in the US. Detailed estimates of the size and trends in the foreign-born resident population have many potential applications, and tabulations of foreign-born population stocks are routinely provided by the US Census Bureau. However, these estimates may not stratify results according to all the dimensions required, or may be reported in categories that are overly broad for a particular use. Public-use microdata are also available for many of the large population surveys conducted in the US, but while these datasets allow a high level of disaggregation, the sampling uncertainty associated with 'raw' estimates calculated directly from these data can be substantial (Bazuin and Fraser 2013). Moreover, evidence on foreign-born immigration flows is substantially weaker than evidence on current stocks, with direct estimates of individuals legally admitted to the US (U.S. Department of Homeland Security 2016) excluding undocumented migrants, and indirect estimates back-calculated from the time series of population stocks affected by uncertainty around emigration volume.

The best current data on the foreign-born resident population come from the American Community Survey (ACS), which collects data on a large population-based sample of the US resident population on an ongoing basis, with annual data releases (Mather, Rivers, and Jacobsen 2005). For the ACS, the foreign-born population will include legal immigrants ('green-card' holders), legal non-immigrants (temporary migrants), asylees and refugees, and undocumented migrants, as long as they meet survey criteria for current residence. Early variants of the ACS were introduced from 2000-2004, and the survey has been conducted in a standardized format from 2005. For earlier years, the 5% sample of the decennial census provided similar information on country of origin and other basic demographic data.

Using a mechanistic model of entry, exit, and membership of the US foreign-born population, this study combined data on foreign-born individuals included in the 2001-2016 ACS public-use microdata samples as well as the 5% sample from the 2000 decennial census. By pooling data across samples it is possible to observe successive immigration cohorts at multiple points in time, stratified in a number of dimensions. By placing logical and probabilistic constraints on how individuals enter and exit the foreign-born population, the analytic approach was able to reduce the sampling uncertainty associated with the survey estimates and adjust for sources of bias in these data. The estimation framework also allowed for short-term future population projections, by extrapolating trends estimated for the components of population change.

This analysis provides precise estimates of the number of foreign-born individuals living in the US, for calendar years 2000-2016, stratified by multiple individual-level characteristics (country of origin, year of age, year of entry to the US, and calendar year), as well as projections for the years 2017 and 2018. Moreover, as the estimation approach explicitly modelled entries and exits from the foreign-born population, the analysis also provides annual estimates of foreign-born individuals newly entering the population covered by the ACS and census, a measure of annual immigration volume that may have advantages compared to other approaches. This paper describes the statistical approach used to model population change and generate results, report tests of in- and out-of-sample predictive performance, provide the population estimates, and highlight applications for these estimates.

## Methods

### Data

The primary data sources used for this analysis were the public-use microdata samples (PUMS) for the America Community Survey from 2001 to 2016, and the 5% sample of the 2000 decennial census. These data are publically accessible, and were obtained from the IPUMS-USA repository (Steven Ruggles et al. 2015). Records for foreign-born US residents were extracted from this dataset. 'Foreign-born' was defined as reporting a place of birth outside of the US or US territories, excluding individuals born to US parents. To be a US resident (by virtue of being included in the ACS sample) an individual must have lived at their address for >2 months, or anticipate living at that address for >2 months (the census adopts a slightly different definition based on "usual place of residence" (Mather, Rivers, and Jacobsen 2005), but this difference was ignored for the purpose of creating national-level estimates). Variables for place of birth, year of entry into the US, and current age were extracted from survey data, in addition to analysis weights provided to inflate the individual samples to obtain national population estimates.

Year of entry was bottom-coded at 1949, and current age top-coded at 91. ISO 3166-1 alpha-3 codes (ISO3 codes) were used to identify countries. Individuals were assigned to a single ISO3 code according to their survey country of origin codes. For survey country of origin codes describing a dependent territory without an ISO3 code, individuals were assigned to the governing state of the territory. All other individuals that could not be mapped to a unique ISO3 code were pooled into a residual category. In addition, countries becoming independent states after 2000 (Kosovo, Montenegro, Serbia, South Sudan, Timor-Leste, respectively) were pooled with the state they were part of as of 2000 (Yugoslavia, Sudan, Indonesia). This was necessitated by the estimation procedure, which relied on consistent reporting of country of origin between successive survey years. Using the country of origin variable, an additional classification was created to group individuals into world region of origin, using the World Bank regional classification (East Asia and Pacific, Europe and Central Asia, Latin America and the Caribbean, Middle East and North Africa, North America, South Asia, and Sub-Saharan Africa). Using these classifications, a combined dataset of raw survey estimates was created from the individual-level PUMS data, for all possible combinations of country or region of origin, current age, year of entry, and survey year.

## Analysis

Analyses were undertaken to obtain population estimates for highly-disaggregated population strata, both for the period covered by the survey data (2000-2016), as well as for a 2-year future projection (2017-2018). These represented 64,549 and 8,164 unique values respectively, for each country or region of origin. Results were estimated for the overall foreign-born population, for each world region of origin, and for each of the top 100 countries of origin (ranked according to US resident population size for each country of origin, averaged over the period 2000-2016).

### *Immigration cohort model*

A compartmental stock-flow model was developed to represent the foreign-born US population from 1950 to 2018. This model allowed for (i) an existing stock of foreign-born individuals in 1950, (ii) yearly additions to the foreign-born population in each calendar year from 1950 to 2018 due to immigration, and (iii) yearly exits from the foreign-born population due to emigration or death. In this model, individuals resident in the US in a given calendar year are stratified by year of age and entry year. The number of individuals in a particular immigration cohort (defined by year and age of entry) present in the US in a given year was assumed to be equal to the number of individuals in the same immigration cohort in the previous year, minus exits to emigration or death. Figure 1 shows a schematic of this stock-flow model for the 1950 immigration cohort.

[Figure 1]

*Initial population*

The initial population for the model is composed of all foreign-born individuals present in the US at the start of 1950, stratified by age but not by year of entry. Data for cohorts immigrating to the US before 1950 were combined into a single cohort, as this group represents a small fraction of the current foreign-born population (~1% of all foreign-born individuals represented by the pooled survey data reported an entry year before 1950).

*Immigration*

The log of total immigration volume each year from 1950 to 2018 was modelled as a random walk. By estimating this time series in log space, it was assumed that yearly variability in immigration volume is proportional to the scale of immigration.

The age distribution of immigrating cohorts was represented using penalized B-splines (Lang and Brezger 2004, Hogan and Salomon 2012). This approach constrains the age distribution to be smooth but imposes no additional structure on the shape of the distribution. This age distribution was composed of two components: a one-dimensional spline representing the average age distribution across all entry years, and a two-dimensional spline surface representing temporal deviations from this average pattern. This specification allowed for an overall age distribution as well as smooth changes in this distribution between successive immigration cohorts.

*Emigration and mortality*

Individuals exit the resident foreign-born population through mortality and emigration. It was assumed that mortality rates would be primarily determined by individual age, and evidence on mortality rates by single year of age was drawn from recent US life tables (National Center for Health Statistics 2016). Separate life tables are not available for foreign-born residents, and so life tables for the general population were used. Mortality rates derived from these life tables were applied as exit rates based on current age in the model. Variation in these age-based exit rates was modelled via a penalized B-spline, to allow for deviations in mortality rates between the general population and individual immigrant groups. In addition, age-specific mortality rates were assumed to follow a log-linear decline as a function of calendar year, based on trends reported in decennial life tables 1950-2010 (National Center for Health Statistics 2016).

Exits due to emigration were assumed to be primarily determined by time since entry, with higher emigration rates for recent immigrants (Bhaskar et al. 2013, Ahmed and Robinson 1994). A function was specified to allow the emigration rate to decline smoothly up to 15 years after entry to the US, after which it was held fixed. The total exit rate for a given year and age stratum was calculated as the sum of age-based and year-of-entry-based exit rates.

*Misclassification of reported age and entry year*

Population estimates calculated directly from the survey data show periodic spikes in the population distribution as a function of reported age and entry year (Supplementary Materials, Figure S4). For both of these variables, large positive deviations are apparent on the decades, and smaller positive deviations apparent at mid-decade. In other situations, implausible patterns in ACS results have revealed systematic biases due to misreporting by survey respondents (Gates and Steinberger 2009, Van Hook and Bachmeier 2013), and it was hypothesized that the periodic effects observed in the raw estimates resulted from misreporting by survey respondents, with the true value for age and or entry year being rounded to the nearest decade or mid-decade. A measurement-error model was introduced to correct for this misclassification, allowing that some fraction of respondents would round their response to these questions, either to the decade or the mid-decade. These fractions were estimated from the data.

*Undercounting of foreign-born populations*

Previous research suggests that raw estimates derived from the ACS underestimate true population size for foreign-born populations (Jensen, Bhaskar, and Scopilliti 2015). While analysis weights provided for the ACS are adjusted to account for under- or over-reporting, the ACS is only controlled by age, sex, race, and Hispanic origin, not nativity. The extent of undercount is thought to be greater for recent immigrants, undocumented migrants, immigrants of Hispanic origin, younger age groups, and older ACS survey years (Martin 2007, Jensen, Bhaskar, and Scopilliti 2015, Massey and Capoferro 2004). The magnitude of this bias cannot be identified from the survey data alone, and evidence of undercounting is generally derived from comparison to other data sources. Using estimates of the size of the ACS undercount reported by the US Census Bureau (Jensen, Bhaskar, and Scopilliti 2015), the analysis allowed for underreporting in the PUMS data, so that final analytic estimates would provide an estimate of true population size. These adjustments were specified as inflation factors that varied with time since entry (higher for more recent immigrants), country of origin (higher for countries in Latin America and the Caribbean), and survey year (higher for earlier survey years with smaller sample size). For countries in Latin America and the Caribbean, an average undercount rate of 5.0% was assumed for survey years 2005-2016, and an average undercount of 2.0% was assumed for other countries over the same period, consistent with recent Census Bureau estimates (Jensen, Bhaskar, and Scopilliti 2015).

Estimation

A Bayesian approach was used to implement the analysis. First, formulae describing the relationship between model parameters and population totals were defined. Secondly, a likelihood function was constructed for the survey data, accounting for the sampling uncertainty in survey responses. Thirdly, probability distributions were defined to express prior knowledge (or lack of it) for model parameters. In general, weakly informative priors (Gelman 2006) were specified, except where substantial prior information was available, such as with age-specific population mortality rates. The posterior distribution of the model parameters was calculated as the normalized product of the prior distribution and the likelihood function. This posterior parameter distribution was used to calculate the population estimates.

The estimation model was fitted separately for each country or region of origin. Model estimates for the surveyed population (reflecting any misreporting and underreporting) were compared to the raw survey values to validate the estimation results (next section). For each country or region of origin ‘true’ values were also produced, which are adjusted to remove the effects of misreporting and underreporting. In addition to these population estimates, the analysis produced estimates of annual entries to the population covered by the ACS and census, as a measure of immigration volume.

The model was fitted using adaptive Hamiltonian Monte Carlo sampling, as implemented by the Stan probabilistic programming software (Carpenter B et al. 2017, Stan Development Team 2016). Numerical Bayesian methods produce results as a sample from the posterior distribution of model parameters and other quantities of interest. The mean of these posterior samples was used to create point estimates for each population group of interest, and equal-tailed 95% posterior intervals calculated to quantify uncertainty in estimates. Data processing was conducted in R (R Core Team 2016). Full details of the technical specification of this analysis, including model equations, prior distributions, and likelihood function, are provided in the Supplementary Material.

Validation

Assessing predictive performance by comparing estimates to the data used to create them ('in-sample fit') will reward over-fitting, with the optimal procedure being one that simply returns the observed values. Comparing the model predictions to data not used to fit the model ('out-of-sample fit') provides a more rigorous validation of the estimation procedure. As a test of out-of-sample predictive performance the model was re-estimated for a sample of seven countries with selected survey years removed, testing the ability of the estimation procedure to reproduce these held-out values. The countries—Fiji, Mexico, Pakistan, Peru, Poland, Somalia, and Viet Nam—were chosen to represent a range of world regions and resident population sizes. For each of these countries the model was re-estimated three times, holding out data for the years 2005, 2010, and 2015 respectively, and fitting the model to the remaining data. Results from these analyses was compared to raw data, to assess the ability of the model to predict the data that had been excluded from estimation.

Estimated values and held-out survey data were compared visually, plotting the estimated population as a function of each dimension of interest (year of entry, age at entry, and current age) in order to identify systematic deviations that might suggest a problem with the estimation approach. In addition, standardized residuals were calculated by dividing the estimation residuals (raw survey estimate for a unique combination of place of birth, year of entry, and current age, minus the modelled estimate for the same value) by the standard error estimates provided by the survey methodology (US Census Bureau and US Department of Commerce 2014). Theoretically, these standardized residuals would have a standard deviation of 1.0 for a model that perfectly predicted the mean of each observation. The fraction of instances in which the survey estimate were predicted to be zero (i.e. no individuals with a given set of characteristics included in the sample) was also calculated, and compared this to the empirical distribution from the survey. With a well-performing model, the modelled probabilities should reproduce the observed frequencies. Finally, logged values for modelled vs. raw population estimates were plotted to assess any estimation errors associated with the magnitude of population estimates. With a well-performing model, the points on these scatterplots should cluster on the diagonal.

## Results

### Comparison of modelled results to raw survey estimates

Figure 2 shows total population estimates for 2016, comparing modelled estimates for the true population (i.e. controlling for undercount and misclassification of demographic data by survey respondents) to 'raw' survey estimates, which were estimates derived directly from the PUMS data by weighting observations by the analysis weights provided. As might be expected, the modelled population estimates for countries of origin with large resident populations (Figure 2, lower panel) are very similar to the raw survey estimates, as for these populations the raw estimates will have low sampling uncertainty relative to the size of the population. For countries-of-origin with smaller resident populations (Figure 2, upper panel), there are more substantial differences between modelled and raw estimates, reflecting greater sampling uncertainty in the results of an individual ACS round. As an example, Yemen has one of the largest differences between modelled and raw population estimates for 2016. Compared to the average population estimate from the three prior survey years, the raw 2016 population estimate for Yemen is 73% higher, an increase of 29,000 compared to the year before. While this could signal a major uptick in new immigrants compared to death and emigration, the number of individuals from Yemen in their first year in the United States was only 6,400 in the 2016 ACS, implying a much lower immigration rate than would be needed to explain this result. This suggests the high 2016 population estimate for Yemen is simply a consequence of sampling uncertainty. The modelled estimate, by incorporating evidence across the 17 survey years, is robust to sampling uncertainty in an individual survey. This effect, of smoothing implausible deviations in the time trend, results in a lower modelled population estimate for 2016, a 6.3% increase on the modelled estimate for the preceding year. Figures comparing the time series of modelled population estimates to raw survey values for 2000-2016 are provided in the Supplementary Materials, plus projections to 2018. The Supplementary Materials also include graphs comparing modelled to raw estimates for the 2016 resident population distribution by current age, age at entry, and year of entry (Figures S7).

[Figure 2]

While the modelled estimates appear to have some advantages for estimating total population size for countries with smaller resident populations, the utility of these estimates is more obvious when comparing estimates for more highly-disaggregated population strata.

Figure 3 compares modelled and raw 2016 population estimates for three countries (Somalia, Mexico, and Viet Nam), stratified by single year of age and single year of entry. Panel A shows heatmaps of population density created from raw survey values, in which the effects of sampling uncertainty can be seen. In this figure, each cell represents a single population stratum, and warmer colors indicate greater population within that stratum. An empty cell indicates no one from that stratum was included in the survey for that year. Even for Mexico—the country of origin with by far the largest resident population—there are multiple strata where no individuals were included in the survey, and sampling uncertainty can be observed in the variation in color between adjacent cells. Viet Nam has a smaller resident population, and while major features of the population distribution can be seen, there is substantial noise apparent in the raw values. For Somalia—a country whose resident population is less than 1% of Mexico's—there is little that can be learnt from the raw survey values, with the majority of strata being empty. Panel B shows heatmaps of population density created from the modelled estimates. By incorporating evidence from across the 17 survey years, these modelled estimates provide lower variance estimates of the population distribution, demonstrating patterns that could be difficult to observe in the raw values.

There are alternative approaches that could reduce sampling uncertainty in these small-group estimates. In particular, more precise estimates could be obtained by pooling categories of interest together (Citro and Kalton 2007) and therefore using a coarser grid to stratify the population, such as by using 5- or 10-year wide bins for age and years since entry. However, such coarsening could obscure relevant features of the distribution, such as the spike in the population from Viet Nam reporting 41 years since entry. This corresponds to an entry year of 1975, with over 10 times the number of individuals estimated to share this entry year in 2016 compared to 1974 or 1976. Using wider bands to categorize age and years since entry would not fully obscure such a major immigration cohort, but potentially useful information would be lost. Population estimates for these highly disaggregated population strata, for calendar years 2000 to 2018 and for each country or region of origin, are included in the Supplementary Materials.

[Figure 3]

<u>Out-of-sample validation</u>

Out-of-sample predictive performance was used to assess the quality of the estimation approach. Figure 4 compares modelled and raw 2015 population estimates for three example countries (Somalia, Mexico, and Viet Nam). For each country, estimates were produced by a model fitted to the time series of data excluding 2015. Two sets of modelled estimates are shown—estimates of the 'true' resident population (dashed blue lines), which adjust for biases introduced by misreporting of demographic characteristics by survey participants as well as under-coverage of the ACS), and estimates of the survey population (solid lines and shaded regions), which do not adjust for these effects and so are directly comparable to raw survey results. These figures show the modelled estimates closely following the systematic patterns apparent in the raw survey estimates (points). The prediction intervals—which should contain approximately 95% of the survey estimates if the estimation approach is working correctly—also appear to be well calibrated. In the Supplementary Materials (Figures S6), similar comparisons are shown for seven countries (Fiji, Mexico, Pakistan, Peru, Poland, Somalia, and Viet Nam) and each of three hold-out years (2005, 2010, 2015).

[Figure 4]

In addition to comparing out-of-sample performance graphically, standardized residuals were calculated to quantify differences between raw and modelled population estimates. Across the 21 country-year comparisons considered in the out-of-sample validation, the standard deviation of these standardized residuals ranged from 0.23 to 1.67, with a mean value of 1.09. For these standard deviations to fall below 1.0 (theoretically optimal predictive performance) would normally imply over-fitting of the observed data, yet this is not possible as the validation data were not used to fit the models. Values lower than 1.0 were only observed for countries with small resident populations (Fiji, Somalia), and so the most likely explanation for this finding is that the method provided by the ACS to calculate standard errors is conservative in the context of small population counts. Similarly, other approaches used to assess model performance (log-log scatterplots of modelled vs. raw population estimates, comparison of modelled vs. predicted probabilities of an individual is a given stratum being included in the survey) did not reveal any major problems with the estimation approach (Supplementary Materials, Figures S6).

Volume of migration

With the estimation approach used, the relationship between population size in successive survey years was explicitly modelled via the components of population change. In the analytic model, entries (immigration) and exits (death, emigration) from the resident foreign-born population were allowed to vary as a function of current age, years since entry, country or region of origin, and calendar year. With time-series data on total population, the absolute entries and exits from the population totals cannot be uniquely identified, with only the year-on-year difference between these quantities known (entries and exits from the population could be increased or decreased by matching amounts, for the same net change). However, using data on years since entry allows these two processes to be distinguished—by definition, individuals can only enter an immigration cohort with years since entry equal to zero, and for subsequent years (i.e. with years since entry > 0) the only change in the cohort will be through exits to death or emigration. This was exploited to derive estimates of the number of individuals newly entering the foreign-born population each year, as a measure of immigration volume for each country and region of origin. By assuming that emigration rates were only influenced by current age, country or region of origin, and years since entry, and that age-specific mortality rates followed a well-defined time trend, it was possible to estimate immigration totals for the full period from 1950 to the present. Figure 5 presents estimated annual immigration volume for the seven example countries used for out-of-sample predictive checks, as well as for the total foreign-born population. As these estimates are derived from models fit to ACS data, these estimates represent the number of individuals entering the country to become a 'resident' according to the definitions used in the ACS (i.e. current or anticipated residence at current address for more than two months).

[Figure 5]

The modelled immigration estimates can be compared to other approaches for estimating immigration flows. In addition to the modelled estimates, Figure 5 presents the numbers of new Legal Permanent Resident (LPR) admissions per year (red line). These data are reported in a standardized format by the Department of Homeland Security Office of Immigration Statistics and are commonly used to describe immigration rates. However, using these data to make inferences about the level and trend in total immigration faces several challenges. The first of these is the potential for delays between year of entry to the country and year of obtaining LPR status, with approximately half of all LPR immigrants entering the United States several years before obtaining LPR status (termed 'change-of-status' LPR applications). This time lag can distort estimated immigration trends, particularly where there are events that produce fluctuations in approved 'change-of-status' LPR applications that are unrelated to actual immigration rates. A prominent example of this is the Simpson-Mazzoli Act of 1986, which led to a large number of previously-undocumented residents gaining LPR status over a short period of time. This produced a large spike in the LPR time series around 1990-1991, visible for several of the countries shown in Figure 5. Another drawback of using LPR data to track immigration flows is the exclusion of undocumented migrants, as for some countries-of-origin, particularly those in South and Central America, a substantial fraction of migrants are thought to be undocumented. Others will enter and reside in the country legally, but hold non-immigrant visas and so not be included in the LPR data. As the modelled estimates are based on the ACS, they will include legal immigrants (LPR), legal non-immigrants (temporary migrants), humanitarian migrants, and undocumented migrants, as long as they meet survey criteria for current residence (Mather, Rivers, and Jacobsen 2005). The difference between LPR immigration and the more inclusive modelled estimates can be seen most prominently in Figure 5 for Mexico and All immigrants, where the LPR time trend is substantially lower than the modelled estimate. For other countries, particularly Viet Nam, the LPR time series and modelled estimates track each other closely.

The raw ACS estimate for individuals who reported entering the country in the same year as the survey was conducted provides another approach to estimating immigration flows (Figure 5, green line). A major challenge for this approach is the need to adjust for the fact that this population is only partially observed in the ACS—as the ACS is conducted continuously throughout the year, interviews conducted earlier in the year will exclude individuals entering the country later in the year. In Figure 5 this undercount is adjusted for by simply doubling the raw values, which would be appropriate if migrants enter the country at a constant rate throughout the year and immediately meet the ACS inclusion criteria. However, it is not clear that these assumptions hold. Moreover, the raw ACS values exhibit increased variance compared to the modelled values—particularly apparent for smaller countries of origin like Fiji—and provide no information on immigration flows in years prior to the observed survey period.

Immigration trends were calculated for each country and region of origin, and Figure 6 reports recent immigration trends for all immigrants grouped into world regions (according to World Bank regional groupings). These results show differential trends between world regions, with progressive growth from South Asia and Sub-Saharan Africa compared to other world regions. Some systematic patterns are likely to result from geopolitical events, with immigration from all world regions declining sharply between 2000 and 2002, possibly as a result of the September 11 terrorist attacks. More recently, a major and sustained decline in immigration from the Latin America and Caribbean region occurred between 2000 and 2005, producing a greater than 50% absolute reduction in annual immigration, with this decline almost completely reversed by growth over the following 5-year period. This major decline and subsequent rebound is not apparent for other world regions, and is not seen in the LPR estimates (Figure 5), although immigration from Europe and Central Asia shows a more modest decline and recovery over the same period.

[Figure 6]

## Discussion

This paper reports highly-disaggregated estimates of the foreign-born population residing in the United States for the period 2000-2018, stratified by country and region of origin, current age, and year of entry to the US. By pooling estimates across several survey years, these estimates are more precise than estimates calculated directly from the raw survey data. The approach used to obtain more precise estimates—following individual immigration cohorts through the time series of survey data, and assuming smooth distributions for various demographic characteristics—differs from variance reduction approaches commonly applied to these data, which involve aggregating data within larger categories (Spielman and Folch 2015), or across multiple survey years via the ACS 3-year and 5-year estimates. While these approaches can successfully reduce variance, they can also obscure important features of the data, in those circumstances where outcomes of interest vary across the categories being aggregated. As well as providing improved precision, the modelled population estimates adjust for biases in the ACS data, including the tendency for some survey respondents to round their demographic data to the nearest decade, and under-coverage of the foreign-born population, particularly in the early years of the ACS. In addition to these population estimates, estimates of new entries to the resident foreign-born population are also reported, as a measure of annual immigration volume. These estimates are inclusive of undocumented and temporary migrants, populations for whom immigration and residency estimates are of substantive interest, but difficult to obtain (Warren and Passel 1987, Passel 2006, Baker 2016b, a).

The results section provides some examples of how these estimates may be used—by improving the precision of stratified population estimates, patterns and temporal trends in these populations can be more easily observed, which may suggest directions for further investigation. In addition to describing patterns and trends, the disaggregated population estimates may also provide inputs into other analyses. The initial motivation for this study was to obtain population denominators for an epidemiological assessment of infectious disease burden in the foreign-born US population, where it was expected that disease burden would likely differ according to several of the demographic characteristics considered in this study. It is likely that other analyses would also benefit from fine-grained data on population distributions for this group, providing population denominators with which measures of the incidence or prevalence of a condition of interest can be calculated (Johnson, Hu, and Dean 2010, Kowdley et al. 2012, Bern and Montgomery 2009).

The population estimates produced by this study have several limitations. Some issues may be inherited from the data source, as the validity of the modelled estimates is contingent on the validity of the ACS data themselves. While two sources of bias were considered in the analysis, there could potentially be other systematic biases in the data used to estimate the model. However, given the large amount of assessment and validation undertaken around the ACS, any remaining biases (Folch et al. 2016) are likely to be small. More major biases may have been introduced by the approach used to smooth survey estimates. For example, while a flexible function was used to describe secular trends in the age distribution of new immigrants, this function still assumed that this distribution would be smooth. If the distribution changed discontinuously over time or age, this sharp change would be captured imprecisely in the model. Similarly, the assumption that emigration rates asymptote to a constant rate after 15 years since entry could also introduce distortions if this assumption does not hold. These are two examples of simplifying assumptions that were necessary to make the model feasible, but that could introduce bias to the modelled estimates if they conflict with survey data. Moreover, poor fit in one part of a model can have downstream effects for other parts of the model, introducing biases that are hard to diagnose and resolve. The steps taken to validate the population estimates showed good out-of-sample predictive performance on a range of test cases, reducing the likelihood that major biases exist, but it is still possible that biases exist for countries or years not included in the validation.

In contrast to the population estimates, the estimates of immigration volume describe a quantity which is not directly observed in the survey data, but is instead computed indirectly as the number of new immigrants needed to populate an observed immigration cohort. As a consequence, these estimates depend more heavily on the validity of modelling assumptions. The comparison data available to validate these estimates are also weaker and have their own biases. If this estimation approach proves useful as a supplement to more direct measures of foreign-born population dynamics, additional testing and validation of the resulting estimates will be valuable.

**Conclusion**

The ACS and decennial census have proven invaluable for understanding population dynamics and societal trends in the United States. This study demonstrates how the evidence from these surveys can be extended by combining raw data with a mechanistic model of population change. Exploiting relationships within the survey data to allow more precise inferences is consistent with the goals of the surveys themselves, by providing the best information for decision-making while minimizing cost and respondent burden.

**Acknowledgements**

I would like to thank Andrew Hill and Suzanne Marks of the CDC Division of Tuberculosis Elimination; Joshua Salomon, Ted Cohen, Meghan Bellerose, Yelena Malyuta, Nicole Swartwood, and Christian Testa of the Prevention Policy Modeling Lab; and other participants of NCHHSTP-NEEMA modelling meetings at which initial versions of this work were discussed. This project was funded by the U.S. Centers for Disease Control and Prevention, National Center for HIV, Viral Hepatitis, STD, and TB Prevention Epidemiologic and Economic Modeling Agreement (NEEMA; #5NU38PS004644).

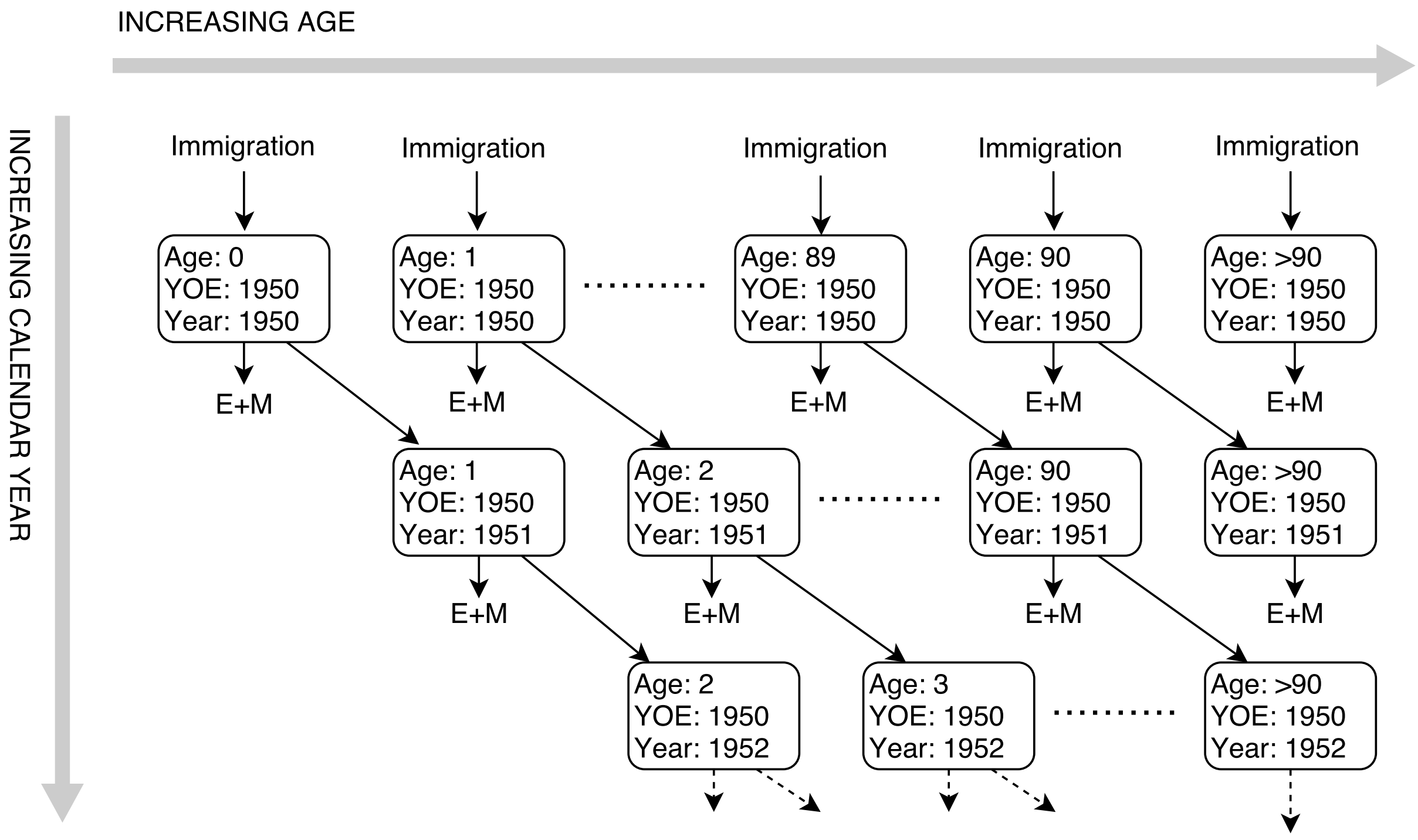

**Figure 1. Schematic of estimation model for 1950 immigration cohort.** 'Age' = current age; 'YOE' = year of entry; 'Year' = current year; 'E+M' = Total exits to emigration and mortality.

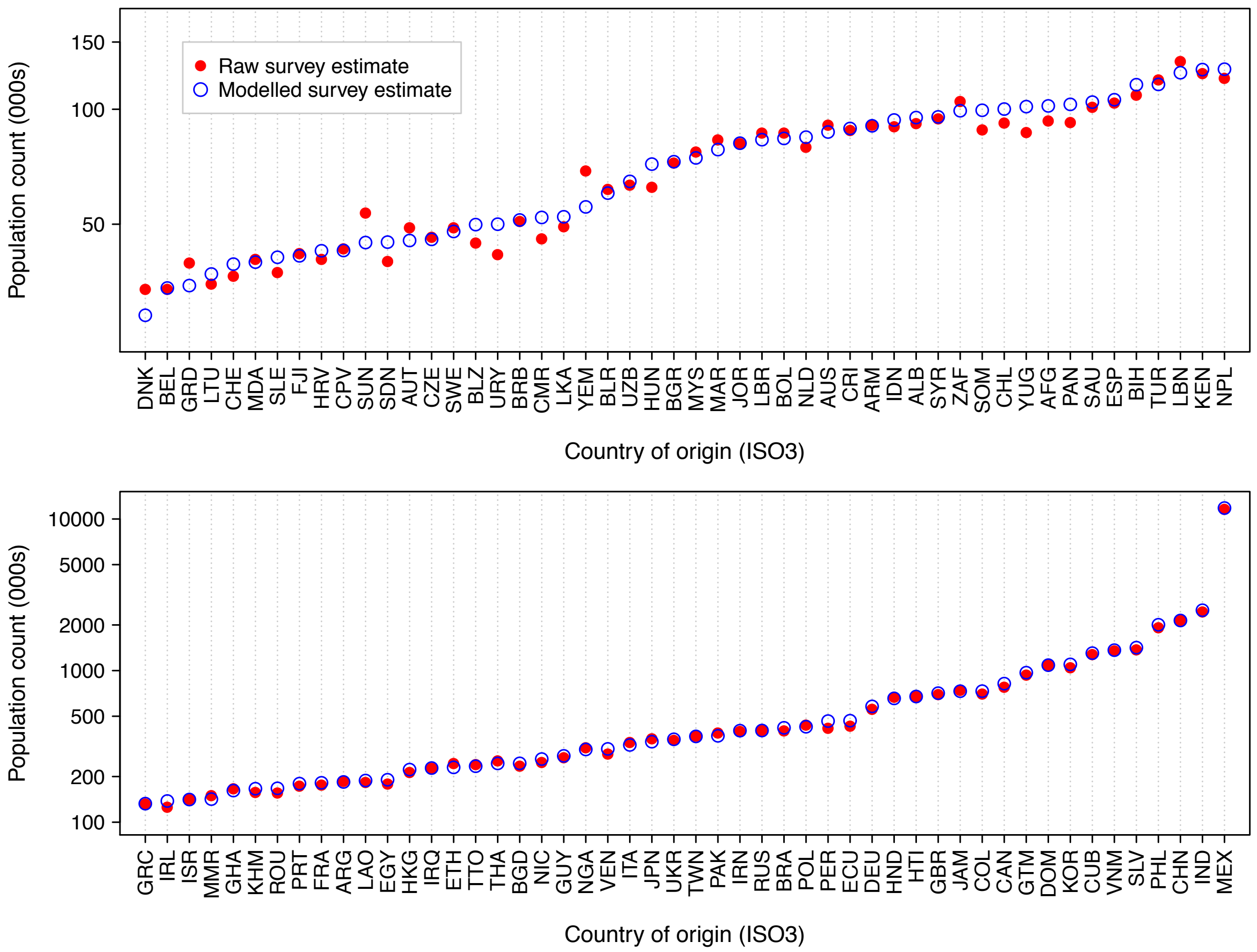

**Figure 2. Total population estimates for 2016, comparing modelled true population estimates to raw survey estimates, for the 100 countries-of-origin by population size.**

Estimates ordered by modelled population size.

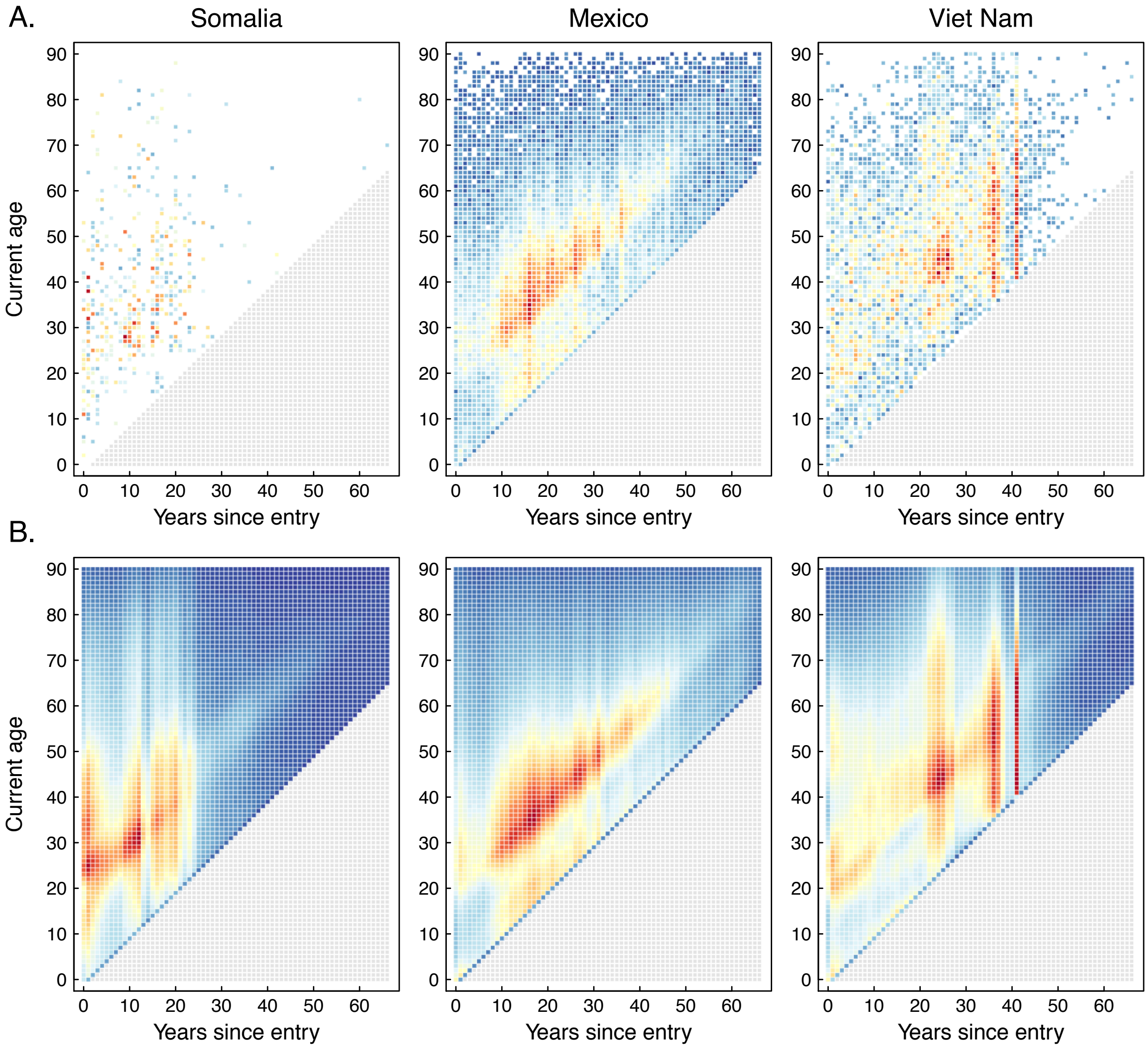

**Figure 3. Distribution of 2016 resident population by single year of age and single year since entry, for Somalia, Mexico, and Viet Nam, comparing raw (Panel A) and modelled (Panel B) survey estimates.** Color gradient demonstrates differences between high population density (warmer colors) and lower population density (cooler colors). Grey cells indicate illogical values (years since entry greater than current age). Empty (white) cells indicate that no one with those characteristics were included in the 2016 ACS sample.

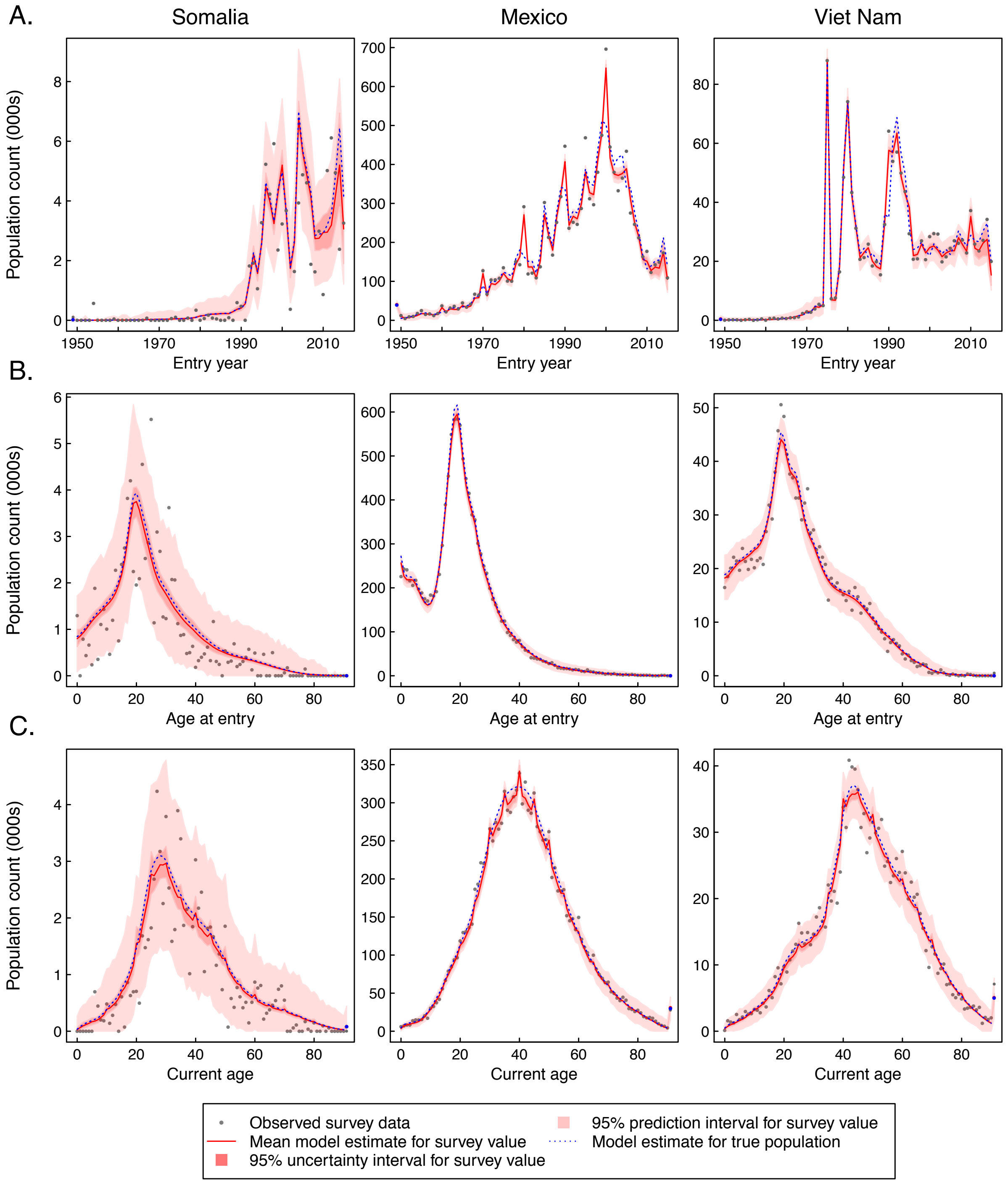

**Figure 4. Predicted distribution of survey estimates for Somalia, Mexico, and Viet Nam in 2015, estimated with 2015 data held out.**

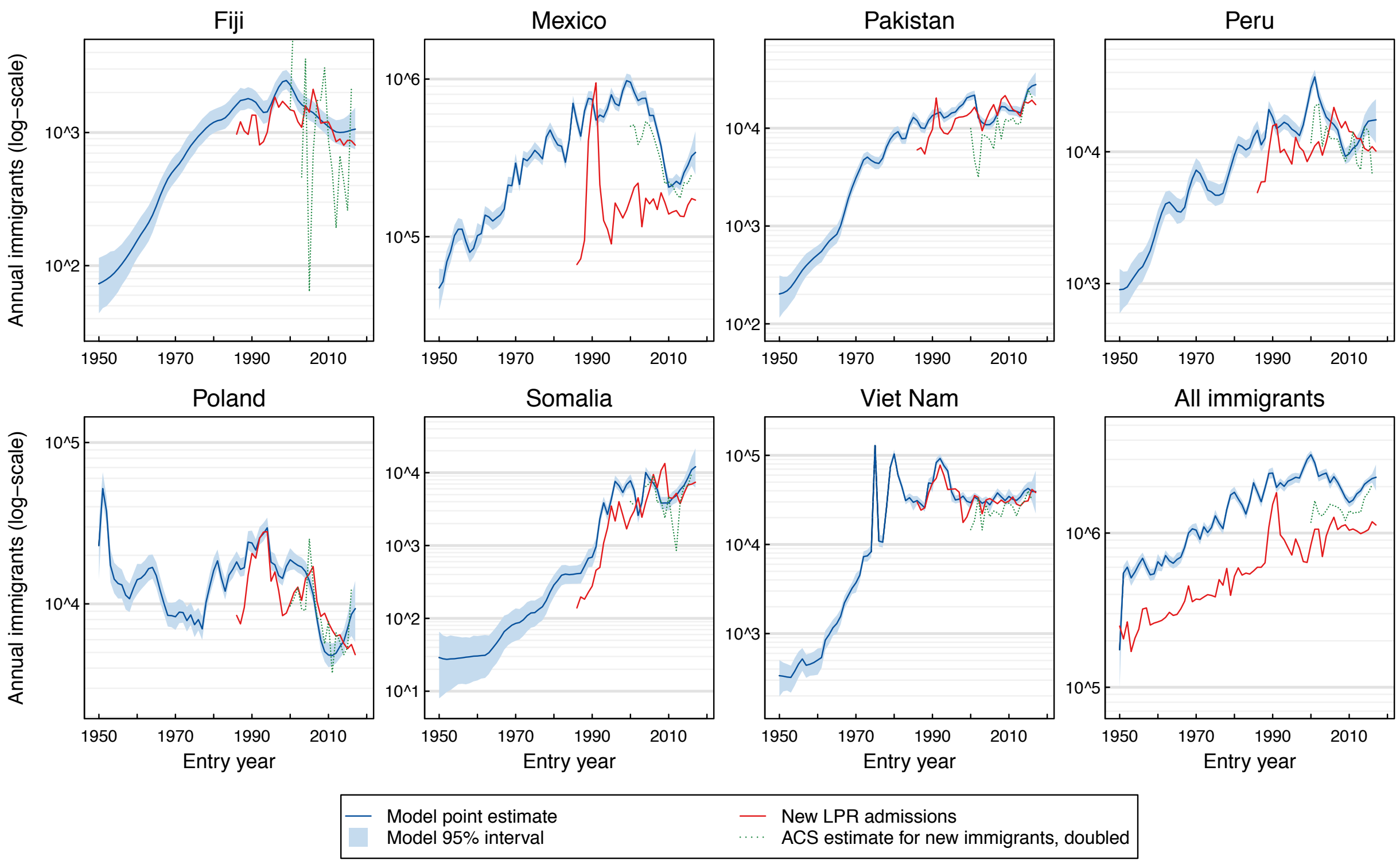

**Figure 5. Estimates of annual immigration rates for seven example countries as well as all countries of origin combined, compared to other measures describing immigration volume.**

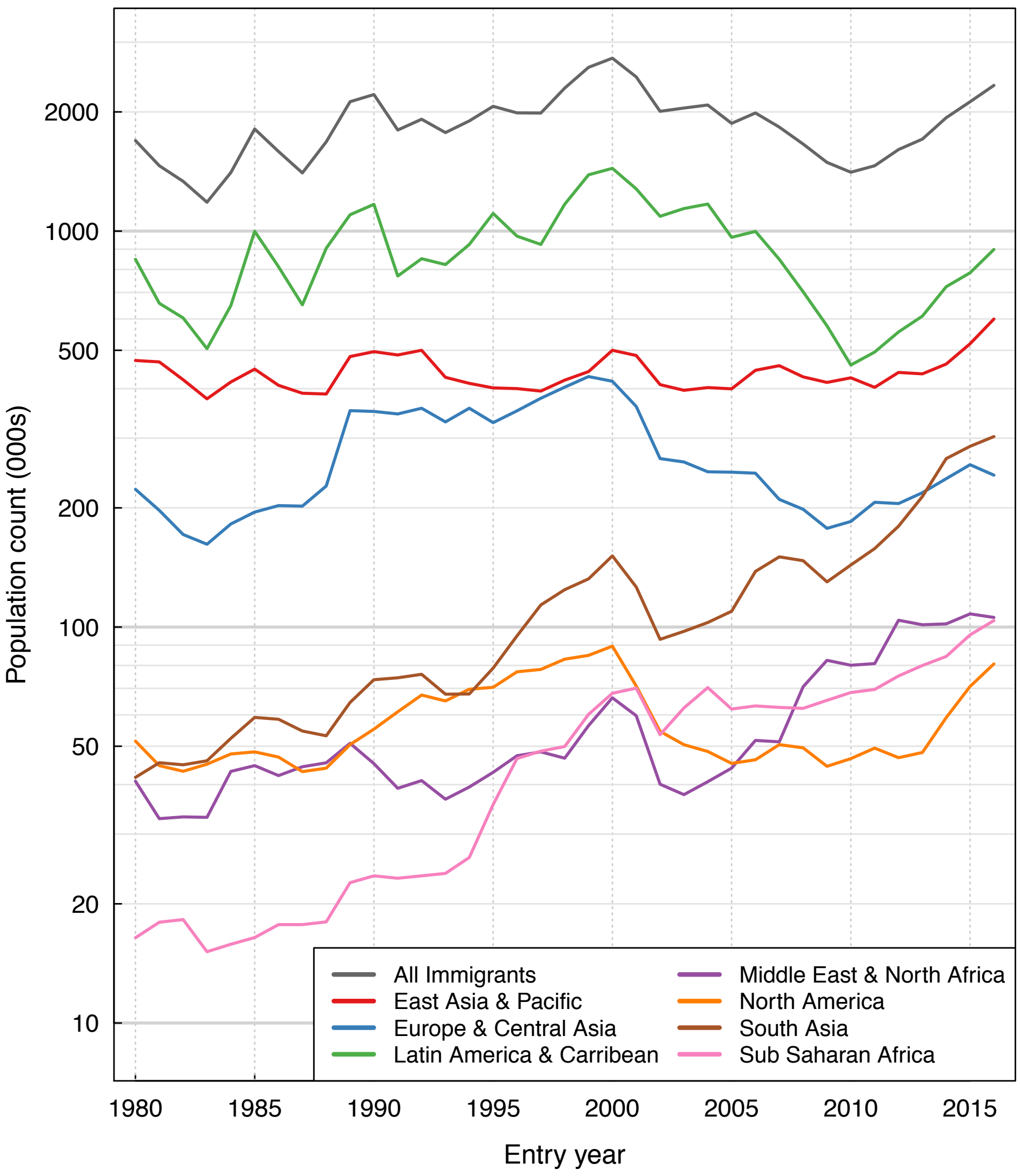

**Figure 6. Estimates of immigration rates from 1980 to 2016 by world region of origin.**

**Supplementary Materials for "High-resolution estimates of the foreign-born population and international migration for the United States".**

Contents

## Technical details of estimation approach

### Notation

Throughout this appendix, the greek alphabet is used to denote model parameters, which are assumed to vary randomly, and which are assigned prior probability distributions. The latin alphabet is used for data and for derived quantities (deterministic functions of model parameters and/or data). In some cases prior distributions are assigned to derived quantities, effectively creating prior distributions for functions of model parameters. Indices for survey year, current age, year of entry, age of entry, and years since entry are denoted $cyr$, $age, yoe$, $aoe$, and $yse$ respectively. These first three of these are reported as part of the microdata sample. The last two are calculated from these quantities:

$$aoe = age - cyr + yoe$$

$$yse = cyr - yoe$$

### Starting population

The natural log of the total resident foreign-born population at the beginning of 1950 is denoted $\alpha_{1949}$, and was given a weakly informative prior distribution (the equal-tailed 95% interval for this distribution ranges from 3.1 * $10^{-9}$ to 3.3 * $10^{8}$):

$$\alpha_{1949} \sim Normal(0, 10)$$

### Annual immigration volume

The natural log of immigration volume 1950-2018 is denoted $\alpha_{yoe}$, for $yoe \in [1950, 2018]$ and was given a weakly informative prior distribution:

$$\alpha_{yoe} \sim Normal(0, 10), \text{ for } yoe \in [1950, 2018]$$

The first difference of $\alpha_{yoe}$ is calculated:

$$\frac{d\alpha}{dyoe} = \alpha_{yoe} - \alpha_{yoe-1}, \text{ for } yoe \in [1951, 2018]$$

These first differences were assumed to follow a Student's *t* distribution centered at zero, with degrees of freedom 10, and dispersion parameter $\sigma_{\alpha 1}$. This dispersion parameter was assigned a prior and estimated simultaneously:

$$\frac{d\alpha}{dyoe} \sim t(10, 0, \sigma_{\alpha 1}), \text{ for } yoe \in [1951, 2018]$$

$$\sigma_{\alpha 1} \sim HalfCauchy(0, 10)$$

By assigning this prior to the first differences of $\alpha_{yoe}$, it was assumed that the time series of logged annual immigration volume follows a random walk. This indirect operationalization was preferable to the direct approach (ie coding parameters for the random walk directly), as it reduces correlation in the posterior distribution and therefore makes this posterior distribution easier for a sampling algorithm to traverse. By assigning a heavy-tailed prior for $\frac{d\alpha}{dyoe}$, the approach allows for occasional changes in immigration volume many times larger than typically observed. The calculation of first differences excludes $\alpha_{1949}$, as there is no relationship assumed between the resident foreign-born population in 1950 and immigration volume that year. To dampen oscillations in the immigration time trend not supported by the data, a penalty function was applied to the second difference of the random walk. The second difference of $\alpha_{yoe}$ is calculated:

$$\frac{d^2\alpha}{dyoe^2} = 2\alpha_{yoe} - \alpha_{yoe-1} - \alpha_{yoe+1}, \text{ for } yoe \in [1951, 2017]$$

As with the first differences, a prior distribution was assigned to these second differences:

$$\frac{d^2\alpha}{dyoe^2} \sim t(10, 0, \sigma_{\alpha 2}), \text{ for } yoe \in [1951, 2017]$$

$$\sigma_{\alpha 2} \sim HalfCauchy(0, 10)$$

**Age distribution of immigrants**

The average age distribution of individuals entering the country was modelled with a penalized b-spline with 17 control points. This spline was implemented with a cubic basis function and a second-order penalty. Knot locations were based on percentiles of the data distribution, allowing greater flexibility for ages with greater relative numbers of new immigrants (20-40 years old). Figure S1 shows the form of these basis functions.

The spline parameters $\gamma_i$ were given a weakly informative prior:

$$\gamma_i \sim Normal(0, 10), \text{ for } i \in [1, 17]$$

The penalty function was applied to the second differences of these spline parameters:

$$\frac{d^2\gamma}{di^2} = 2\gamma_i - \gamma_{i-1} - \gamma_{i+1}, \text{ for } i \in [2, 16]$$

$$\frac{d^2\gamma}{di^2} \sim Normal(0, \sigma_{\gamma 2}), \text{ for } i \in [2, 16]$$

$$\sigma_{\gamma 2} \sim HalfCauchy(0, 10)$$

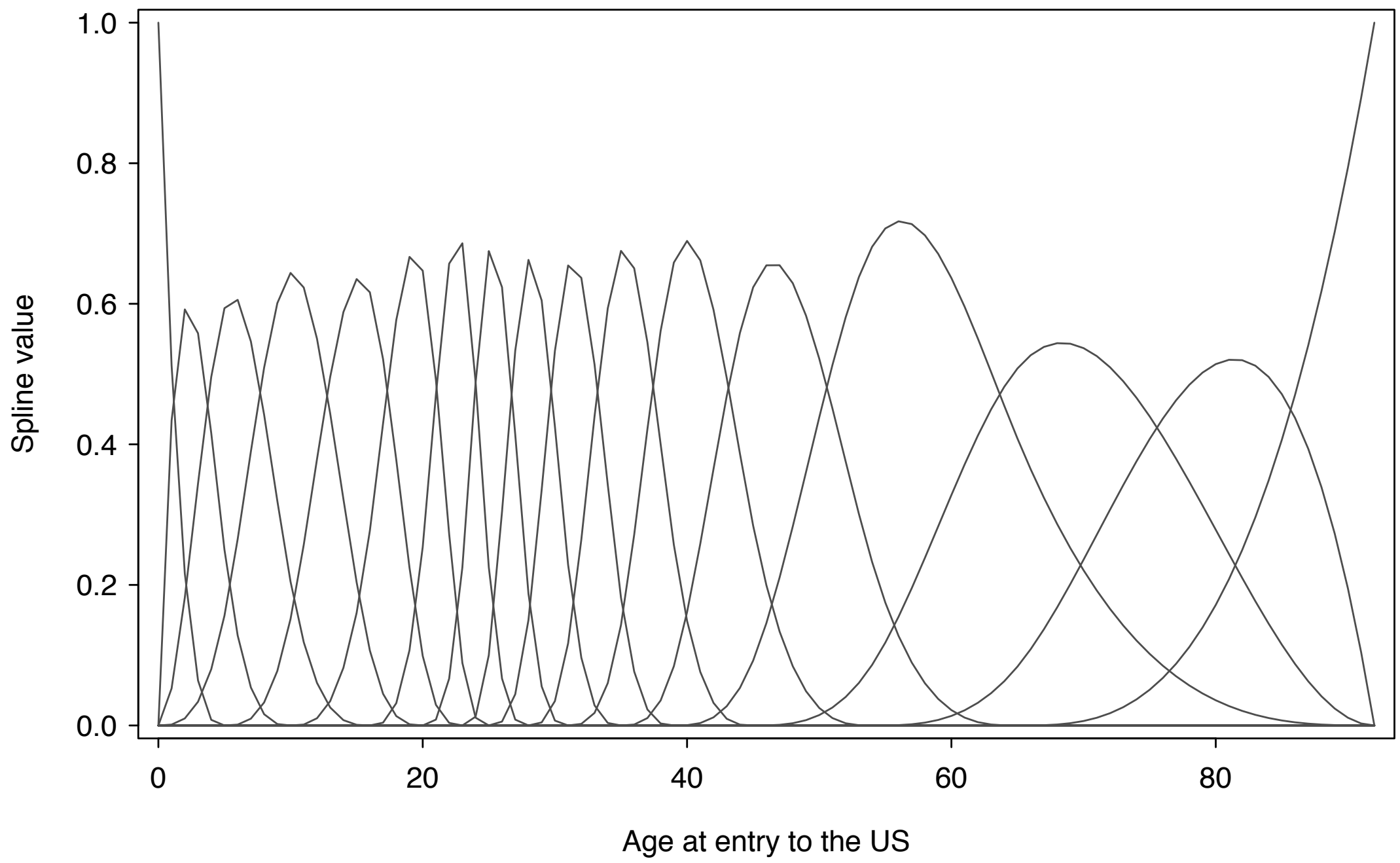

**Figure S1: Cubic b-spline basis functions for age of entry.**

The spline itself ($a_{aoe}$) is calculated as the matrix product of the spline design matrix ($M_{aoe}$) and the vector $\gamma$:

$$a_{aoe} = M_{aoe}\gamma$$

The vector $a_{aoe}$ defines the average age distribution. To allow for deviations from this average distribution over time, a two-dimensional b-spline surface was defined over age (17 control points) and entry year (7 control points). The parameters for this spline surface ($\tau_{i,j}$) were assumed to be drawn from a Normal distribution with mean zero and standard deviation $\sigma_\tau$. This parameter was given its own prior and estimated simultaneously. The prior on $\sigma_\tau$ was more informative than for $\sigma_{\gamma 2}$, under the assumption that temporal deviations in the age distribution would be small relative to the flexibility of the overall age distribution.

$$\tau_{i,j} \sim Normal(\,0, \sigma_\tau), \text{ for } i \in [1, 17] \text{ and } j \in [1, 7]$$

$$\sigma_\tau \sim HalfCauchy(0, 1)$$

The spline itself ($b_{aoe,yoe}$) is calculated as the matrix product of the spline design matrices $M_{aoe}$ and $M_{yoe}$, and the matrix $\tau$:

$$b_{aoe,yoe} \;= M_{aoe}\, \gamma\, M_{yoe}'$$

To restrict the age distribution to positive values, $a_{aoe}$ and $b_{aoe,yoe}$ were defined in log space. The final age distribution ($c_{aoe,yoe}$) was calculated by exponentiating the sum of $a_{aoe}$ and $b_{aoe,yoe}$, and normalizing the resulting values such that the age distribution in any year summed to 1.0. This normalizing constant was given a loose prior ($d_{yoe}$) in order to identify the model:

$$c_{aoe,yoe} = \frac{\exp\left(a_{aoe} + b_{aoe,yoe}\right)}{d_{yoe}}, \text{ for } yoe \in [1950, 2018] \text{ and } aoe \in [0, 91]$$

$$d_{yoe} = \sum_{aoe=0}^{91} \exp\left(a_{aoe} + b_{aoe,yoe}\right), \text{ for } yoe \in [1950, 2018]$$

$$d_{yoe} \sim Gamma(2, 1), \text{ for } yoe \in [1950, 2018]$$

The number of individuals entering the country by age and year was calculated as the product of total annual immigration volume in a given year multiplied by the fraction in a given age group in that year: $exp\left(\alpha_{yoe}\right) * c_{aoe,yoe}$.

**Exit via mortality**

Mortality rates were assumed to vary by age and year. Point estimates for mortality by year of age ($e_{age}$) were drawn from 2012 life tables (most recent available year) for the whole US population (National Center for Health Statistics 2016). These mortality rates were assumed to decline by 1.31% per year over the period 1950-2015, a rate estimated from the decennial life tables reported for this period (Arias et al. 2008, National Center for Health Statistics 1997, 1985, 1975, 1964, National Office of Vital Statistics 1954). Figure S2 shows these point estimates by age and year.

A penalized b-spline with 7 control points was used to model uncertainty in these mortality rates as a function of age. The spline parameters $\delta_i$ were given an informative prior:

$$\delta_i \sim Normal(0, 0.25), \text{ for } i \in [1, 7]$$

The penalty function was applied to the second differences of these spline parameters:

$$\frac{d^2\delta}{di^2} = 2\gamma_i - \gamma_{i-1} - \gamma_{i+1}, \text{ for } i \in [2, 6]$$

$$\frac{d^2\delta}{di^2} \sim Normal(0, \sigma_{\delta 2}), \text{ for } i \in [2, 6]$$

$$\sigma_{\delta 2} \sim HalfCauchy(0, 10)$$

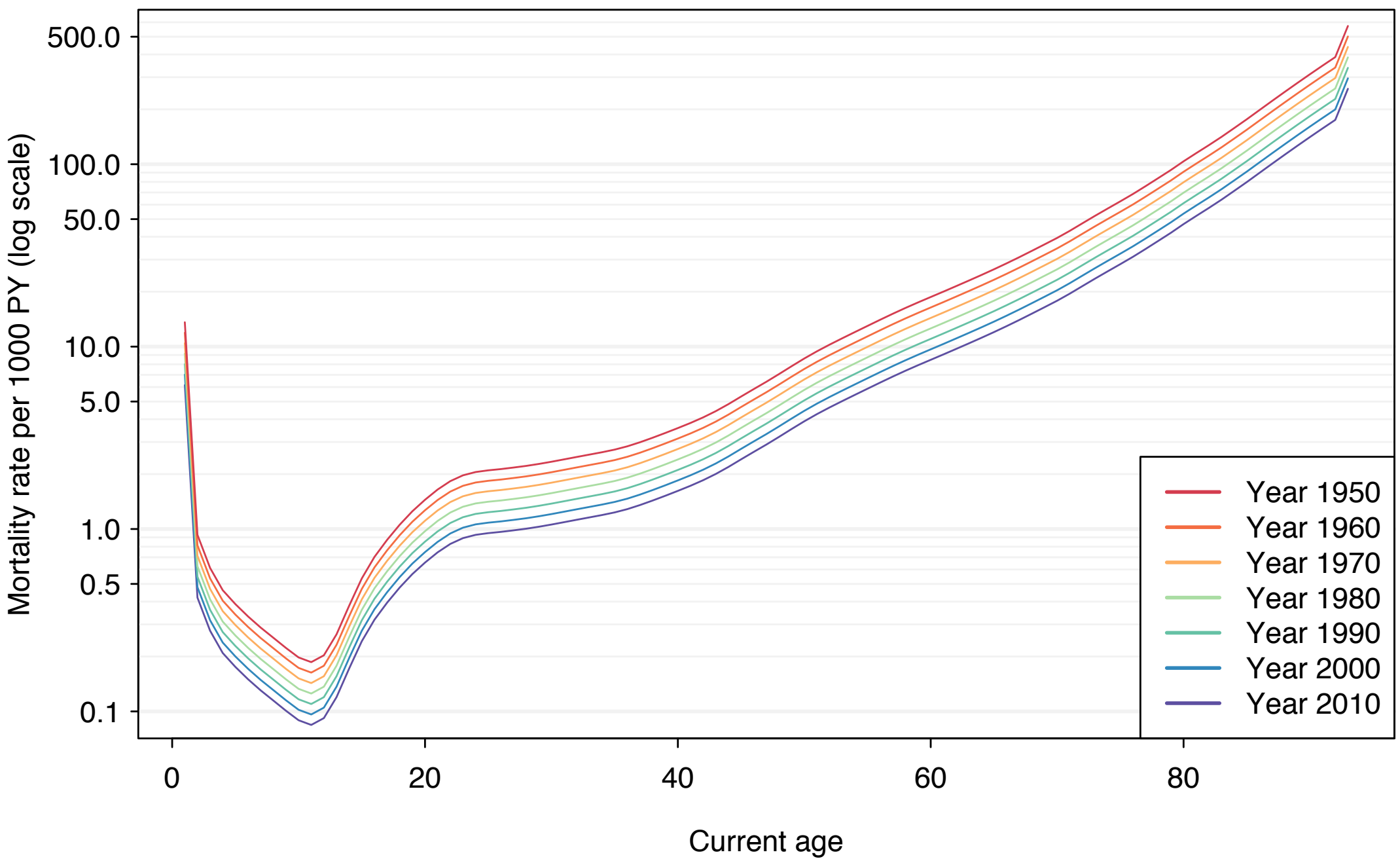

**Figure S2: Point estimates for mortality rates by age and year.**

The spline itself ($f_{age}$) was calculated as the matrix product of the spline design matrix ($M_{age}$) and the vector $\delta$:

$$f_{age} = M_{age}\delta$$

The log annual decline in mortality rates ($\theta$) was modelled with a Normal distribution with an informative prior:

$$\theta \sim Normal(-0.013, 0.0025)$$

The mortality rate for a given month and year ($g_{age,cyr}$) was calculated by combining these age and year effects:

$$g_{age,cyr} = exp\left(e_{age} + f_{age} + \theta(cyr - 2012)\right)$$

### Exit via emigration

The emigration rate was modelled with a simple four parameter spline, allowing the emigration rate to decline with increasing years since entry, leveling off at a fixed value 15 years after entry. The spline parameters $\eta_i$ were given informative priors based on US Census Bureau estimates for emigration rates (Bhaskar et al. 2013), and the last two parameters were held equal to produce a smooth transition to fixed values after 15 years:

$$\eta_1 \sim Normal(ln(0.10), 0.5)$$

$$\eta_2 \sim Normal(ln(0.02), 0.5)$$

$$\eta_3 \sim Normal(ln(0.0025), 0.5)$$

$$\eta_4 = \eta_3$$

The first 15 years of the spline ($h_{yse}$) were calculated as the matrix product of the spline design matrix ($M_{yse}$) and the vector $\eta$.

$$h_{yse} = M_{yse}\eta, \text{ for } yse \in [1, 15]$$

For individuals with >15 years since entry the emigration rate was assumed equal to the value in year 15. Figure S3 shows the implications of these assumptions.

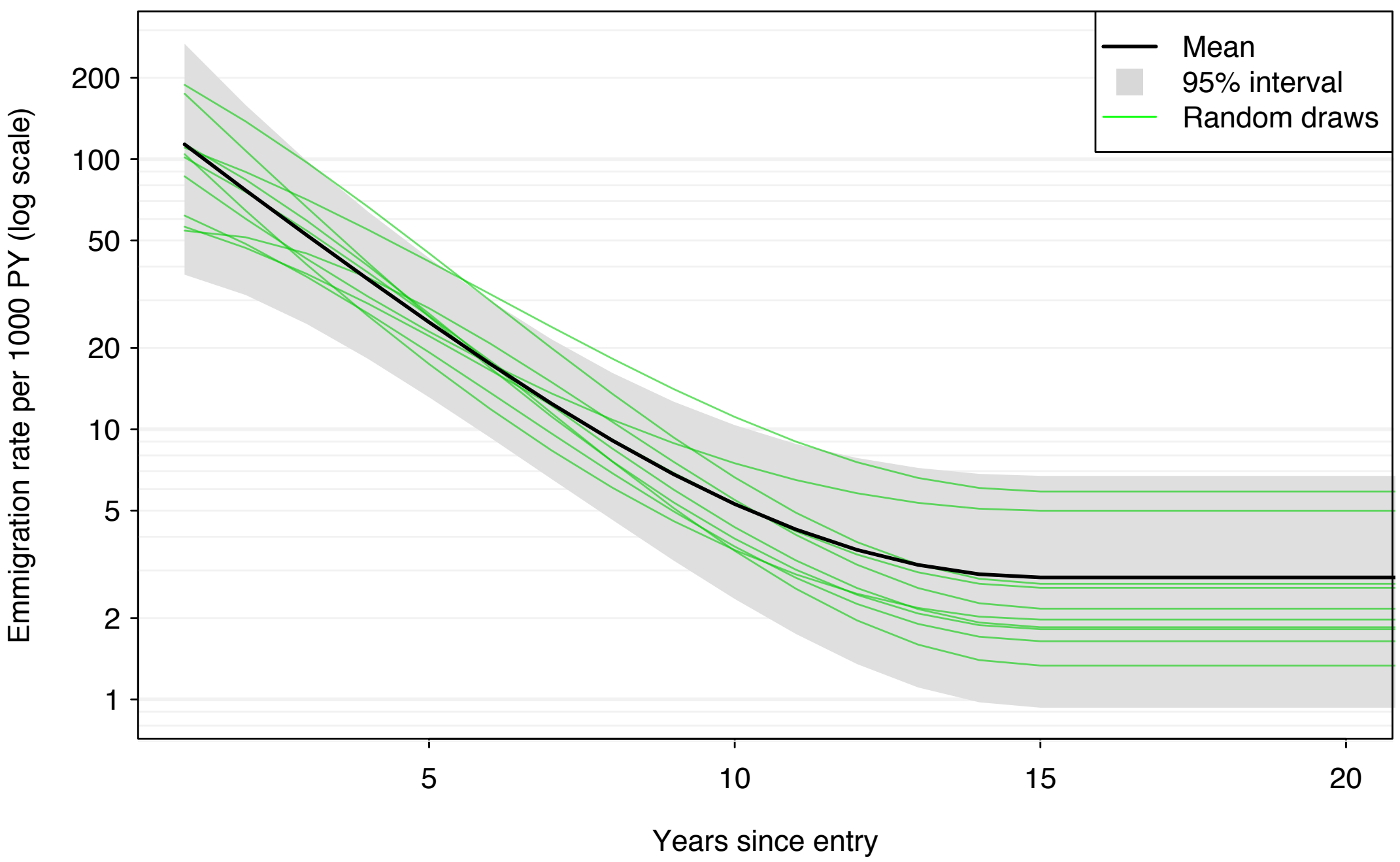

**Figure S3: Prior estimates for emigration rates by years since entry.**

**Cumulative exit probability**

Individual-level risks of emigration and mortality were assumed to be independent, as well as being independent across years. This allowed us to calculate the probability of an immigrant exiting the US resident population ($i_{aoe,yoe,yse}$) by summing mortality and emigration rates across the years since entry, then converting this cumulative exit hazard to a probability:

$$i_{age,cyr,yse} = 1 - exp\left(-\sum_{i=0}^{yse-1} g_{aoe+i,yoe+i} + h_i\right)$$

**Estimates of true population size by age, calendar year, and year of entry**

The equations described above allow an estimate to be calculated for the size of the foreign-born population by age, calendar year, and year of entry ($j_{age,cyr,yoe}$):

$$j_{age,cyr,yoe} = exp(\alpha_{yoe}) * c_{aoe,yoe} * (1 - i_{age,cyr,yse})$$

This general equation was modified for the oldest age group ($age$=91), which includes all individuals >90 years of age. As this stratum includes multiple years of age, it will be larger than the smooth age distribution represented by $c_{aoe,yoe}$. To account for this, an adjustment factor was added that multiplied the value in $c_{aoe,yoe}$. This adjustment was modelled on the logged scale, and allowed to vary smoothly over time, operationalized as a penalized b-spline with 7 control points.

The spline parameters $\rho_i$ were given an informative prior, based on US life expectancy at 90 years of age:

$$\rho_i \sim Normal(1.38, 0.25), \text{ for } i \in [1, 7]$$

This prior produces a mean adjustment factor of 4.1. The penalty function for the spline was applied to the second differences of these spline parameters:

$$\frac{d^2\rho}{di^2} = 2\rho_i - \rho_{i-1} - \rho_{i+1}, \text{ for } i \in [2, 6]$$

$$\frac{d^2\rho}{di^2} \sim Normal(0, \sigma_{\rho 2}), \text{ for } i \in [2, 6]$$

$$\sigma_{\rho 2} \sim HalfCauchy(0, 10)$$

The spline itself ($k_{yoe}$) was calculated as exponentiated value of the matrix product of the spline design matrix ($M_{yoe}$) and the vector $\rho$:

$$k_{yoe} = exp(M_{yoe}\rho)$$

**Estimates of survey population size by age, calendar year, and year of entry**

The expected value of the raw population estimates from the ACS and Census data ($\hat{\jmath}_{age,cyr,yoe}$) was assumed to differ from the true population ($j_{age,cyr,yoe}$) for several reasons.

<u>Misreporting of entry year and age</u>

The raw ACS population estimates show periodic patterns in the population distribution as a function of reported age (Figure S4, left panel) and reported entry year (Figure S4, right panel).

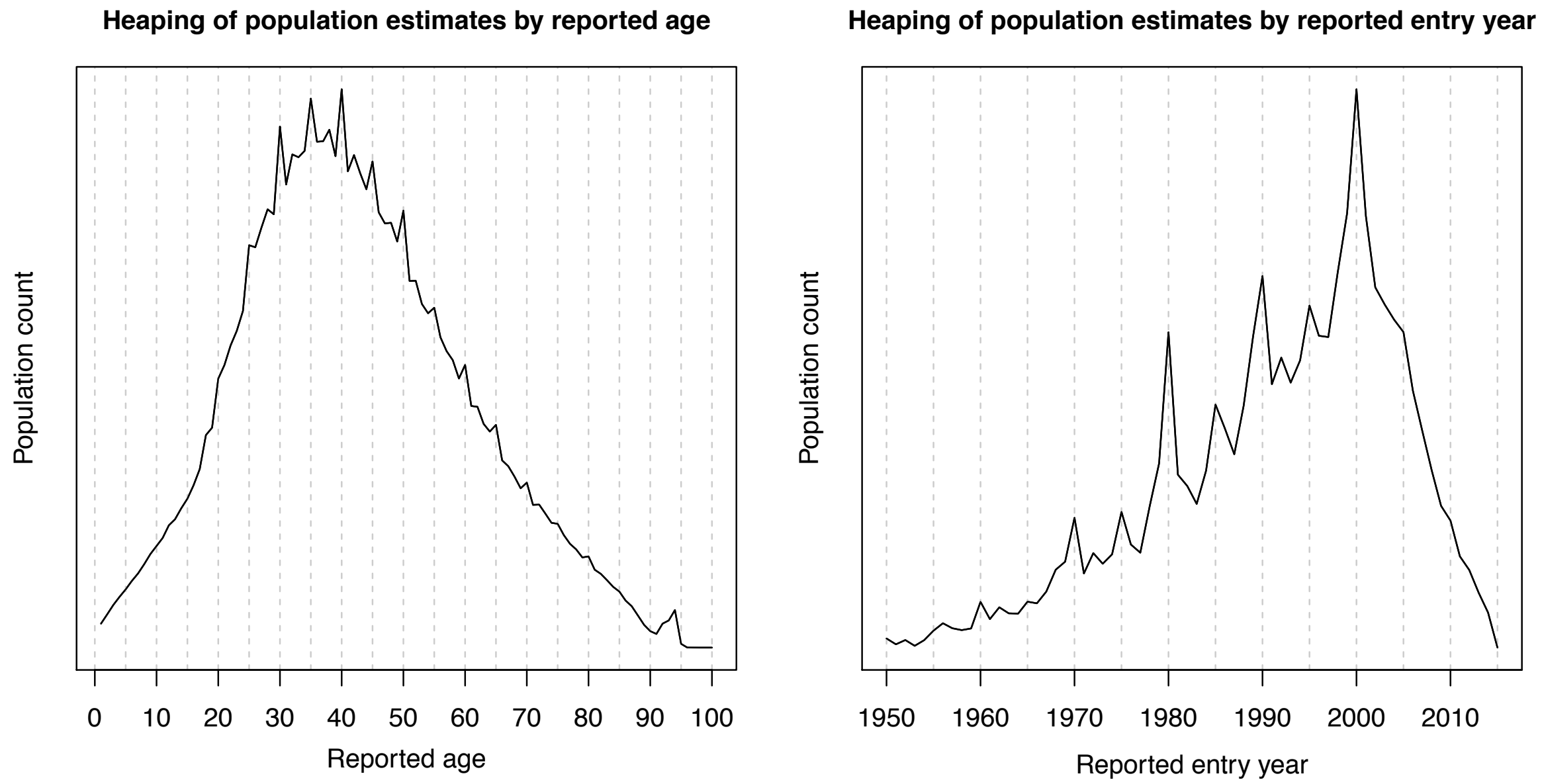

**Figure S4: Distribution of raw survey population estimates by reported age and reported entry year (all immigrants).**

These periodic effects were hypothesized to result from misreporting of responses to age and entry year questions by survey respondents, with the true value being rounded to the nearest decade or mid-decade. To correct for this misclassification, I allowed that some fraction of respondents would round their response to these questions, either to the decade or the mid-decade.

I created parameters describing the fraction of respondents that would round their year of entry to the nearest decade ($\zeta^{yoe10}$, a vector of length 7, with 1 term for each decade since 1950) and half-decade ($\zeta^{yoe5}$, a vector of length 5). Similar terms were created for the fraction of respondents that would round their current age to the nearest decade or half decade ($\zeta^{age10}$ and $\zeta^{age5}$, each of length 9). For all individuals entering within 4 years before or after a given decade, a fraction of these individuals $\zeta_i^{yoe10}$ were assumed to report their entry year as the nearest decade. For all individuals entering within 2 years before or after a given decade, a fraction of these individuals $\zeta_i^{yoe5}$ were assumed to report their entry year as the nearest half-decade. The same approach was used to reassign survey responses for current age. Prior distributions for these fractions were created so that values would only be reassigned if there were clear evidence for this in the survey data. This was done by giving informative priors to the hyper-parameters $\sigma_{\zeta^{yoe}}$ and $\sigma_{\zeta^{age}}$, pulling them towards zero. An additional constraint was added so that each fraction would be restricted to the range [0, 0.25], so that the sum of these values for any individual in the population could never be greater than 1.0.

$$\zeta_i^{yoe10} \sim HalfNormal\left(0, \sigma_{\zeta^{yoe}}\right), \text{ for } i \in [1,7]$$

$$\zeta_i^{yoe5} \sim HalfNormal\left(0, \sigma_{\zeta^{yoe}}\right), \text{ for } i \in [1,6]$$

$$\sigma_{\zeta^{age}} \sim HalfCauchy(0, 0.05)$$

$$\zeta_i^{age10} \sim HalfNormal\left(0, \sigma_{\zeta^{age}}\right), \text{ for } i \in [1,9]$$

$$\zeta_i^{age5} \sim HalfNormal\left(0, \sigma_{\zeta^{ag}}\right), \text{ for } i \in [1,9]$$

$$\sigma_{\zeta^{yoe}} \sim HalfCauchy(0, 0.05)$$

$$0 \le \zeta^{yoe10}, \zeta^{yoe5}, \zeta^{age10}, \zeta^{age5} \le 0.25$$

Undercoverage of foreign-born individuals in the survey data

Prior studies have described undercount of the foreign-born population in the ACS. A function was specified to operationalize these findings, allowing the extent of undercount to vary by country or region of origin, survey year, and years since entry. These effects were operationalized as inflation factors, defined as the ratio of true population to survey population, such that values greater than 1.0 indicate undercounting. It was assumed that survey year effects were proportional to the inverse square root of the sample size of each survey on the log scale (Figure S5), consistent with substantial undercount in the early years of the ACS, followed by gradual improvements from 2005 onwards, and minimal undercounting in the 2000 census. Each of these survey year terms ($\kappa_{svy}$) was given a Normal prior with standard deviation of 0.5.

$$\kappa_{svy} \sim Normal\left(mu_{\kappa_{svy}}, 0.5\right), \text{ for } svy \in [2000, 2016]$$

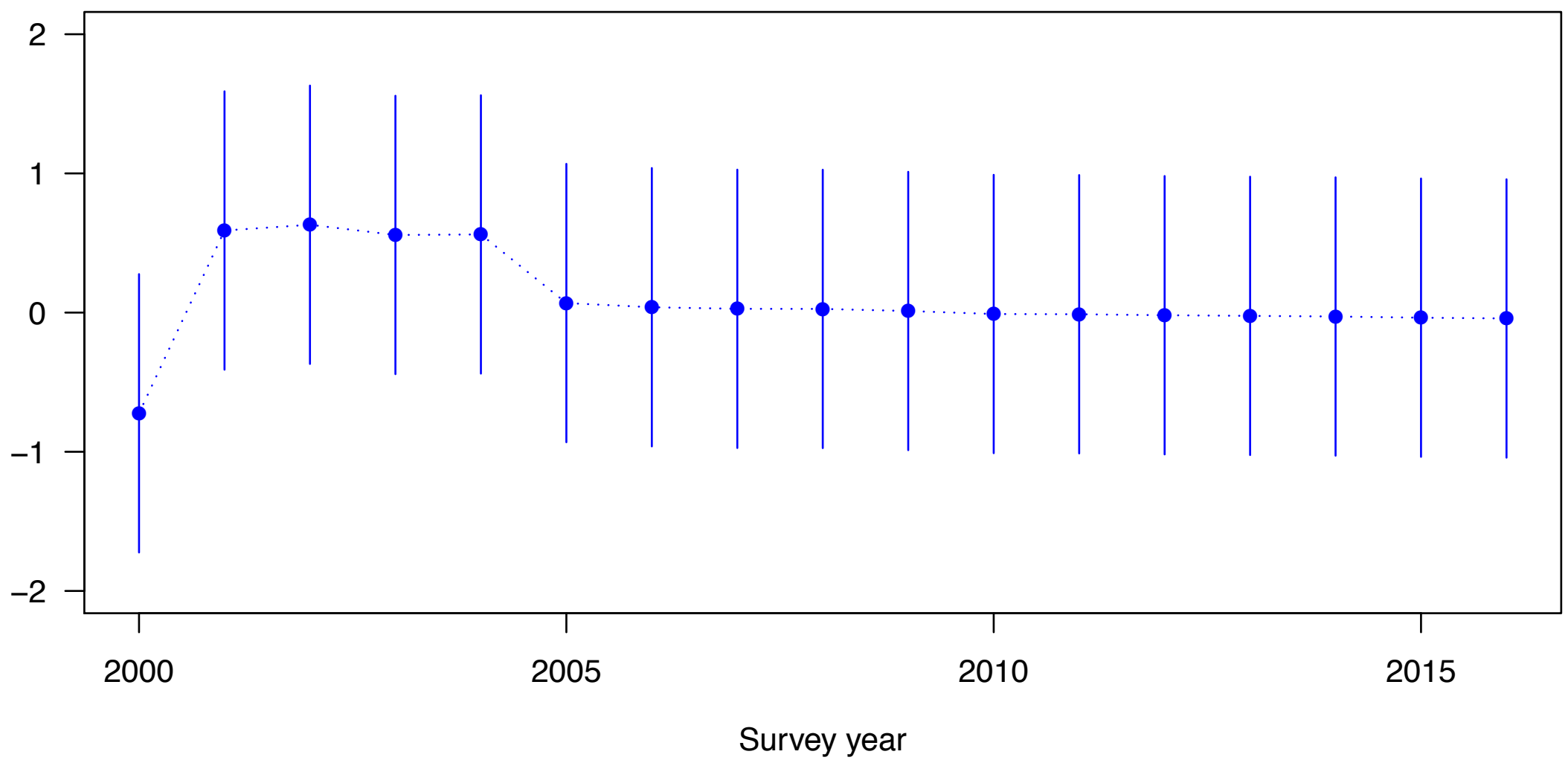

**Figure S5: Prior distribution for κ, with point estimates and 95% intervals**

For years since entry, the change in the extent of undercount was modelled using a simple four-parameter spline, allowing this term to decline with increasing years since entry, leveling off at a fixed value 15 years after entry. The spline parameters $\phi_i$ were given informative priors, and the last two parameters were held equal to produce a smooth transition to fixed values after 15 years, similar to the spline used to model the emigration rate:

$$\phi_1 \sim Normal(ln(0.10), 0.5)$$

$$\phi_2 \sim Normal(ln(0.025), 0.5)$$

$$\phi_3 \sim Normal(ln(0.005), 0.5)$$

$$\phi_4 = \phi_3$$

The first 15 years of the spline ($l_{yse}$) were calculated as the matrix product of the spline design matrix ($M_{yse}$) and the vector $\phi$.

$$l_{yse} = M_{yse}\phi, \text{ for } yse \in [1, 15]$$

For individuals with >15 years since entry the undercount adjustment was set equal to the value in year 15. The overall inflation factor ($m_{cyr,yse}$) was calculated by combining effects for survey year and years since entry:

$$m_{cyr,yse} = 1 + exp(\kappa_{svy} + l_{yse})$$

Finally, the overall undercount for the period 2005-2016 was calibrated to produce an average undercount rate of 5.0% for countries in Latin America and the Caribbean, and 2.0% for countries outside this region, consistent with recent Census Bureau estimates (Jensen, Bhaskar, and Scopilliti 2015).

<u>Partial observation of individuals entering the country during a survey year</u>

Individuals entering the country in a given survey year (ie $yse = 0$) will be underrepresented in the survey results, as data collection is conducted part way through the calendar year, potentially before an individual enters the country. As the ACS is conducted continuously throughout the year, individuals entering the country in a given year will have an approximately 50% chance of being included in the survey round, compared to individuals residing in the country for the full 12 month period. For this reason the population estimates for individuals with $yse = 0$ were multiplied by 0.5, for all years from 2001 onwards. For the year 2000, where data from the 2000 census were used, the population estimates for individuals with $yse = 0$ b$\kappa_{svy}$ was multiplied by 0.25. As the 2000 census was conducted at the start of April, and assuming an approximately constant immigration rate throughout the year, this implies someone entering the country in 2000 would have a 25% chance of being included in the survey relative to someone resident in the country for the full 12-month period.

**Likelihood for survey data**

I used a two-part likelihood for the raw survey estimates. These raw estimates ($y_{age,cyr,yoe}$) were created by summing survey analysis weights for all individuals in a given population stratum and survey year. The population estimates for some strata were zero, indicating no individuals with those characteristics were included in the survey in that year. In the first part of the likelihood, I modelled the probability that a stratum would have a non-zero value ($p_{age,cyr,yoe}$), as a function of the unobserved true population size (already adjusted for the issues described in Section 9), using a Bernoulli likelihood function:

$$p_{age,cyr,yoe} = 1 - exp\left(-\hat{\jmath}_{age,cyr,yoe} * o_{age,cyr,yoe} * \lambda\right)$$

$$\mathrm{I}\left(y_{age,cyr,yoe} > 0\right) \sim Bernoulli\left(p_{age,cyr,yoe}\right)$$

In the equation for $p_{age,cyr,yoe}$, $o_{age,cyr,yoe}$ is the inverse of the average analysis weight calculated from the survey data, such that $\hat{\jmath}_{age,cyr,yoe} * o_{age,cyr,yoe}$ approximates the expected number of individuals from that stratum to be included in the survey. $\lambda$ was an additional term allowing for unmodeled features of the survey design which would lead to over- or under-estimates of the observation probability. $\lambda$ was given a relatively loose prior, centered on 1.0.

$$\lambda \sim Gamma(5,5)$$

In the second part of the likelihood, the non-zero observations were modelled using a Normal likelihood function:

$$y_{age,cyr,yoe} \sim Normal\left(\frac{\hat{J}_{age,cyr,yoe}}{p_{age,cyr,yoe}}, \sigma_y\right) \ \ for\ y_{age,cyr,yoe} > 0$$

$$\sigma_y \sim HalfCauchy(0,1000)$$

The derivation of $\frac{\hat{J}_{age,cyr,yoe}}{p_{age,cyr,yoe}}$ as the expected value of $y_{age,cyr,yoe}$ is as follows:

$$\hat{J}_{age,cyr,yoe} = E\left(y_{age,cyr,yoe}\right)$$

$$= p\left(y_{age,cyr,yoe}\right) * E\left(y_{age,cyr,yoe} \mid y_{age,cyr,yoe} > 0\right)$$

$$+ \left(1 - p\left(y_{age,cyr,yoe}\right)\right) * E\left(y_{age,cyr,yoe} \mid y_{age,cyr,yoe} = 0\right)$$

$$= p\left(y_{age,cyr,yoe}\right) * E\left(y_{age,cyr,yoe} \mid y_{age,cyr,yoe} > 0\right) + 0$$

$$\Rightarrow E\left(y_{age,cyr,yoe} \mid y_{age,cyr,yoe} > 0\right) = \frac{\hat{J}_{age,cyr,yoe}}{p\left(y_{age,cyr,yoe}\right)}$$

**Citations for supplement**

## Description of data structure for population and immigration estimates

Two csv files accompany this paper. The first of these contains the population estimates, while the second contains the immigration estimates.

**Population estimates**: the file "fb_pop_estimates_6-1-2019.csv" includes a table with the following column names:

*survey_year* – survey year of estimate

*current_age* – single year of age, for stratum

*entry_year* – entry year, for stratum

*ALL* – population estimates for all immigrants

*WBregion_EAP… WBregion_SSA* – population estimates for each World Bank region

*AFG… ZAF* – population estimates for the top 100 countries, ordered by ISO3 code

**Immigration estimates**: the file "fb_immig_estimates_6-1-2019.csv" includes a table with the following column names:

*entry_year* – entry year, for stratum

***ALL*** – annual immigration estimates for all immigrants

*WBregion_EAP… WBregion_SSA* – annual immigration estimates for each World Bank region

*AFG… ZAF* – annual immigration estimates for the top 100 countries, ordered by ISO3 code

## Out-of-sample predictive performance for Fiji, Mexico, Pakistan, Peru, Poland, Somalia, and Viet Nam in 2005, 2010 and 2015.

Figures S6-1 to S6-21 present results for out-of-sample predictive performance for seven example countries and 3 years. Each graph contains 5 panels:

*Panel A*: Log-log plot of observed versus predicted population estimates for individual strata.

*Panel B*: Histogram of standardized residuals (yellow) plotted against a unit Normal reference distribution (blue).

*Panel C*: Observed frequency of a stratum having population>0 plotted against the modeled probability of this outcome.

*Panel D*: Out-of-sample model estimates for population distribution by year of entry, plotted against raw survey estimates.

*Panel E*: Out-of-sample model estimates for population distribution by age at entry, plotted against raw survey estimates.

*Panel F*: Out-of-sample model estimates for population distribution by current age, plotted against raw survey estimates.

**Figure S6A-1: Out-of-sample predictive performance for Fiji, 2005.**

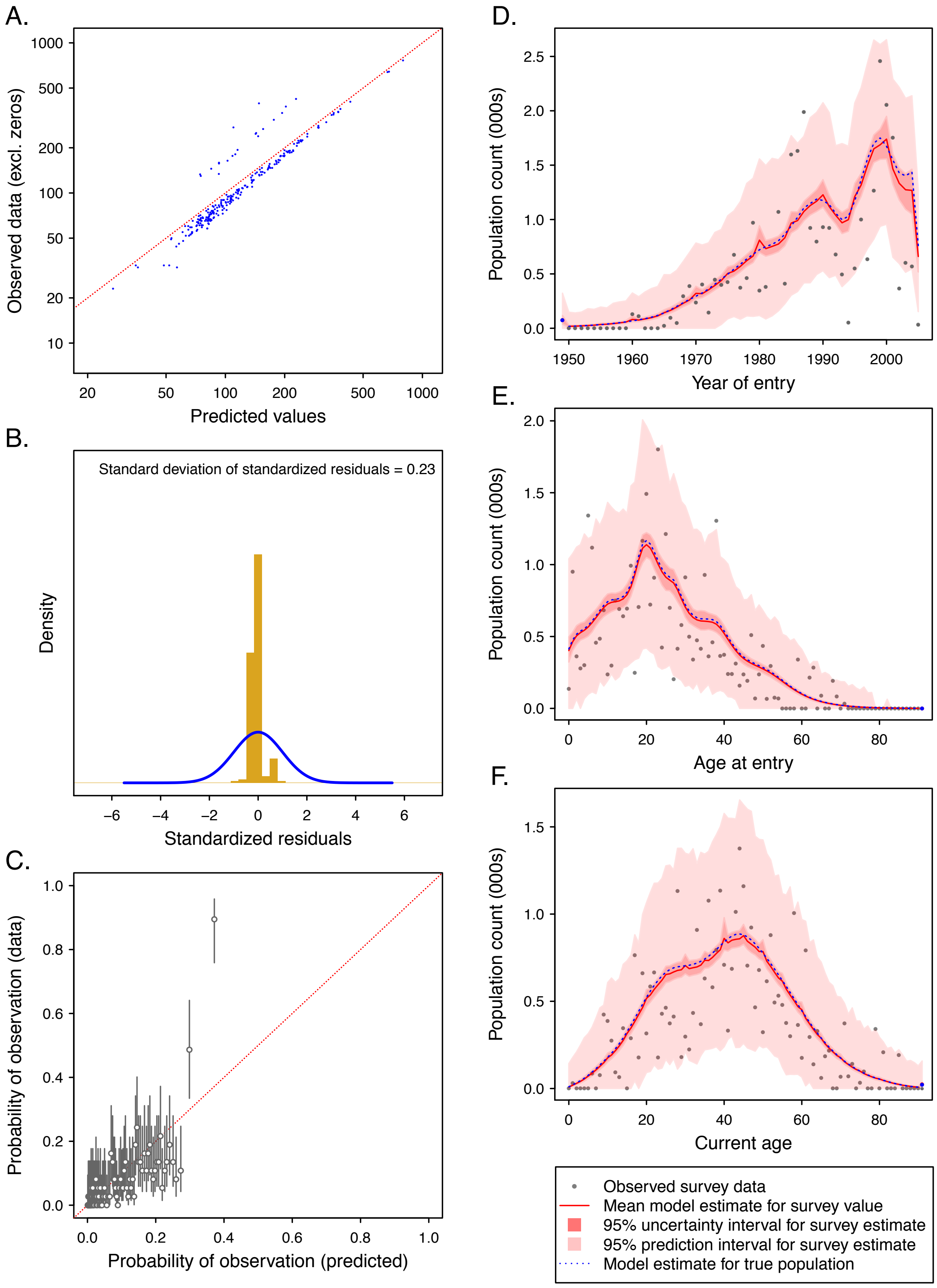

**Figure S6-2: Out-of-sample predictive performance for Fiji, 2010.**

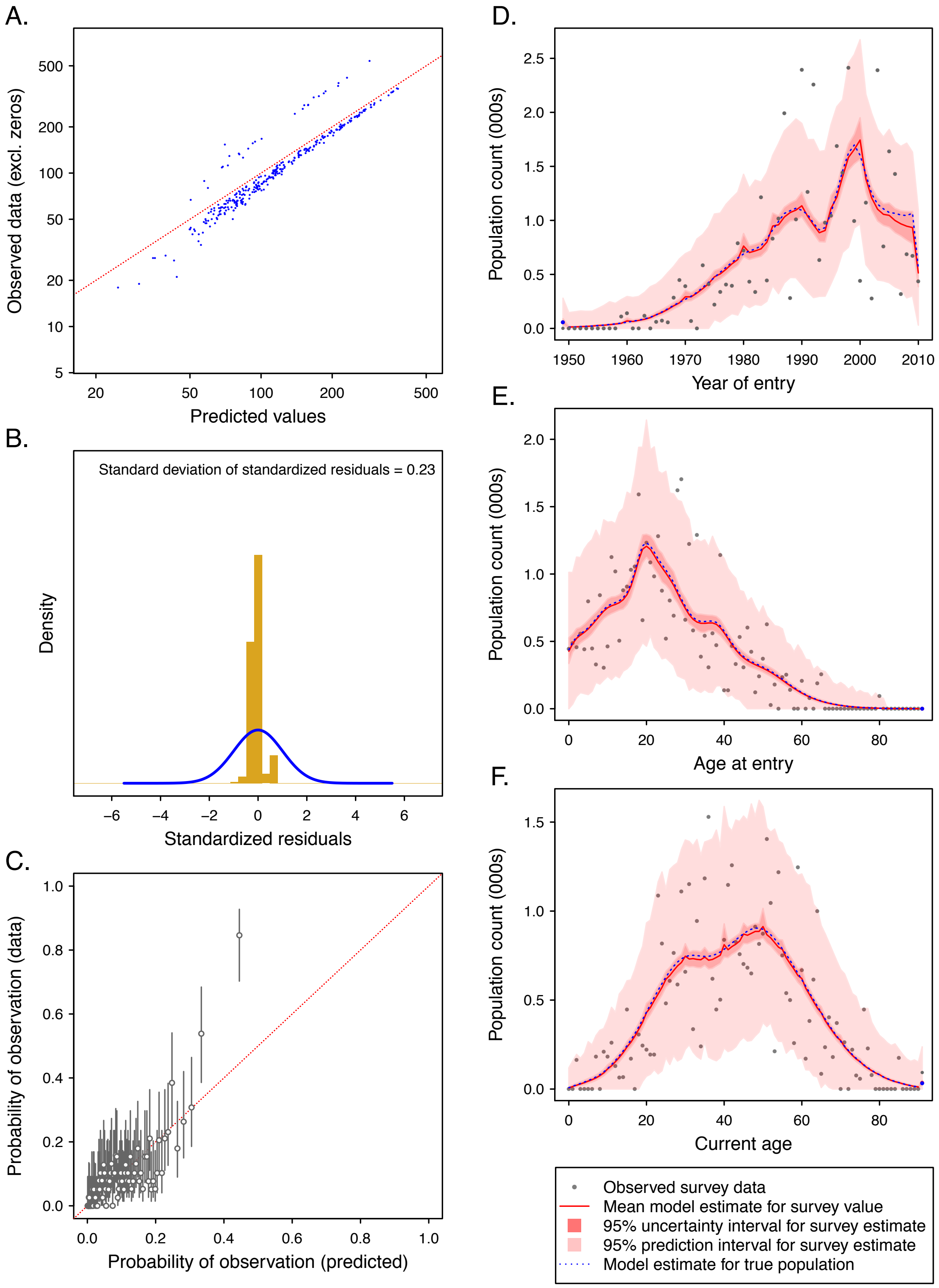

**Figure S6-3: Out-of-sample predictive performance for Fiji, 2015.**

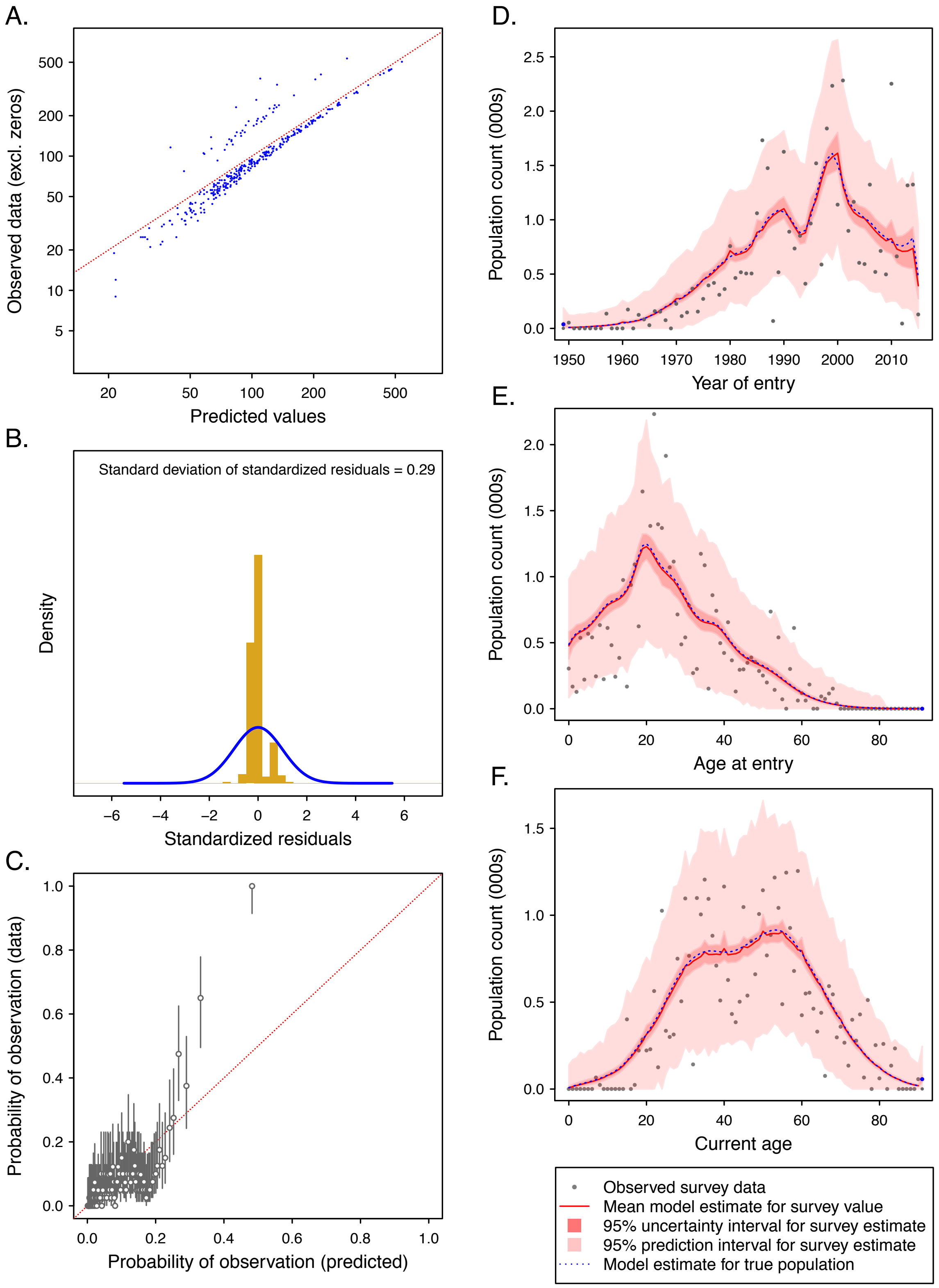

**Figure S6-4: Out-of-sample predictive performance for Mexico, 2005.**

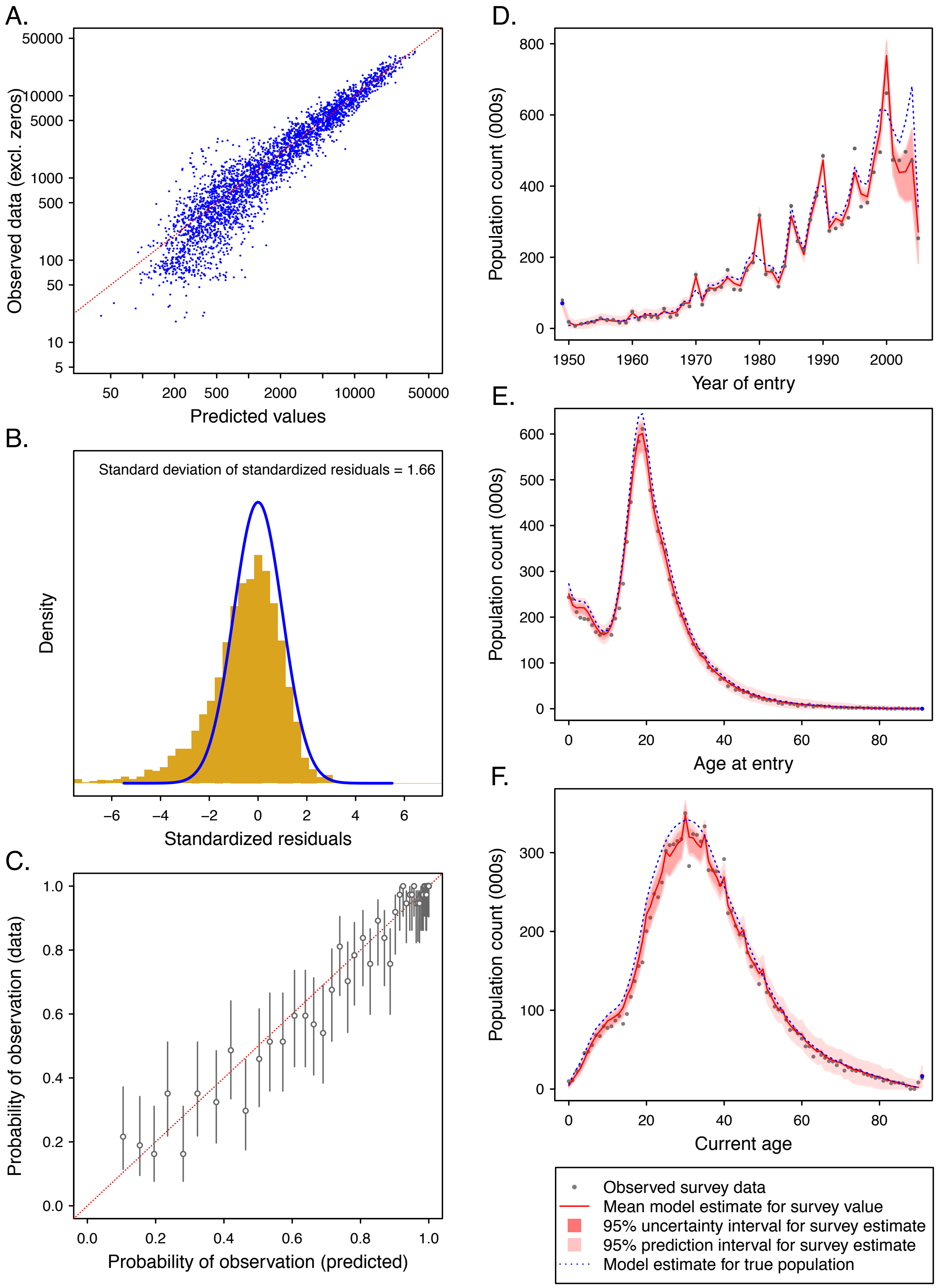

**Figure S6-5: Out-of-sample predictive performance for Mexico, 2010.**

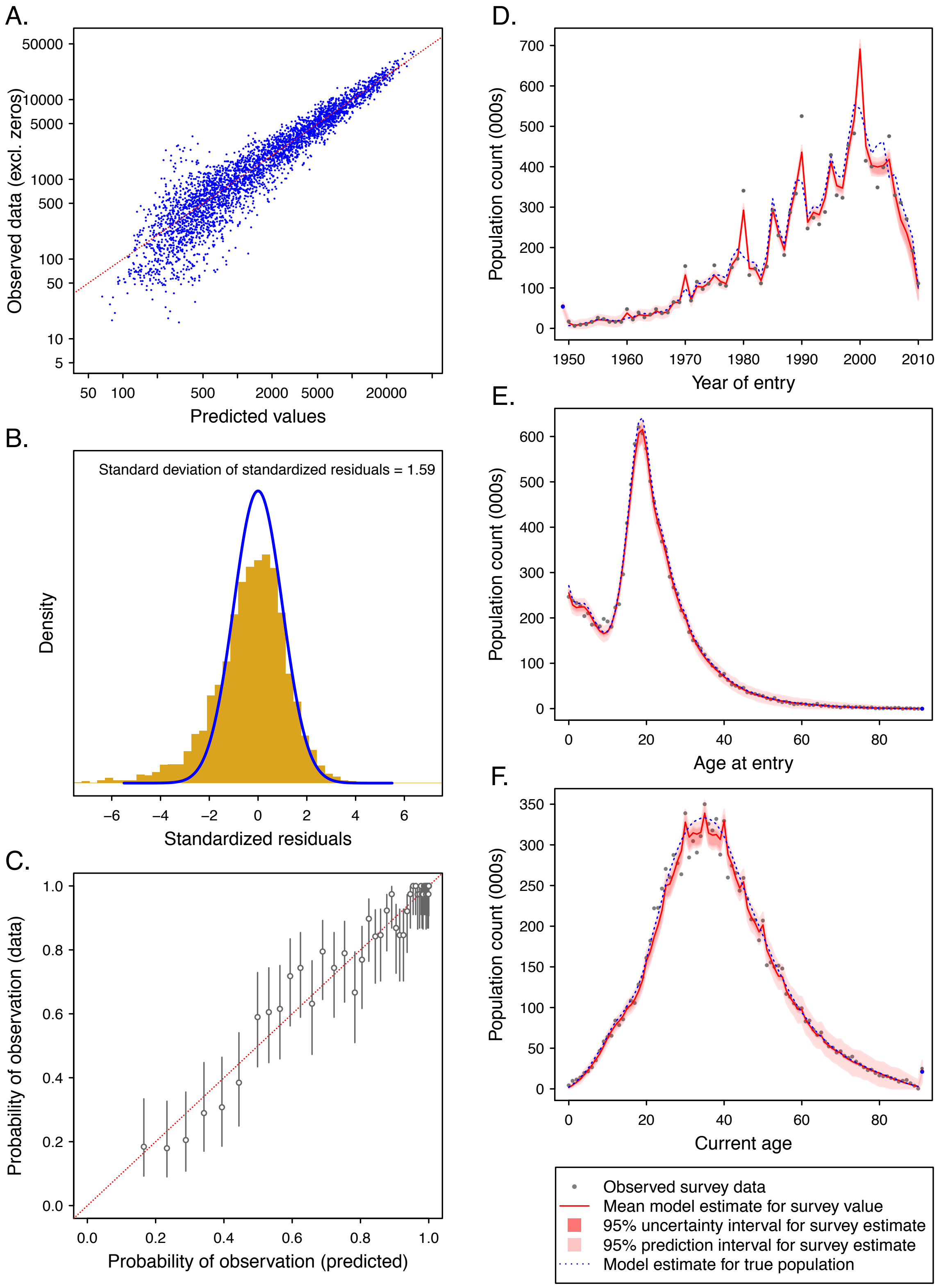

**Figure S6-6: Out-of-sample predictive performance for Mexico, 2015.**

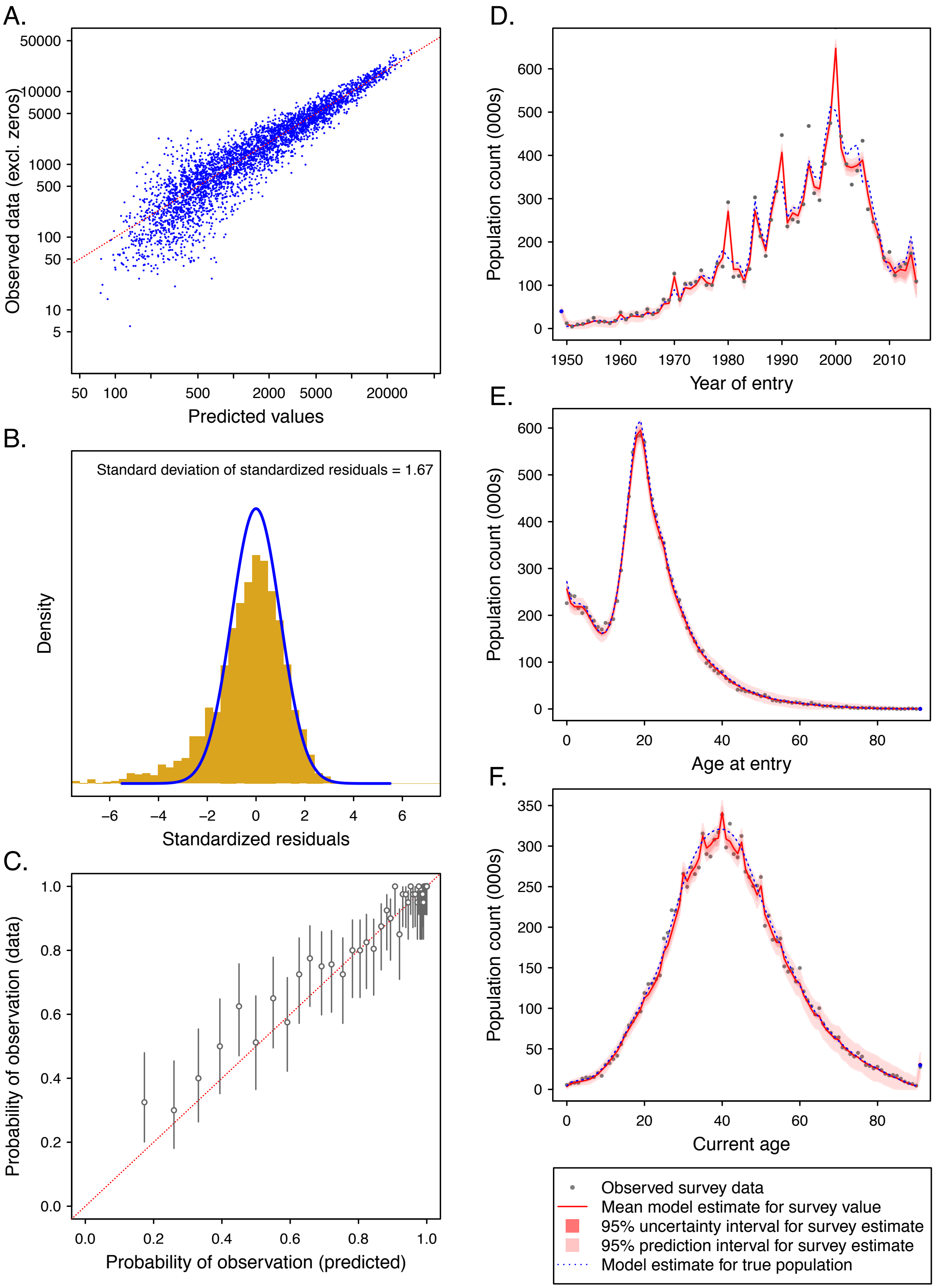

**Figure S6-7: Out-of-sample predictive performance for Pakistan, 2005.**

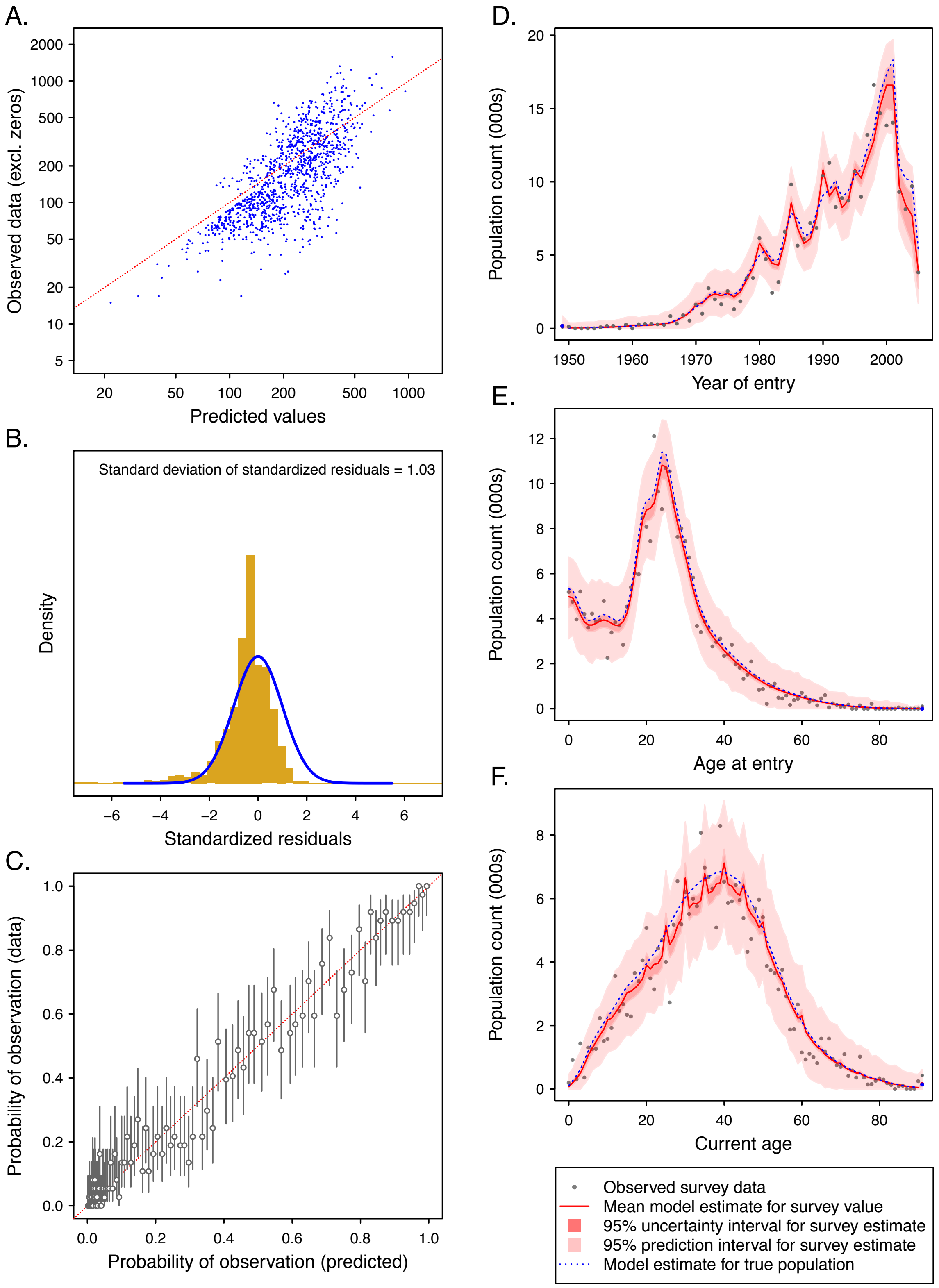

**Figure S6-8: Out-of-sample predictive performance for Pakistan, 2010.**

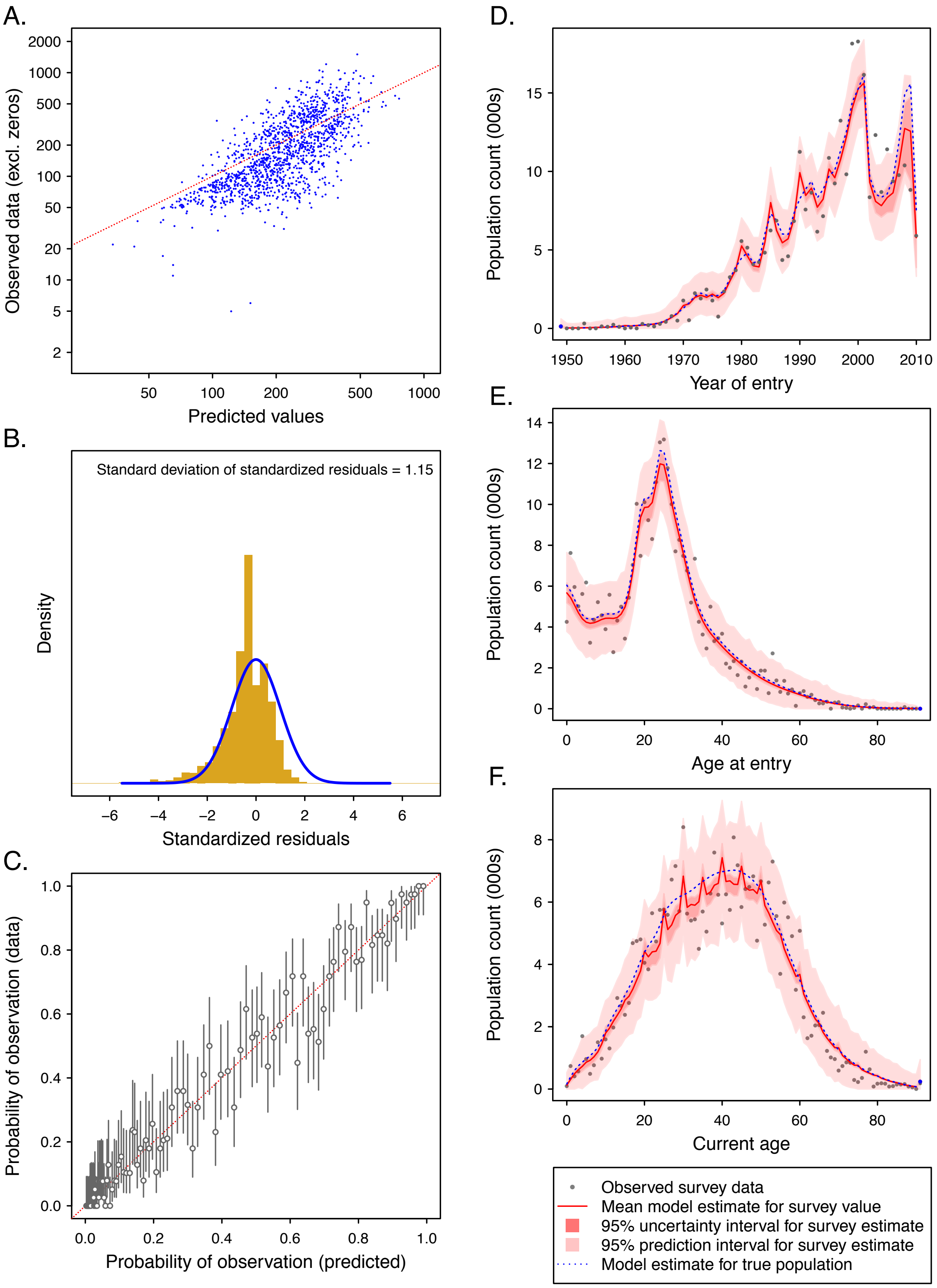

**Figure S6-9: Out-of-sample predictive performance for Pakistan, 2015.**

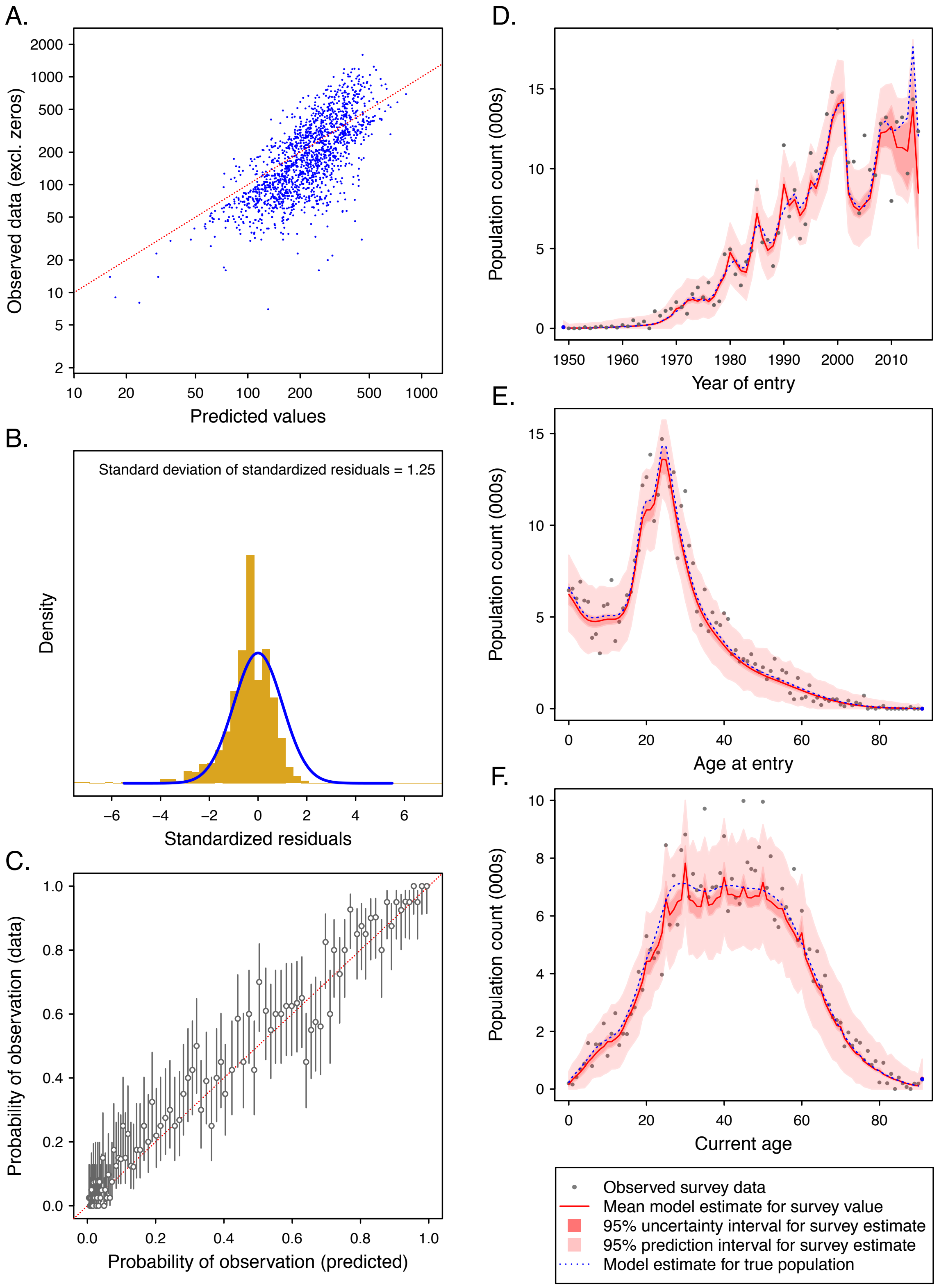

**Figure S6-10: Out-of-sample predictive performance for Peru, 2005.**

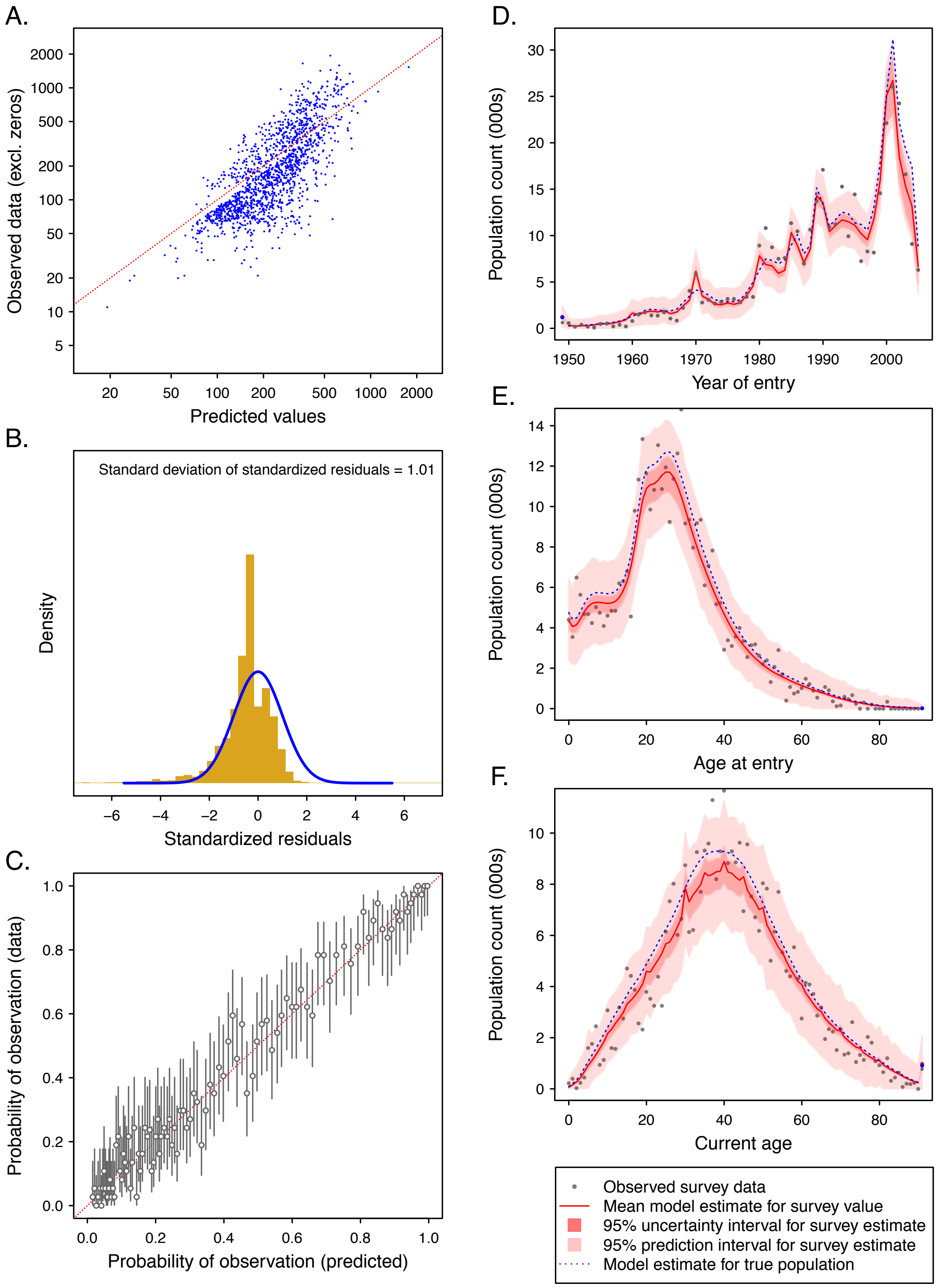

**Figure S6-11: Out-of-sample predictive performance for Peru, 2010.**

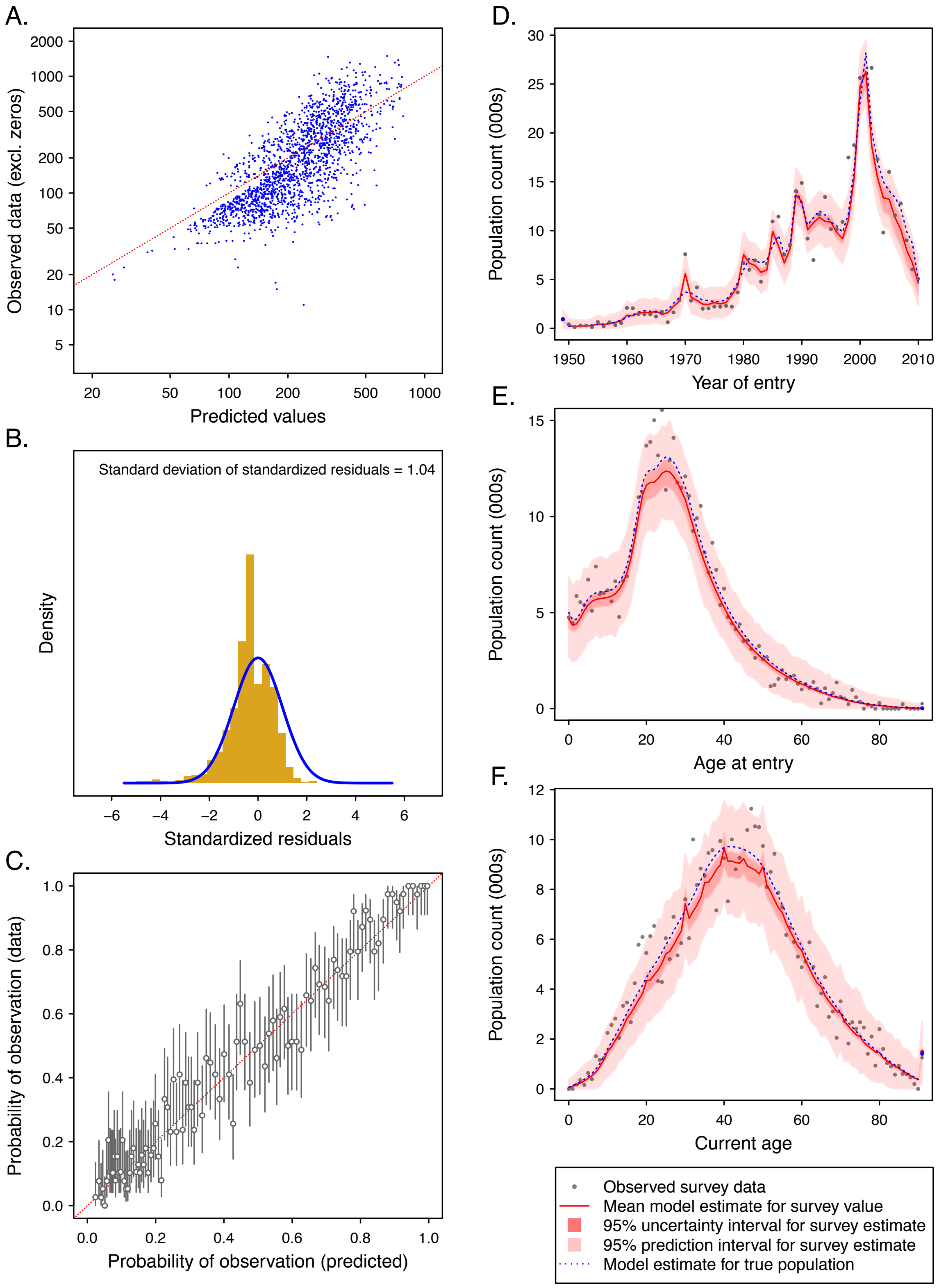

**Figure S6-12: Out-of-sample predictive performance for Peru, 2015.**

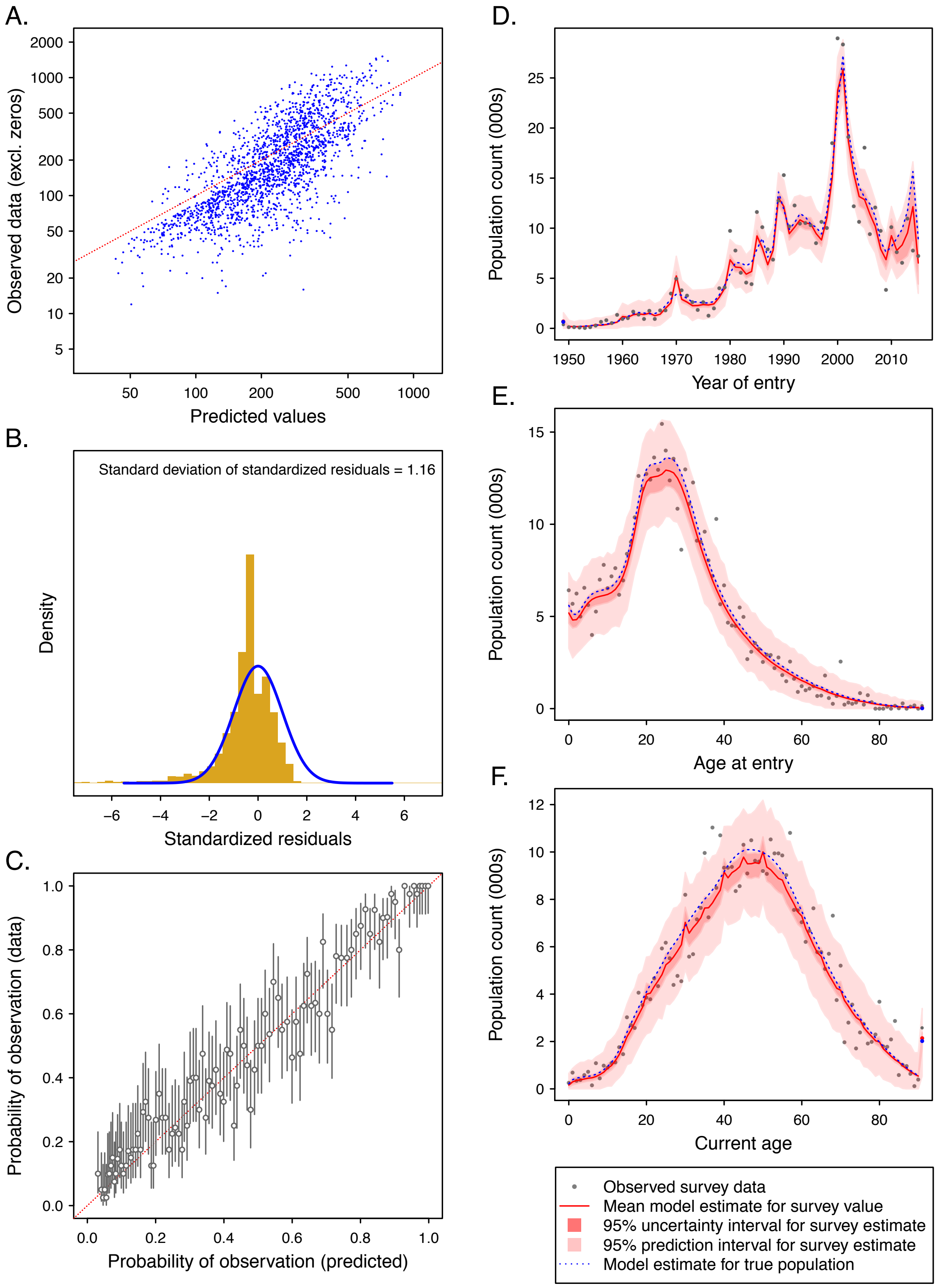

**Figure S6-13: Out-of-sample predictive performance for Poland, 2005.**

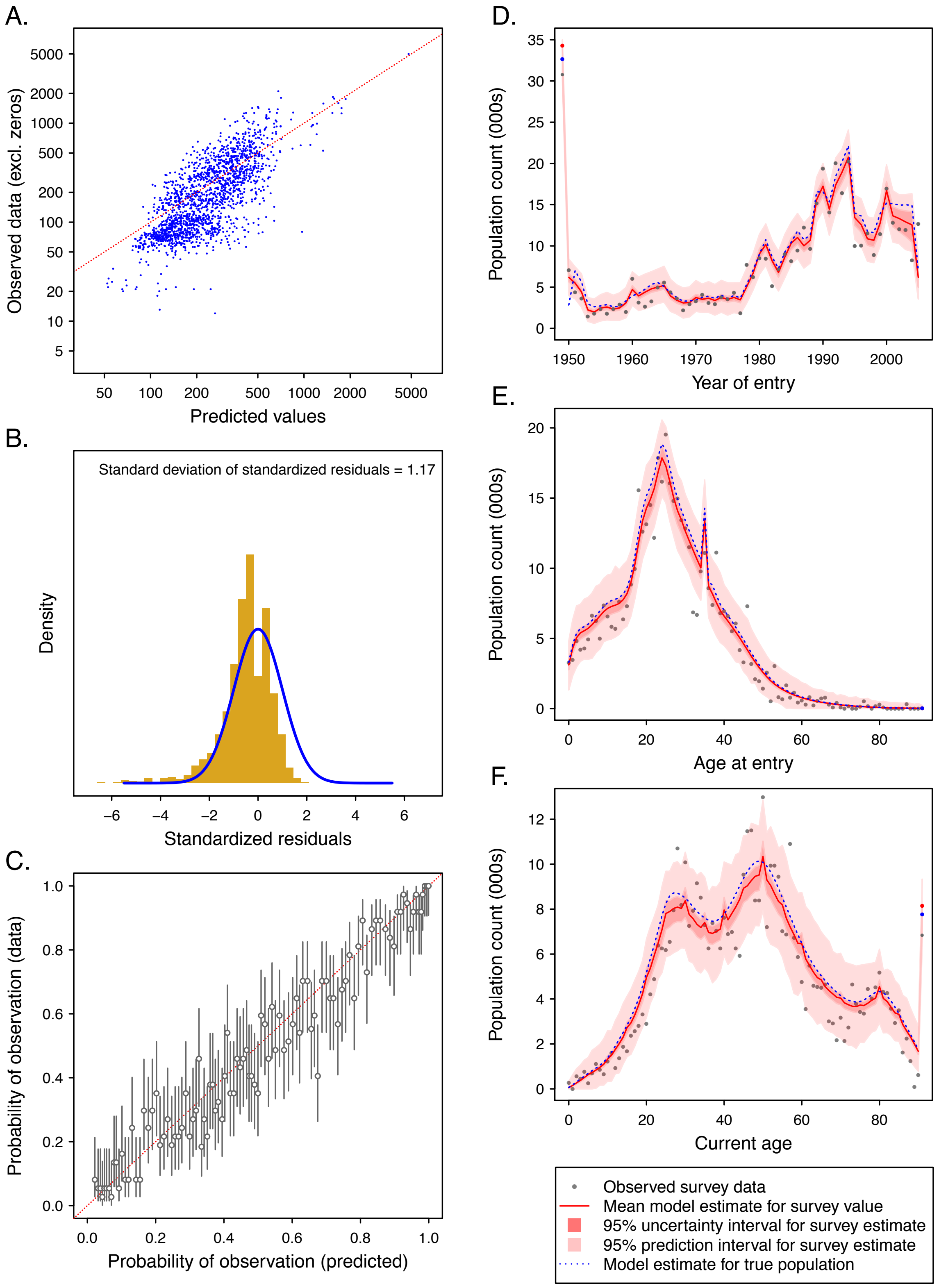

**Figure S6-14: Out-of-sample predictive performance for Poland, 2010.**

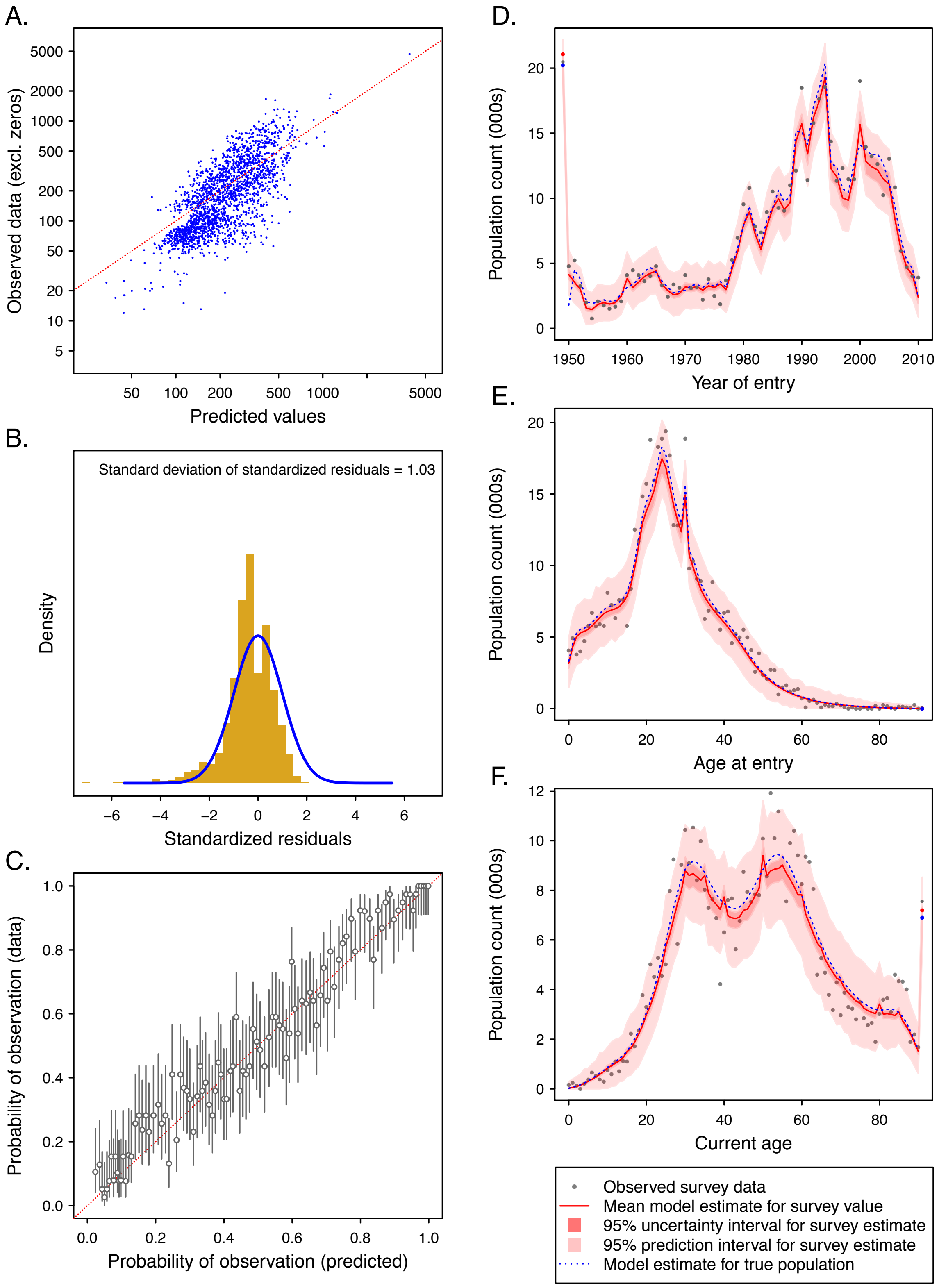

**Figure S6-15: Out-of-sample predictive performance for Poland, 2015.**

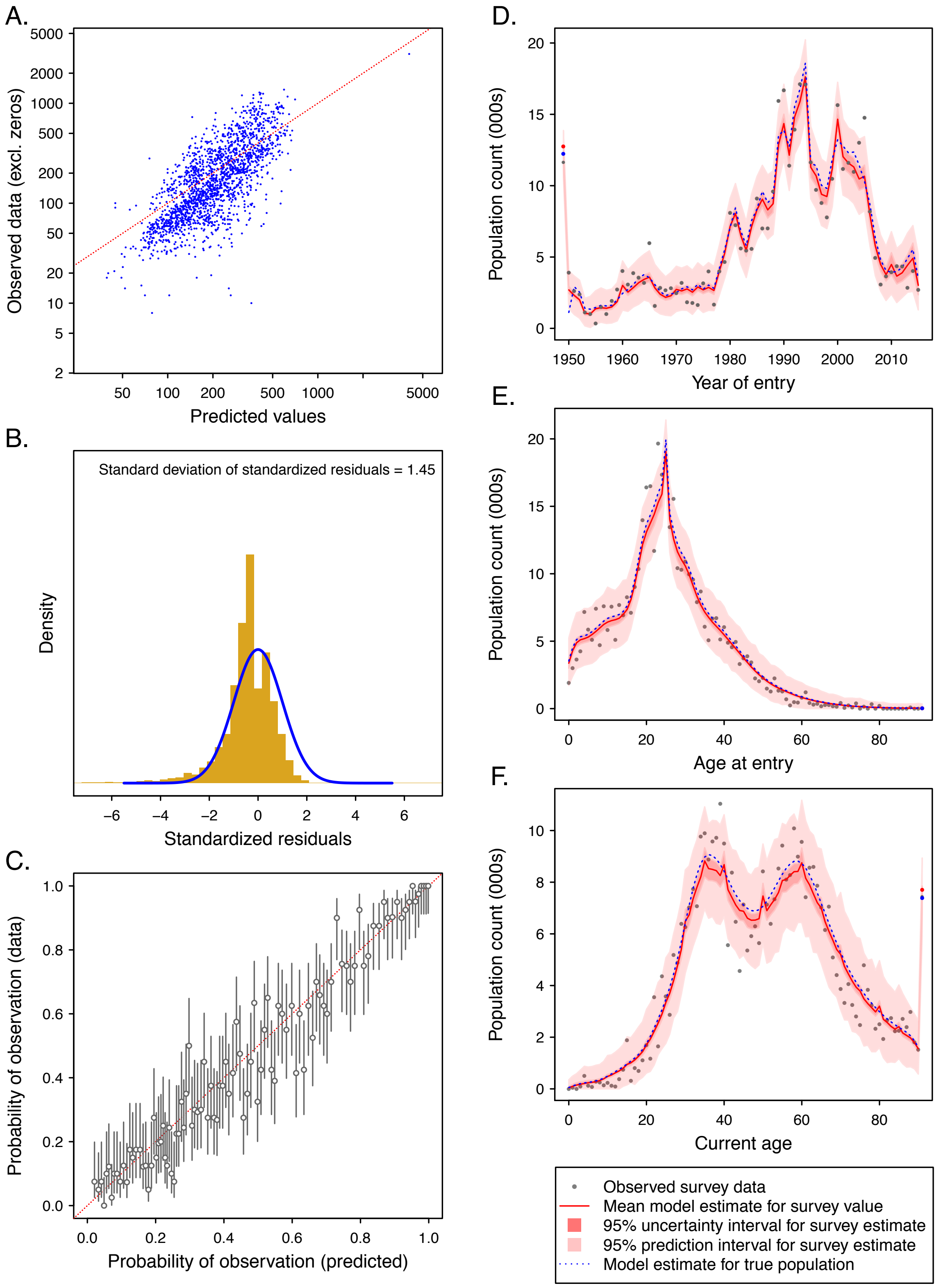

**Figure S6-16: Out-of-sample predictive performance for Somalia, 2005.**

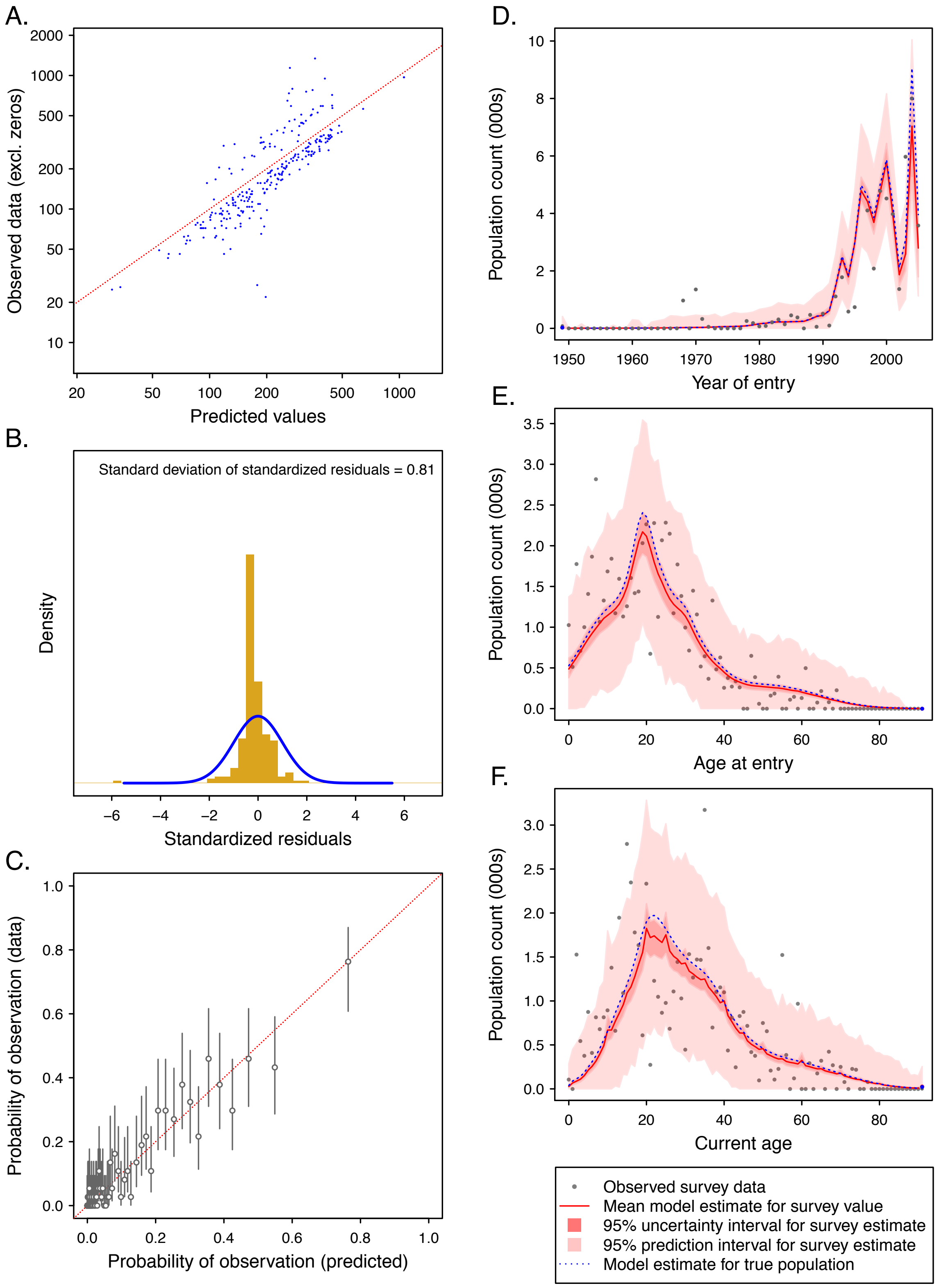

**Figure S6-17: Out-of-sample predictive performance for Somalia, 2010.**

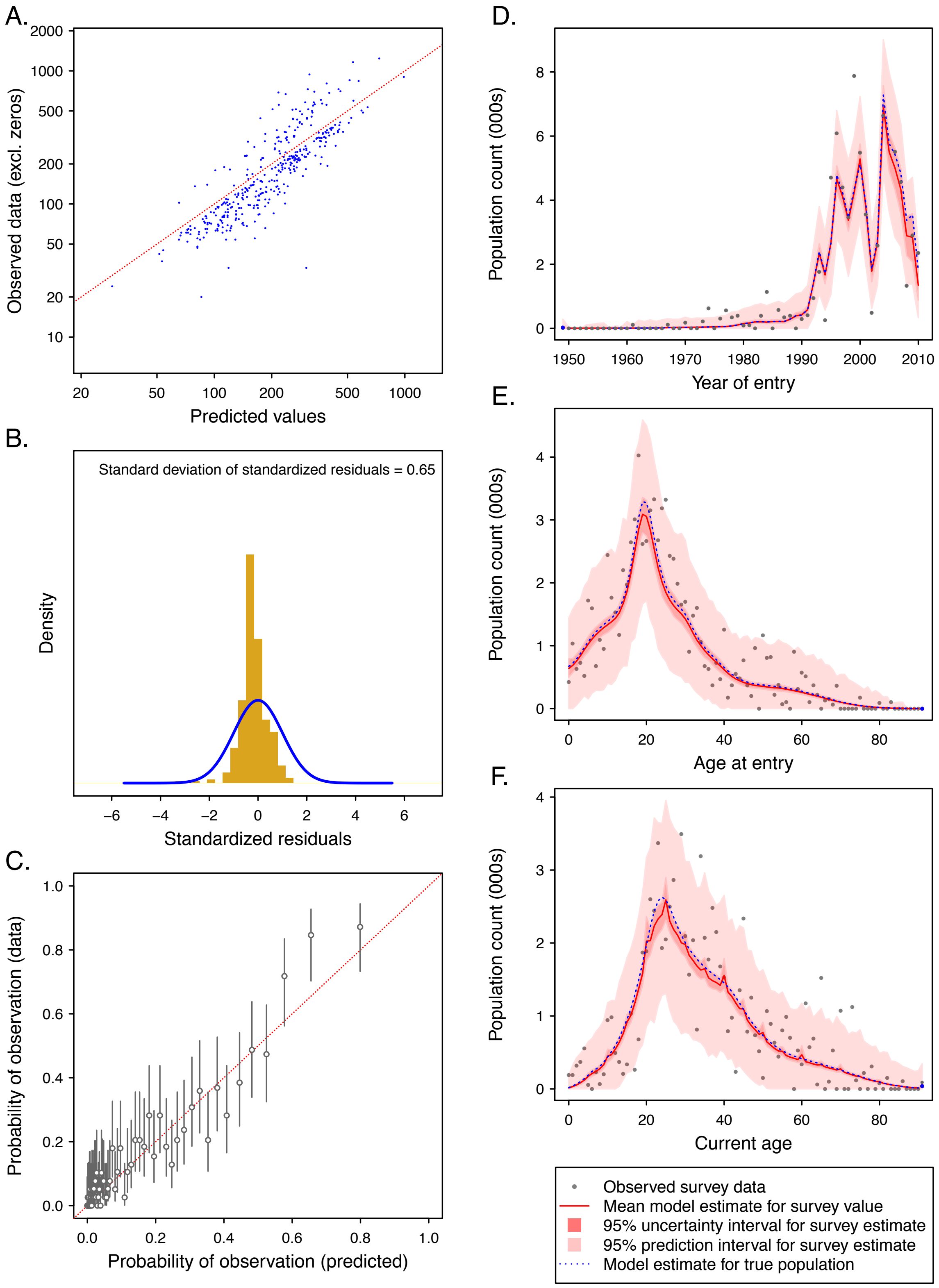

**Figure S6-18: Out-of-sample predictive performance for Somalia, 2015.**

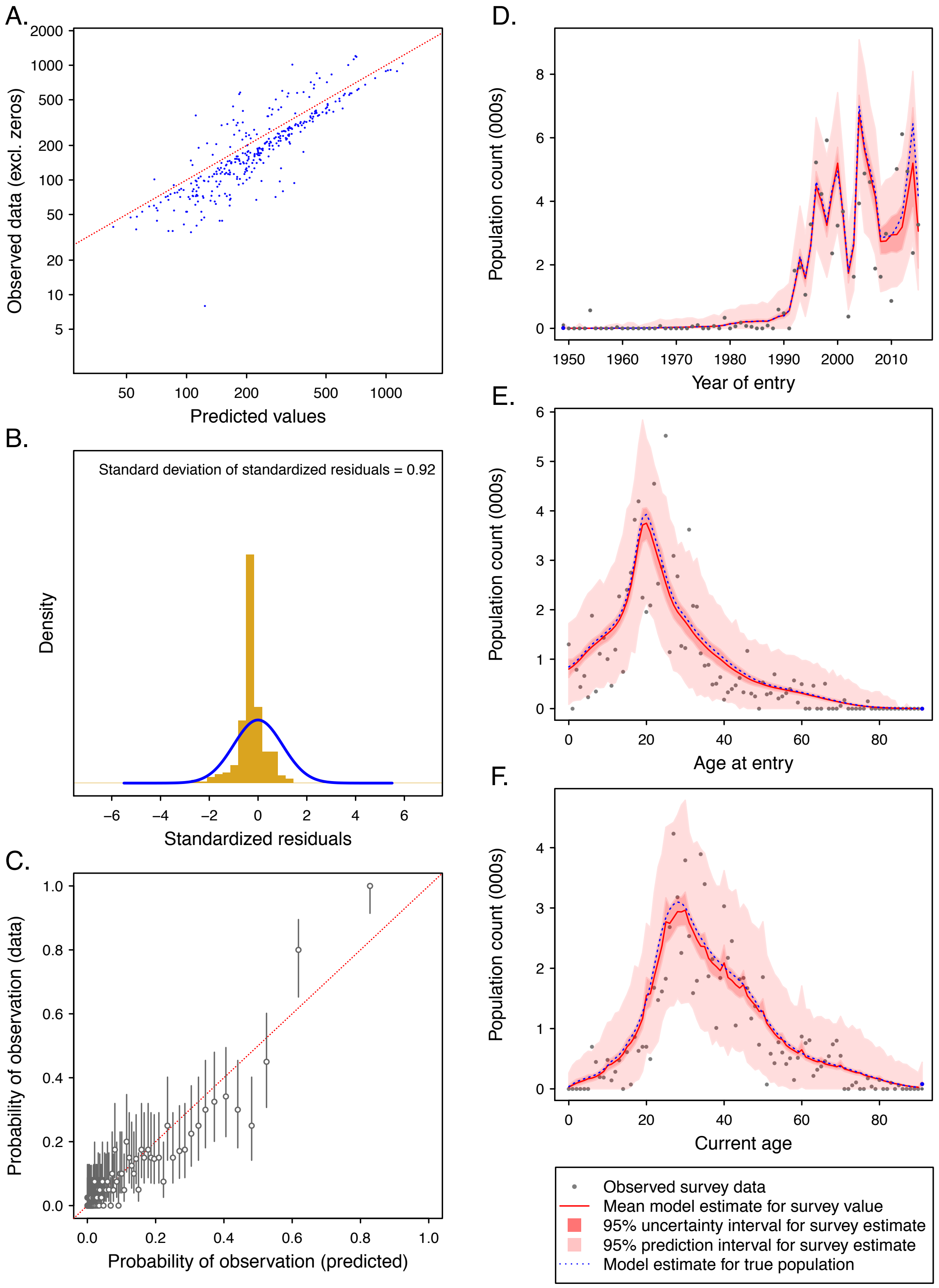

**Figure S6-19: Out-of-sample predictive performance for Viet Nam, 2005.**

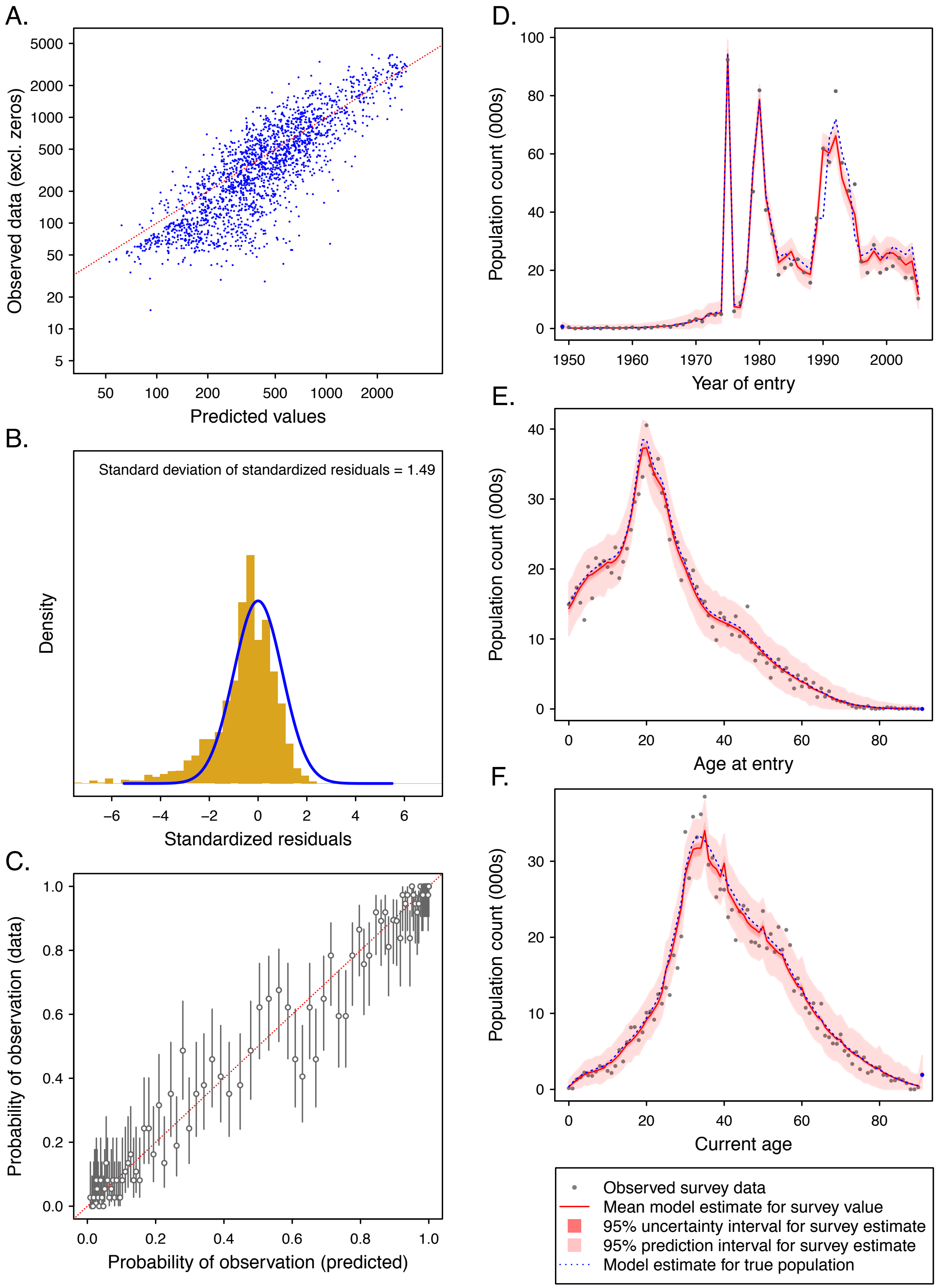

**Figure S6-20: Out-of-sample predictive performance for Viet Nam, 2010.**

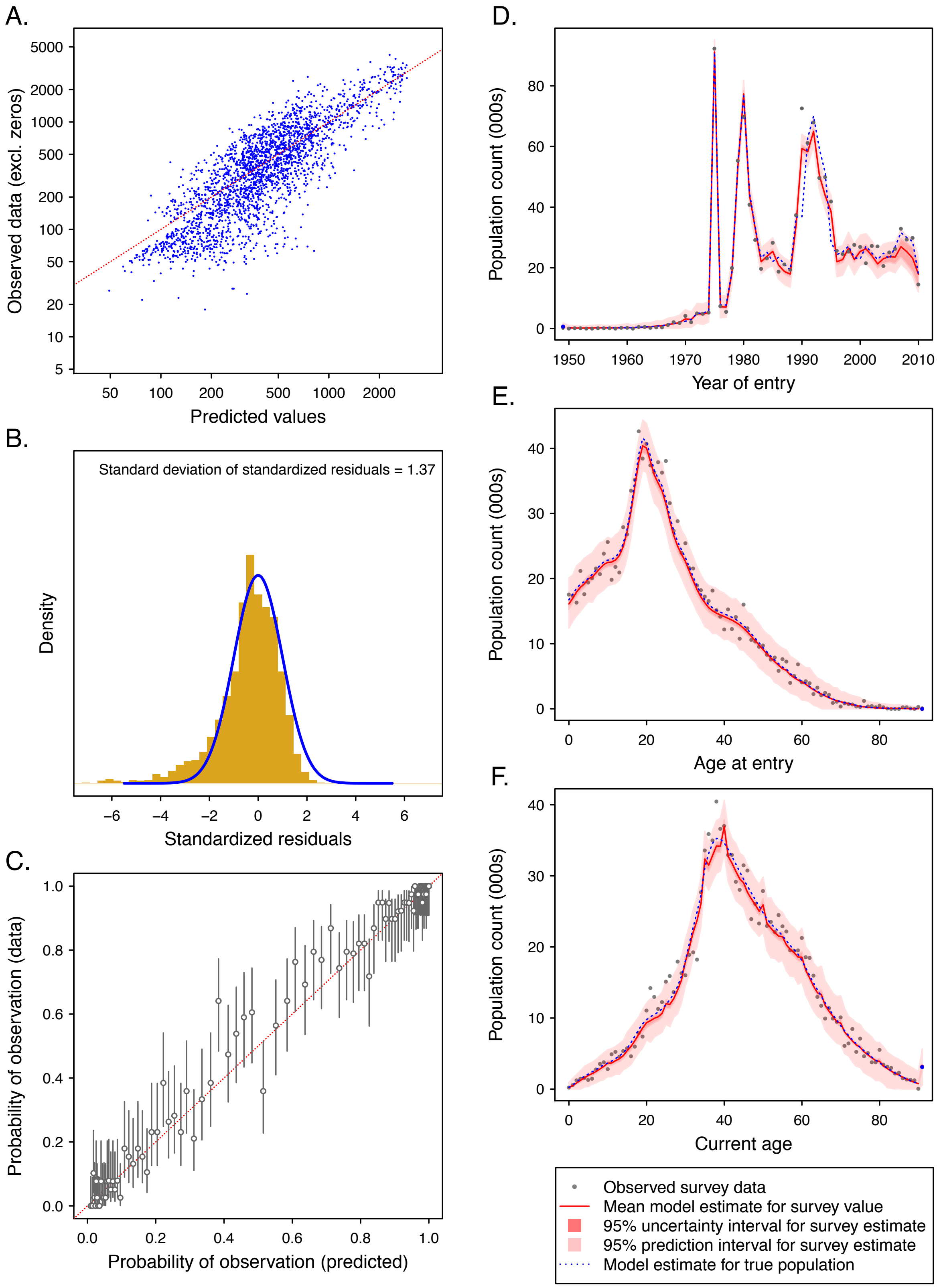

**Figure S6-21: Out-of-sample predictive performance for Viet Nam, 2015.**

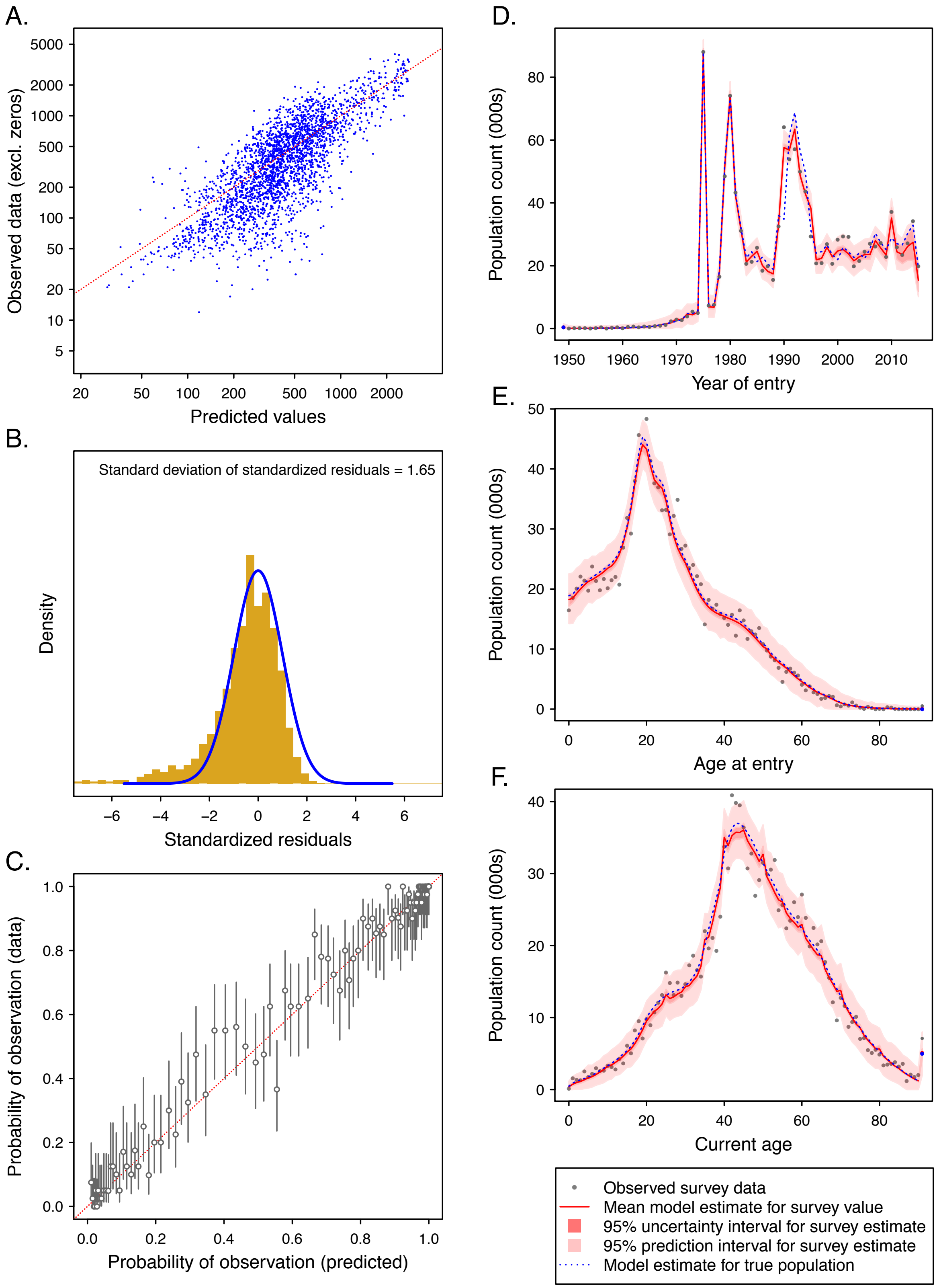

## Comparison of modelled versus raw population estimates for each country and region of origin.

Figures S7-1 to S7-27 compares modeled versus raw population estimates for each modelled geography. These include panels showing the distribution of the 2016 resident population by age at entry (first column), current age (second column), and year of entry (third column. The figures also include panels showing total population estimates for 2000-2016, plus projections to 2018 (forth column).

**Figure S7−1: Comparison of modelled versus raw population estimates for each country and region of origin.**

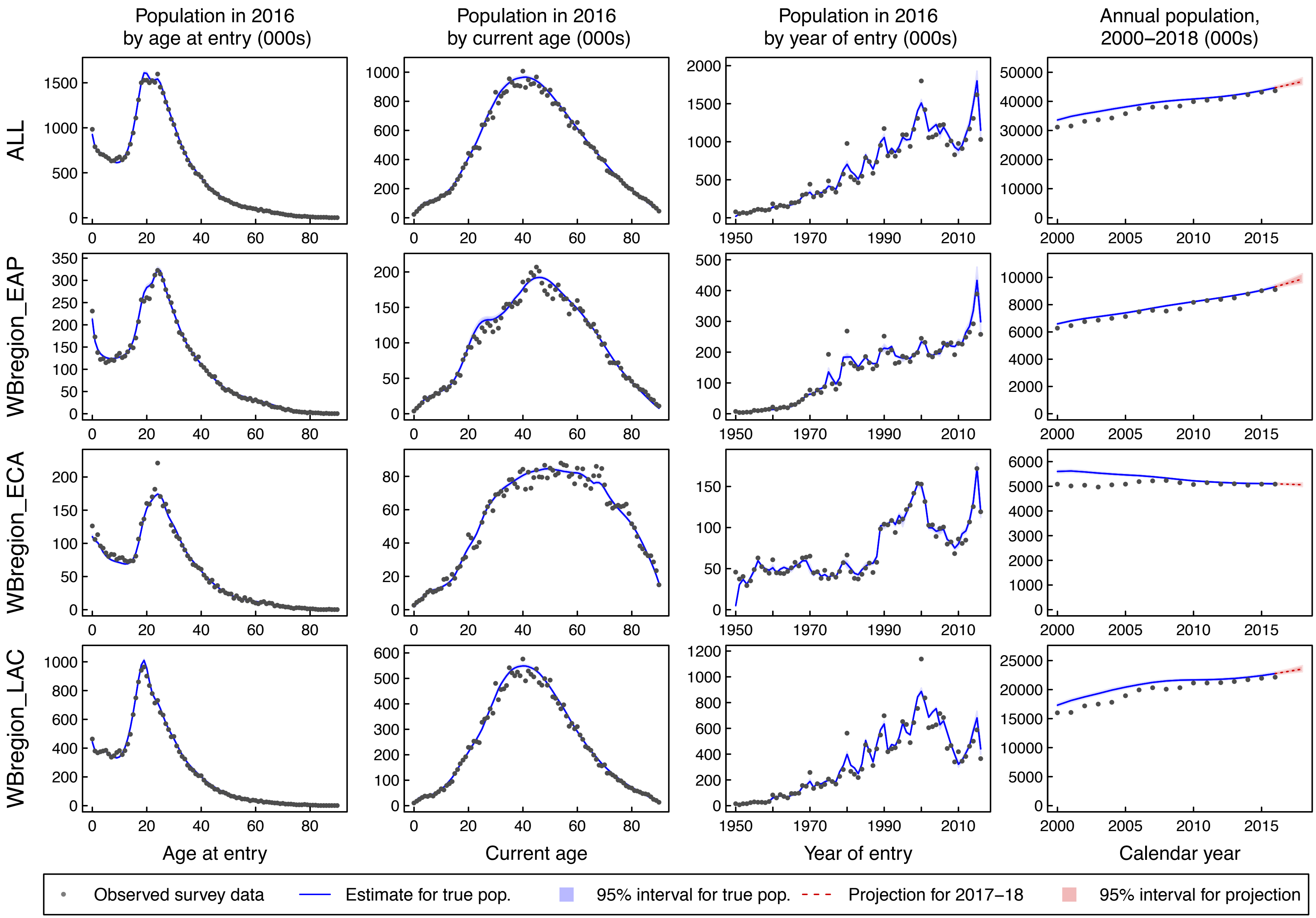

**Figure S7–2: Comparison of modelled versus raw population estimates for each country and region of origin.**

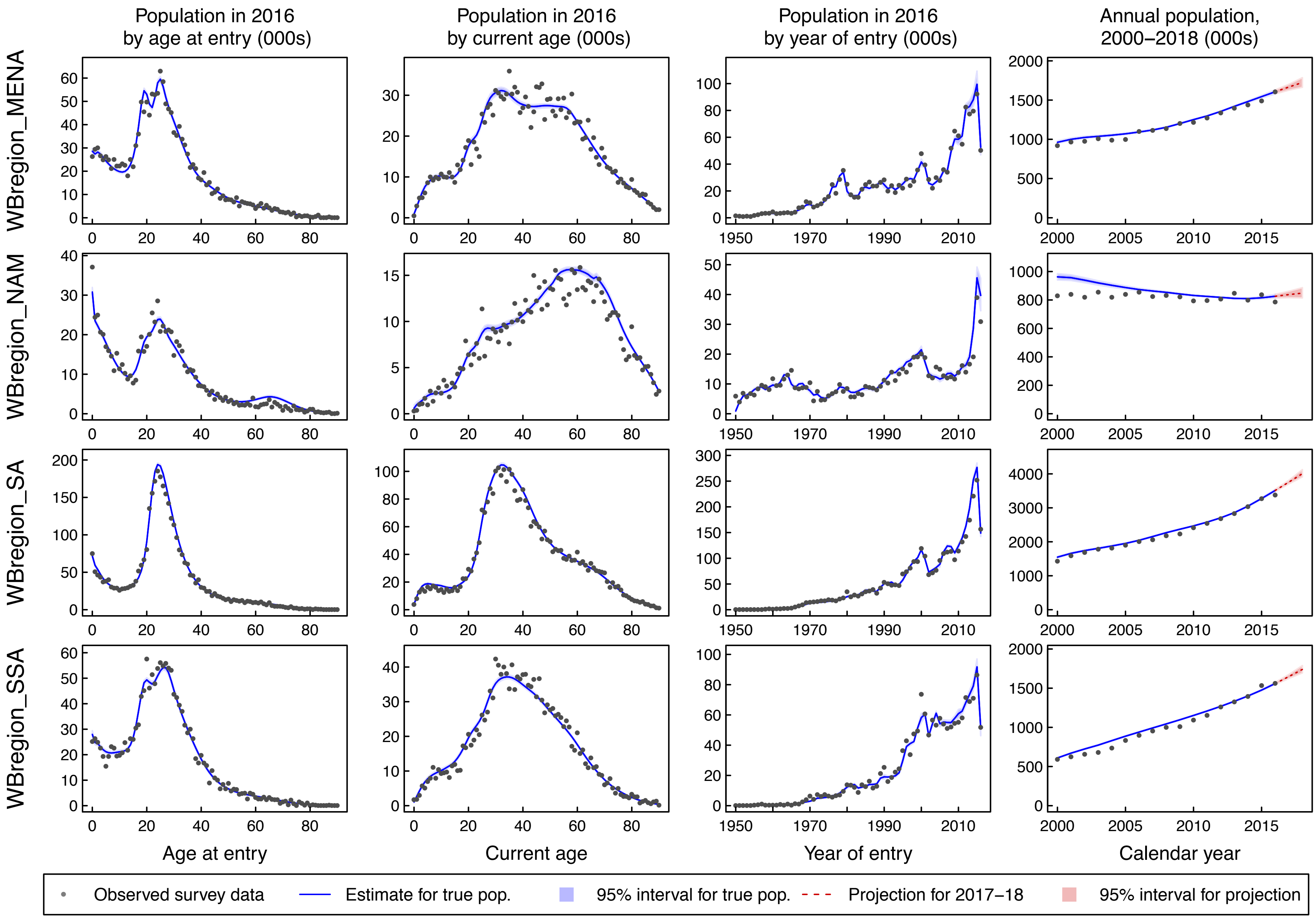

**Figure S7–3: Comparison of modelled versus raw population estimates for each country and region of origin.**

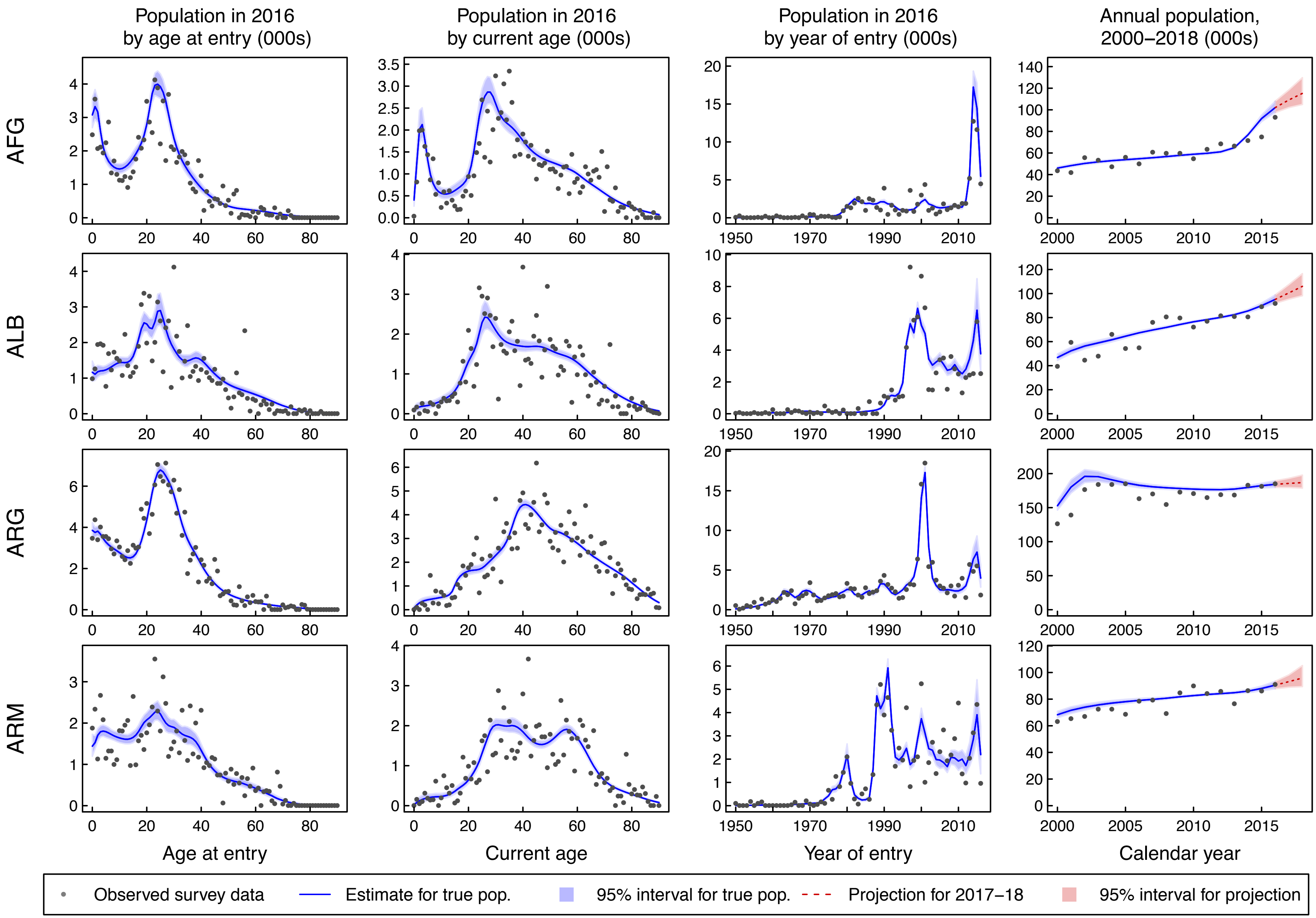

**Figure S7–4: Comparison of modelled versus raw population estimates for each country and region of origin.**

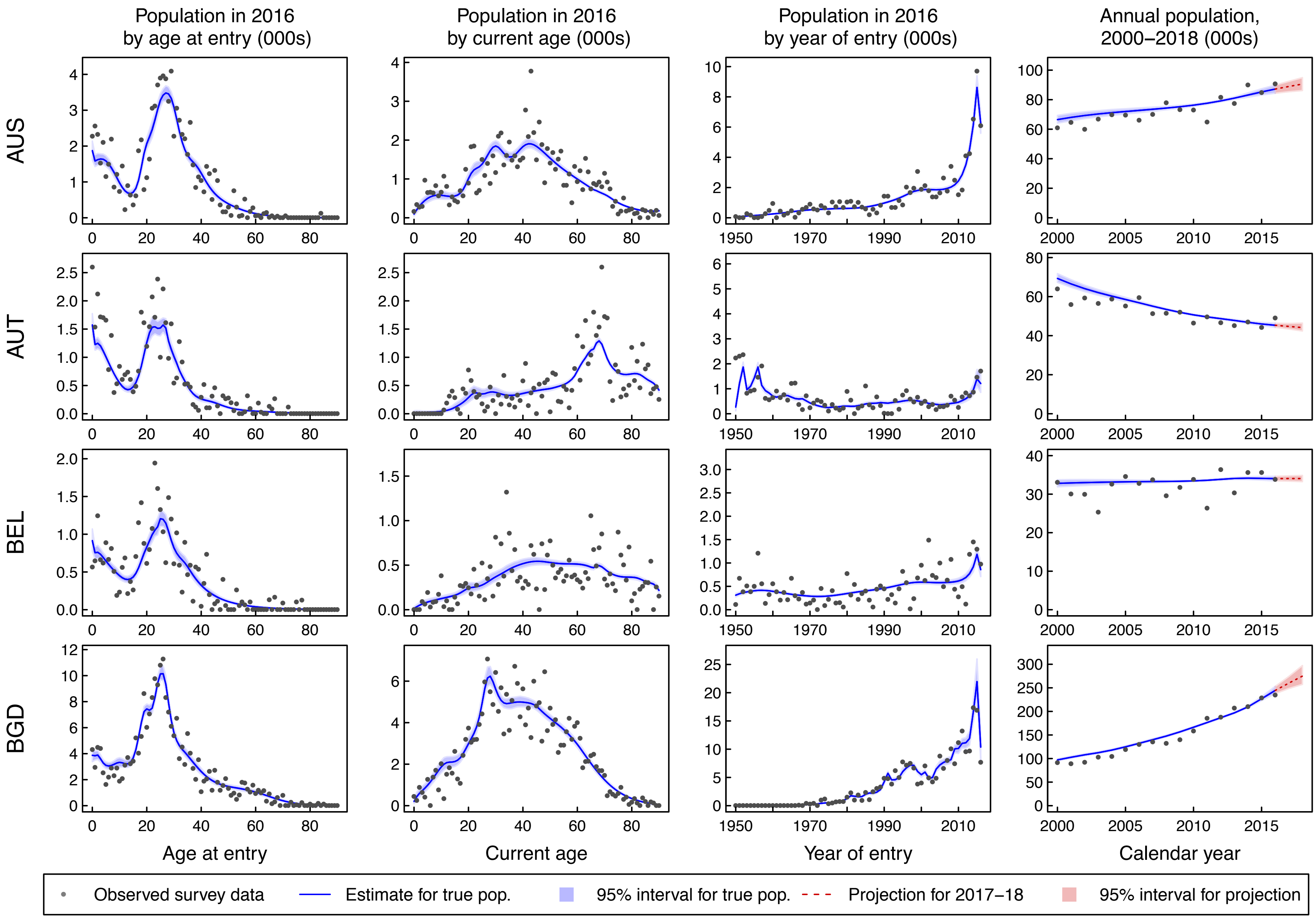

**Figure S7–5: Comparison of modelled versus raw population estimates for each country and region of origin.**

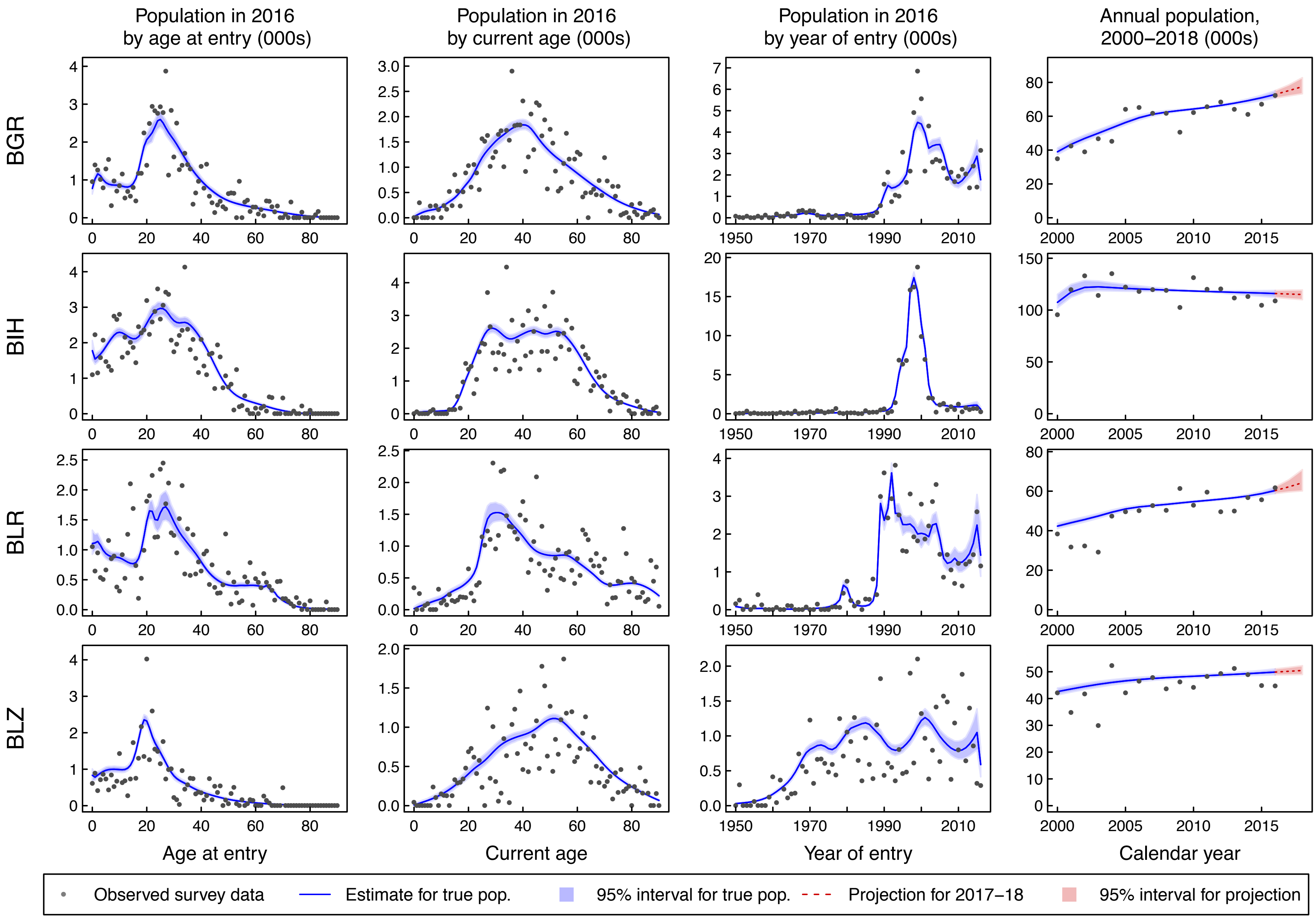

**Figure S7–6: Comparison of modelled versus raw population estimates for each country and region of origin.**

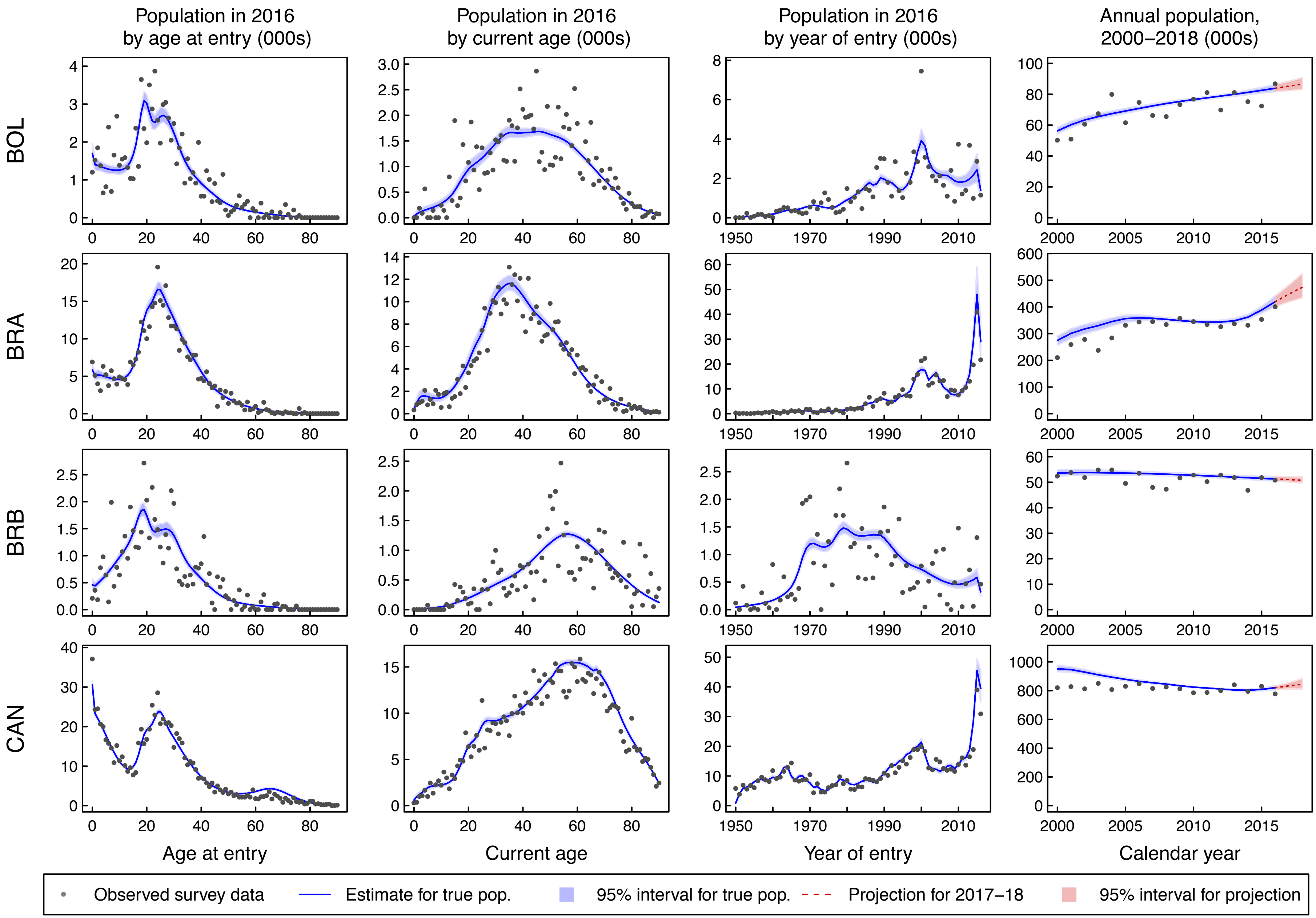

**Figure S7–7: Comparison of modelled versus raw population estimates for each country and region of origin.**

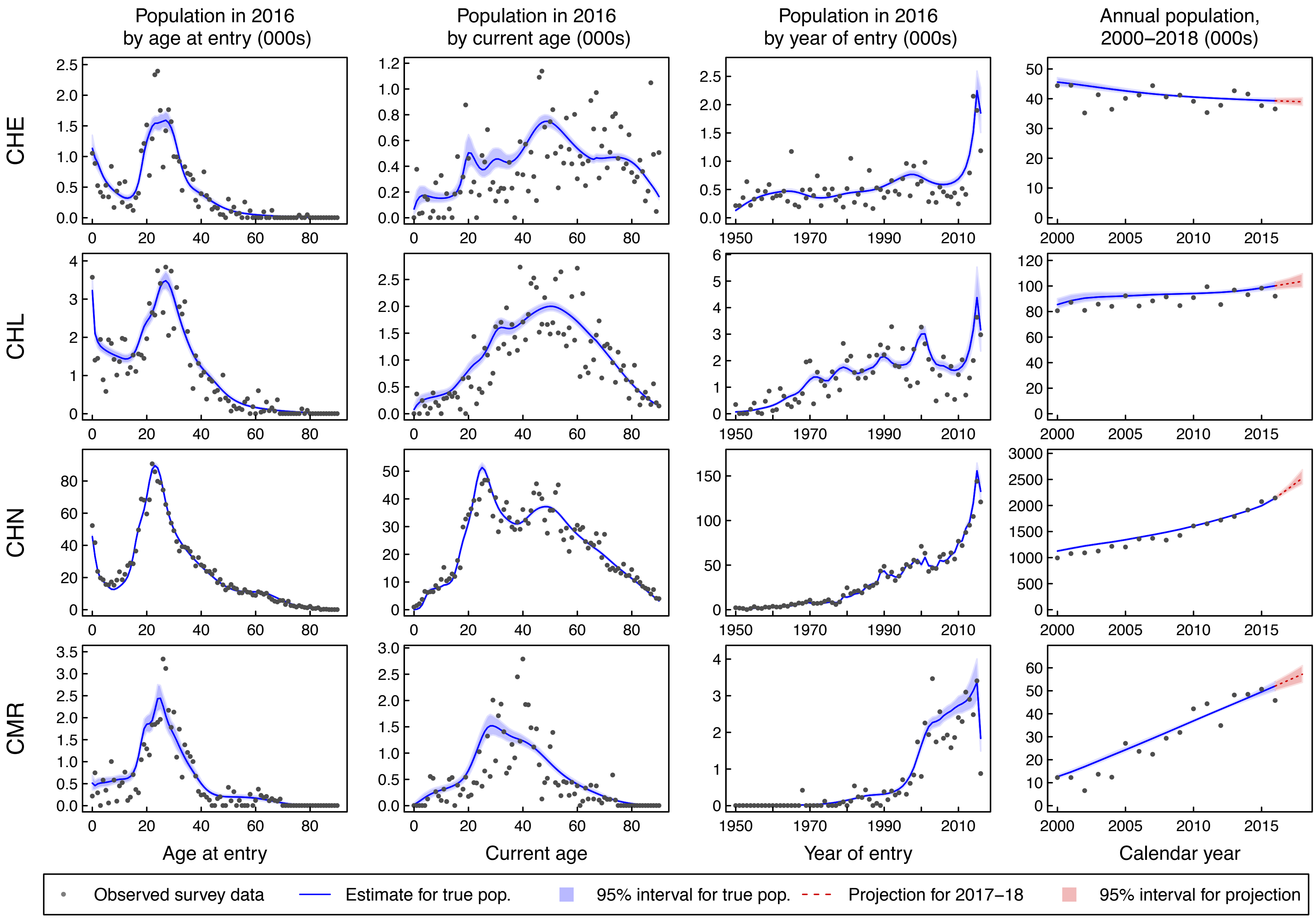

**Figure S7–8: Comparison of modelled versus raw population estimates for each country and region of origin.**

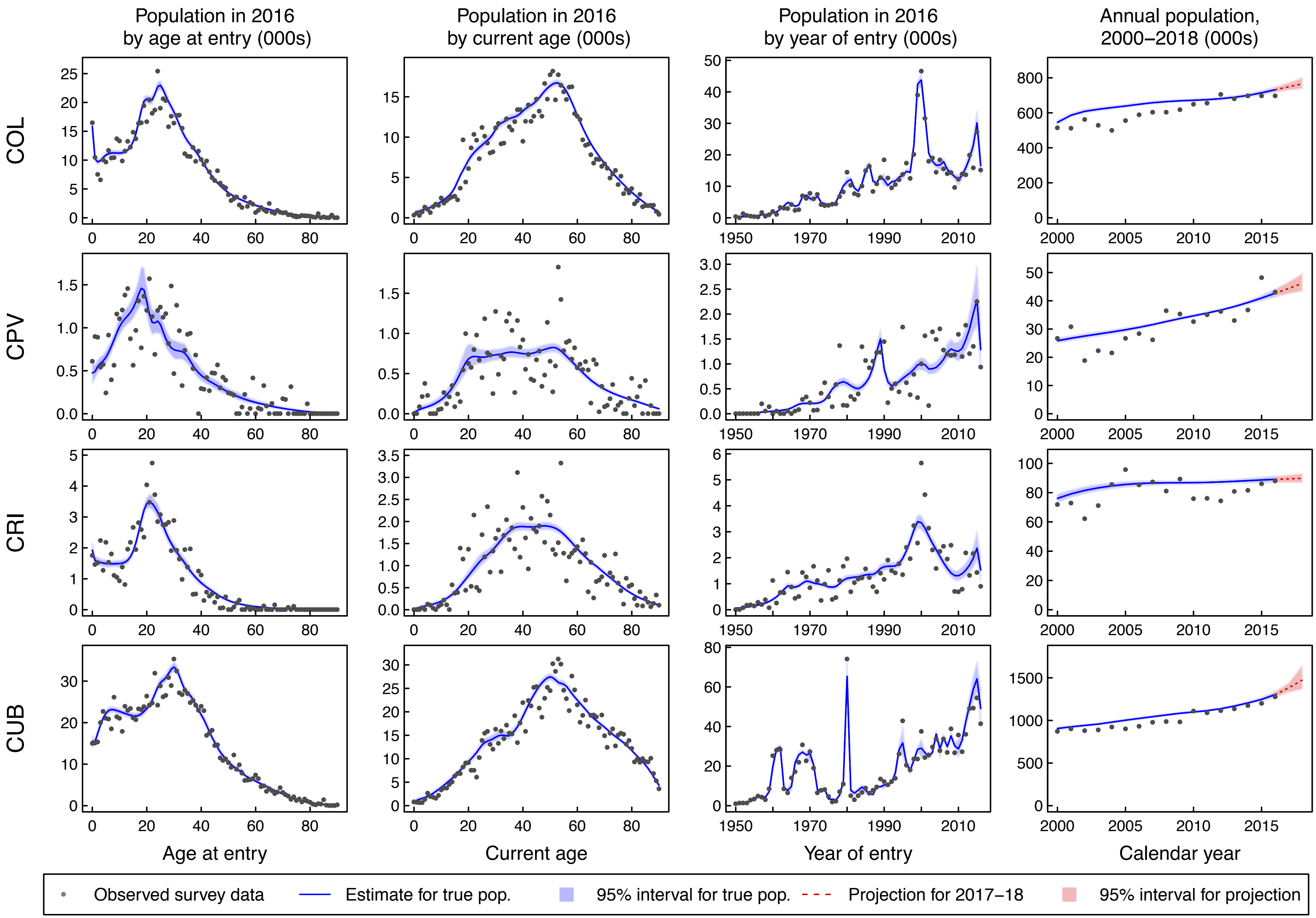

**Figure S7–9: Comparison of modelled versus raw population estimates for each country and region of origin.**

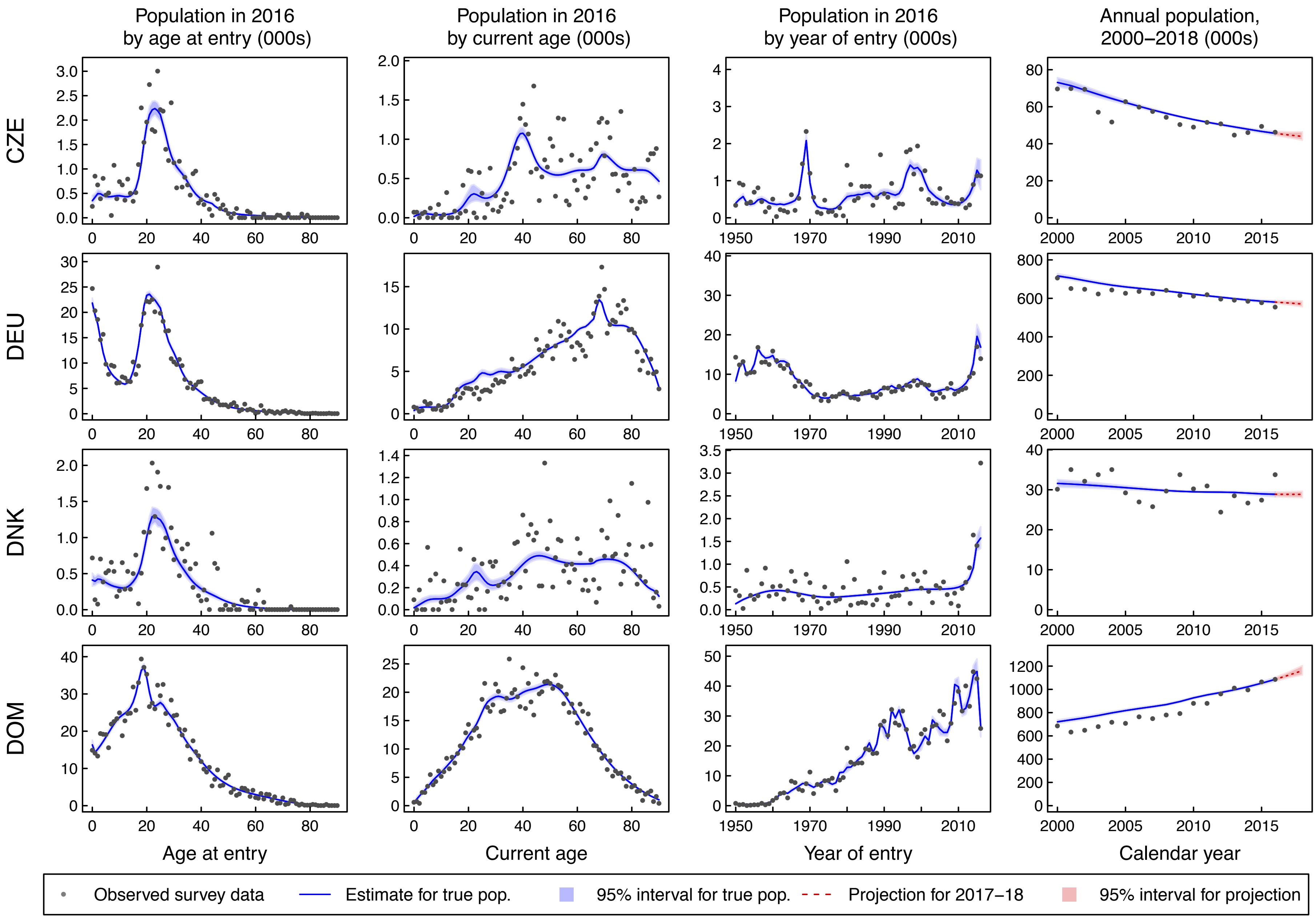

**Figure S7–10: Comparison of modelled versus raw population estimates for each country and region of origin.**

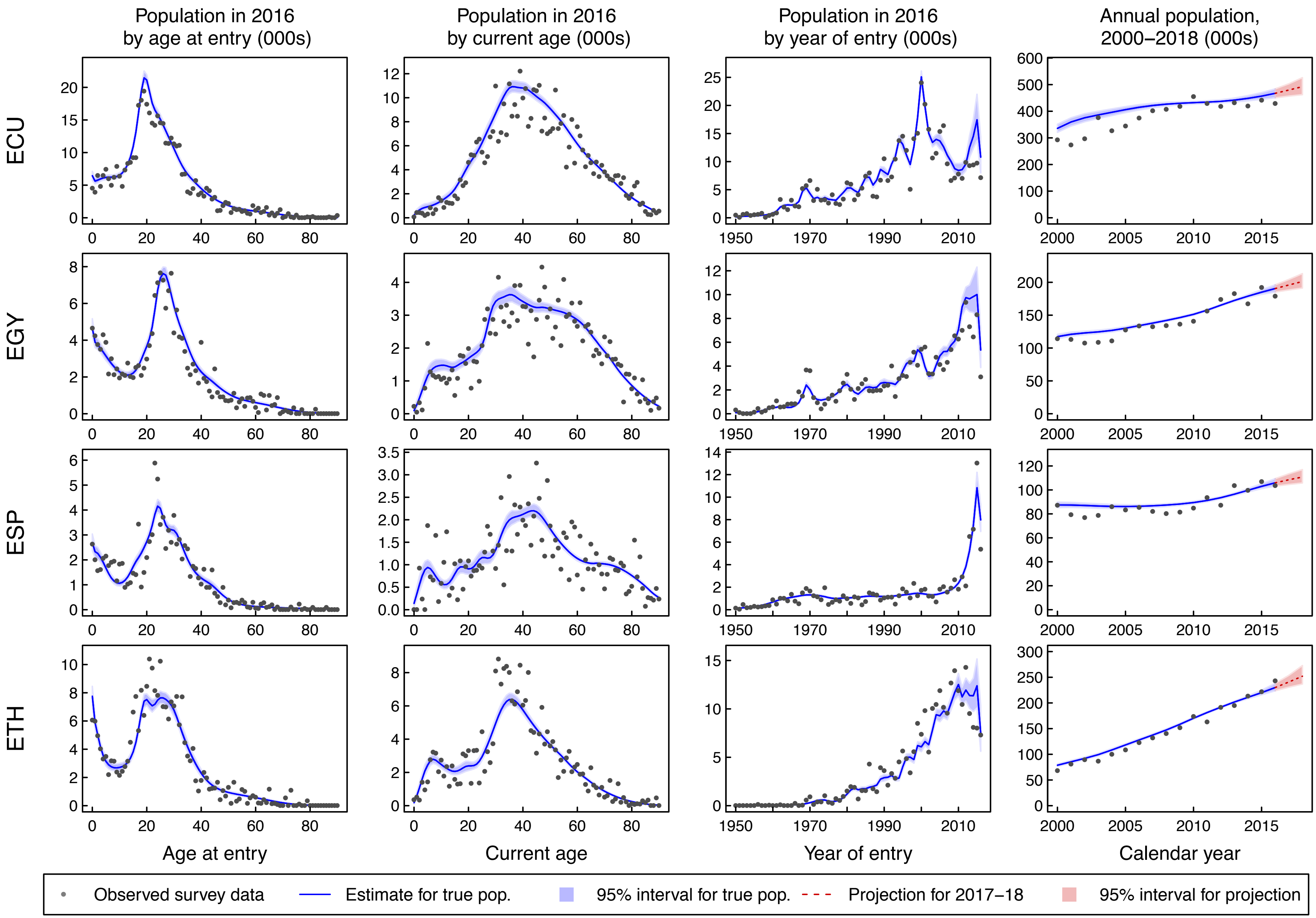

**Figure S7–11: Comparison of modelled versus raw population estimates for each country and region of origin.**

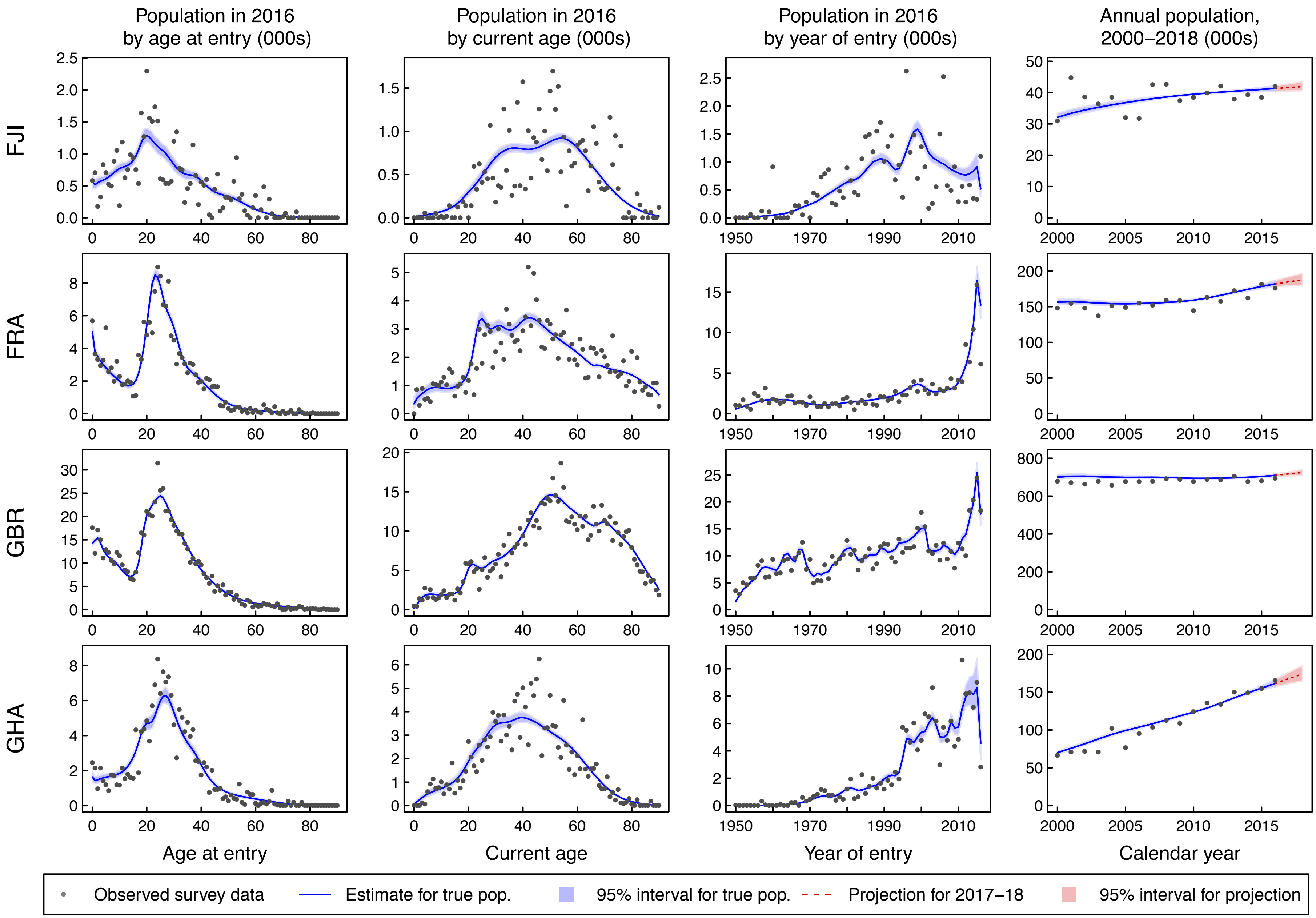

**Figure S7–12: Comparison of modelled versus raw population estimates for each country and region of origin.**

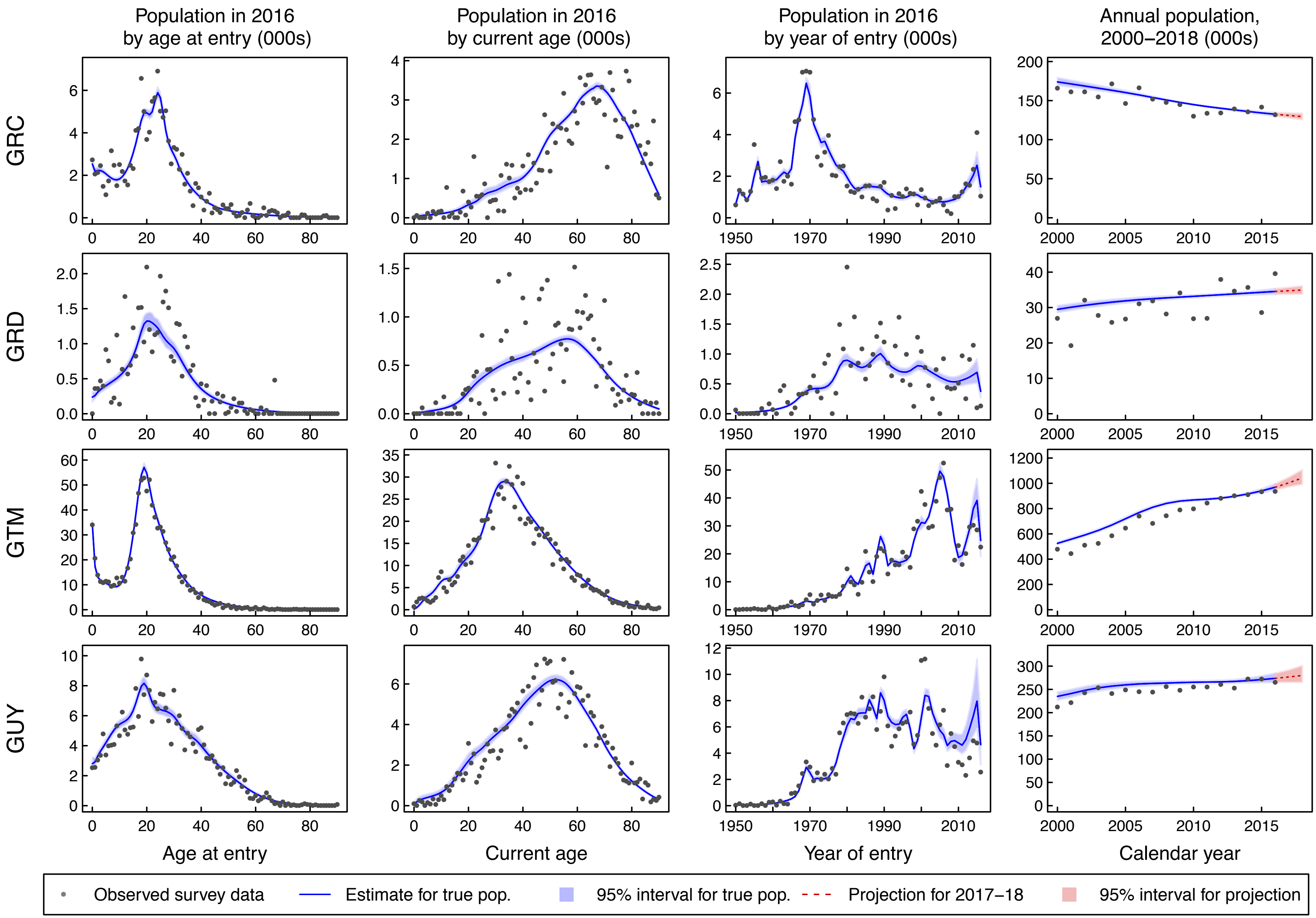

**Figure S7–13: Comparison of modelled versus raw population estimates for each country and region of origin.**

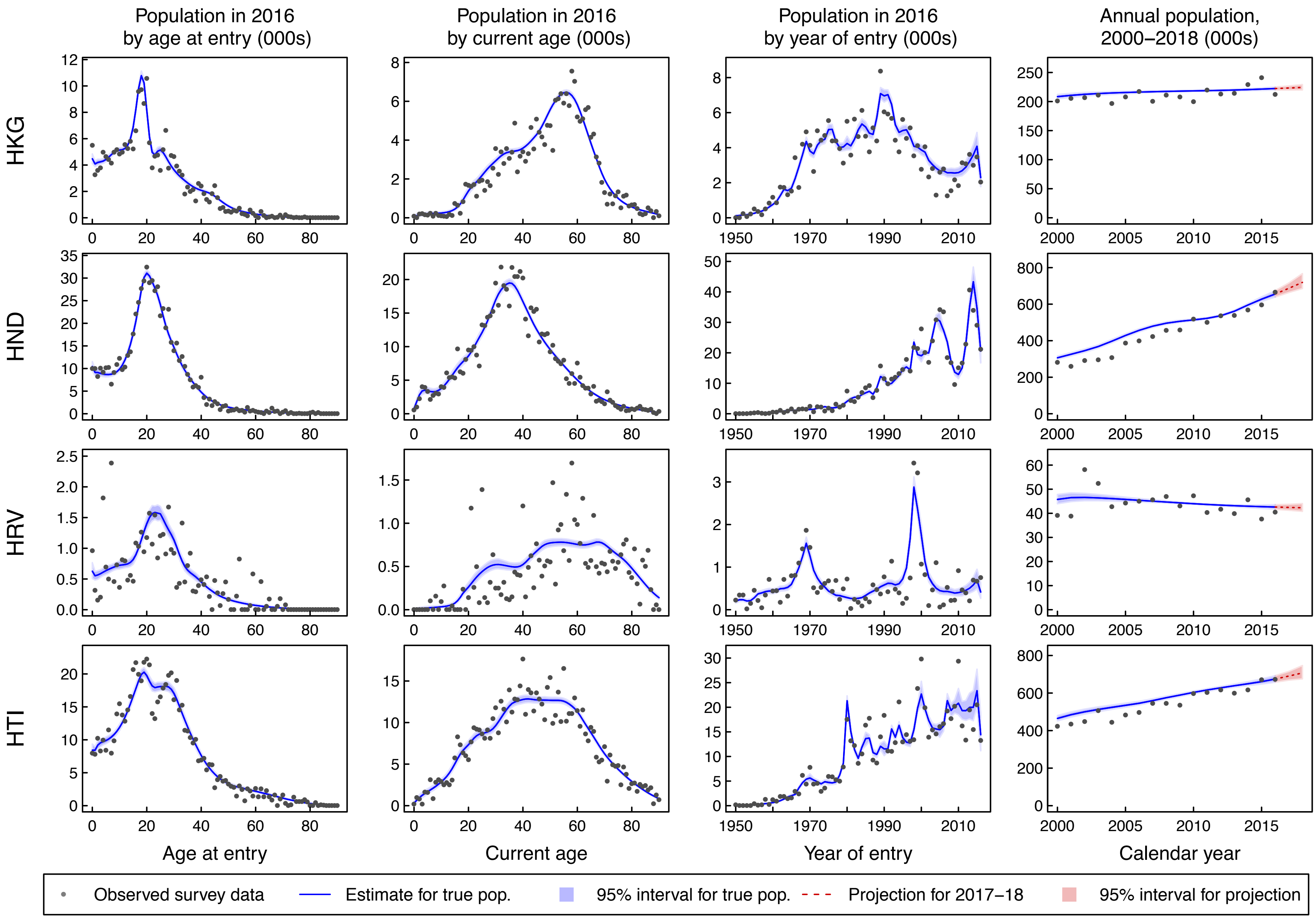

**Figure S7–14: Comparison of modelled versus raw population estimates for each country and region of origin.**

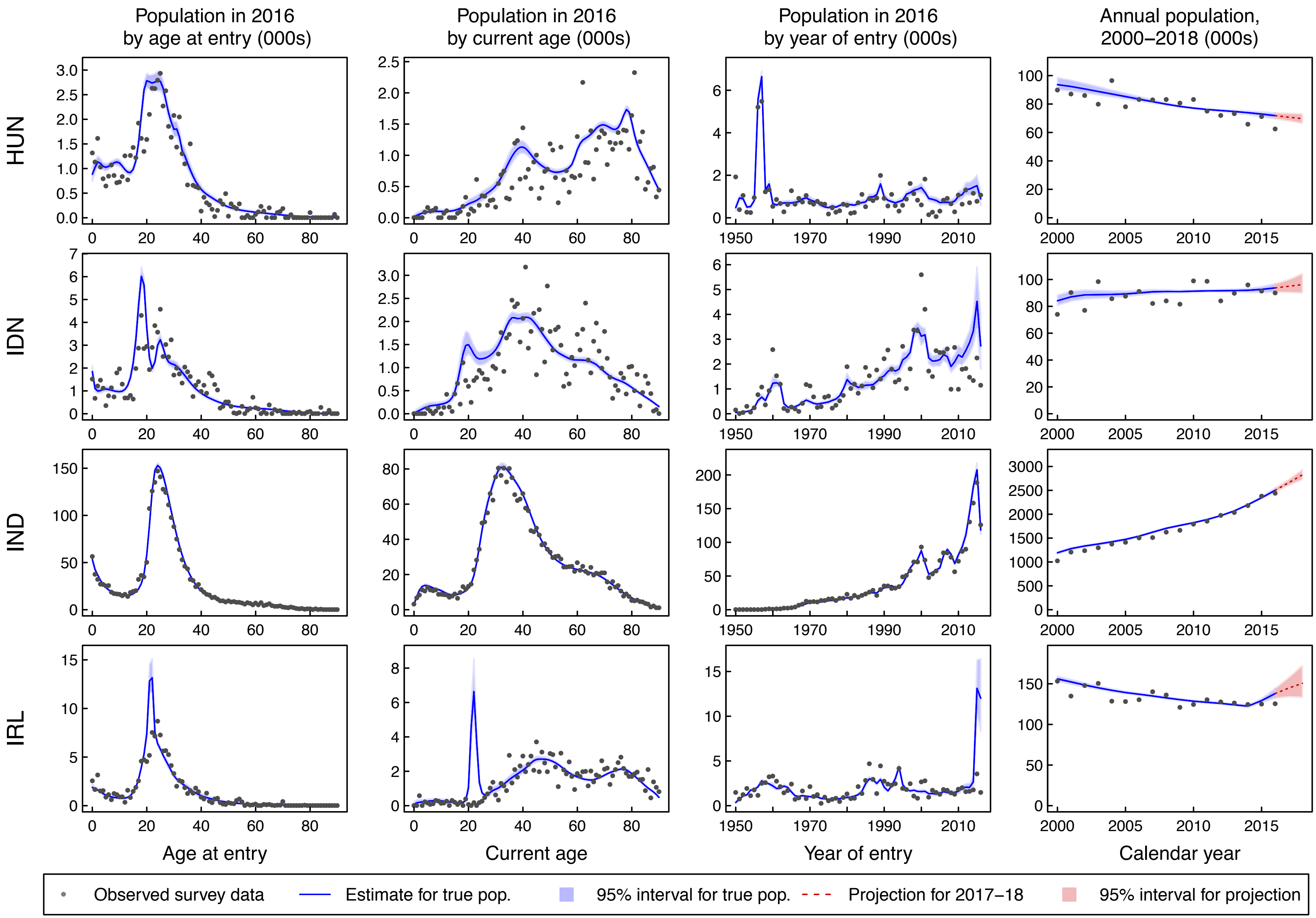

**Figure S7–15: Comparison of modelled versus raw population estimates for each country and region of origin.**

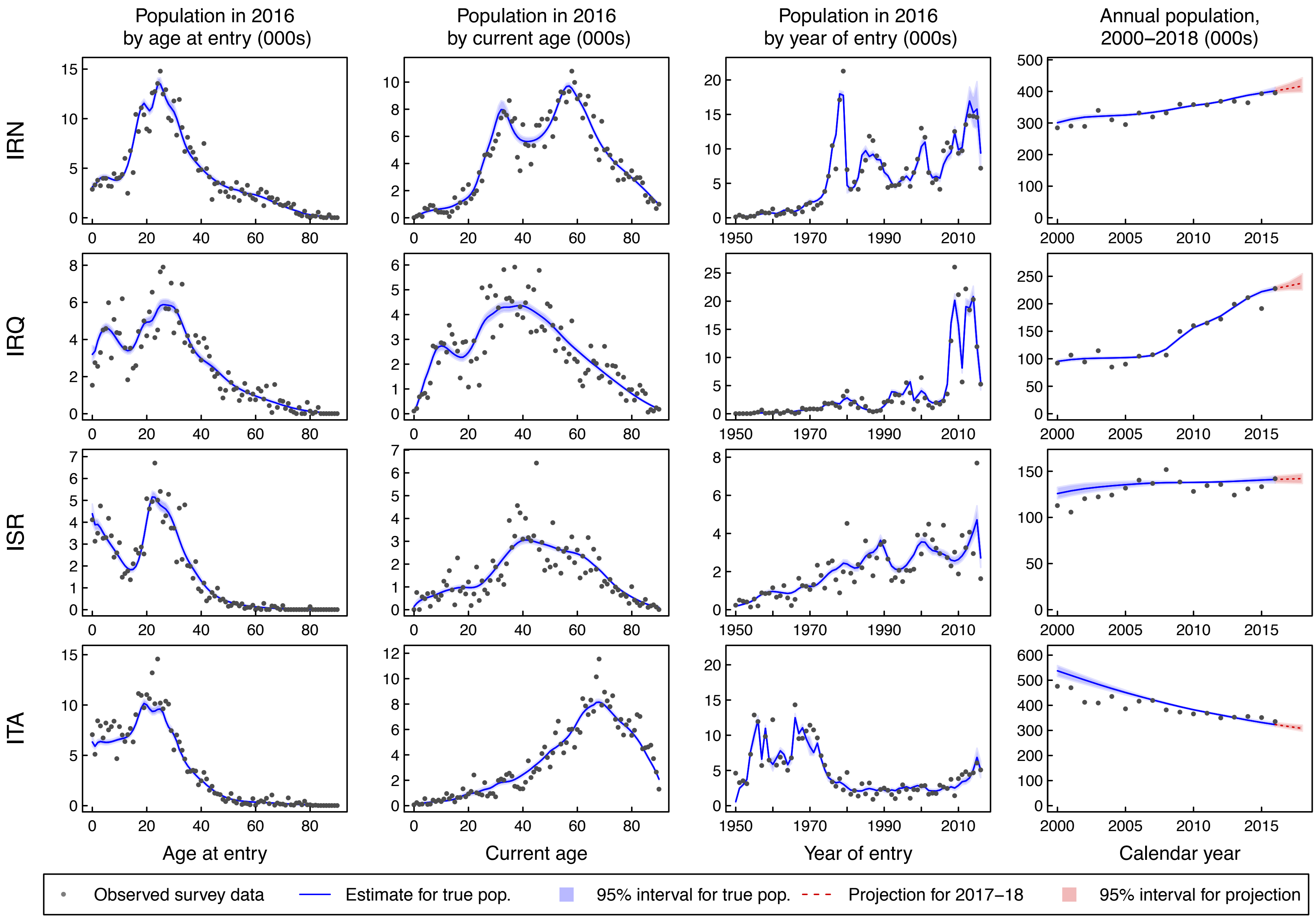

**Figure S7–16: Comparison of modelled versus raw population estimates for each country and region of origin.**

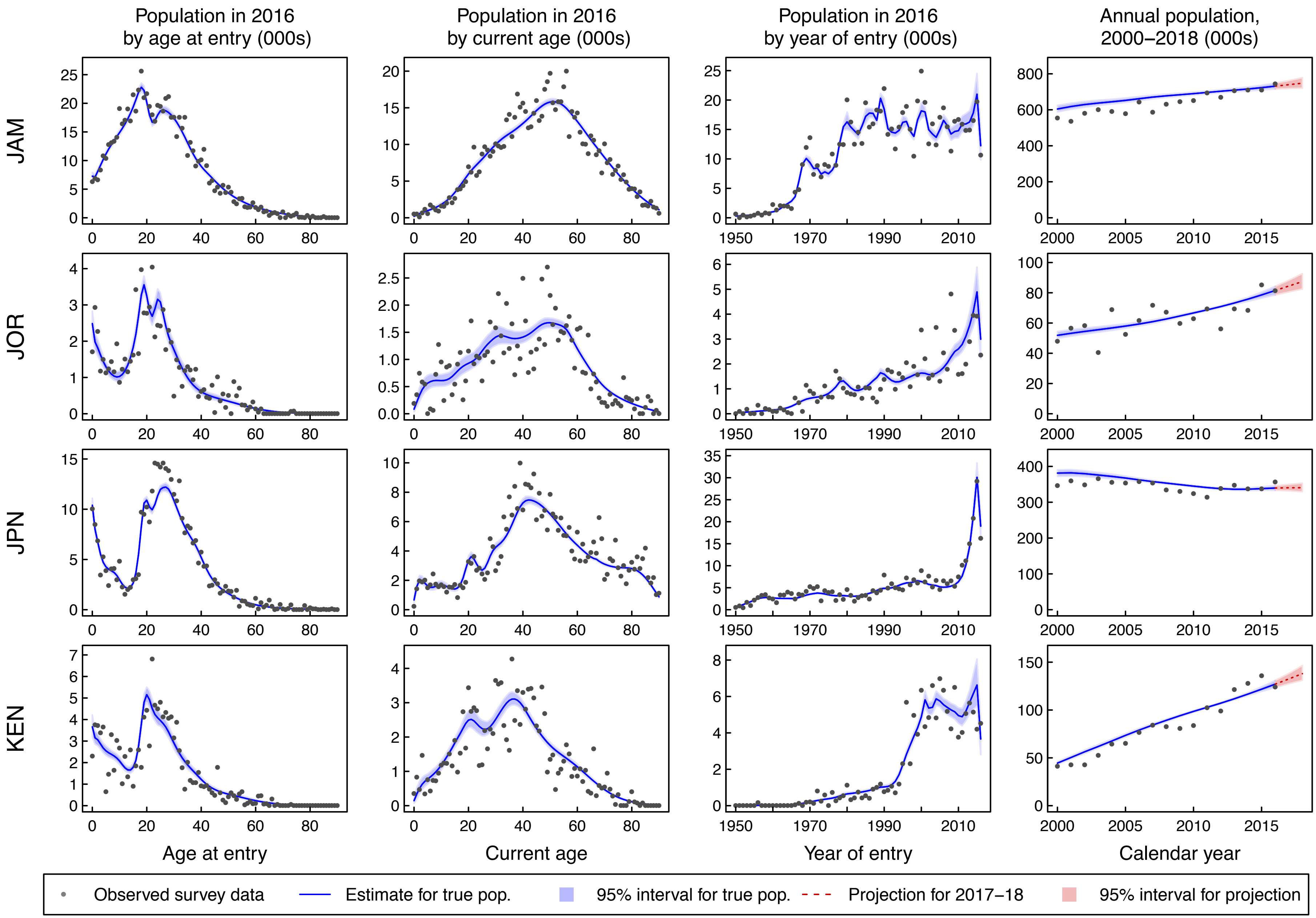

**Figure S7–17: Comparison of modelled versus raw population estimates for each country and region of origin.**

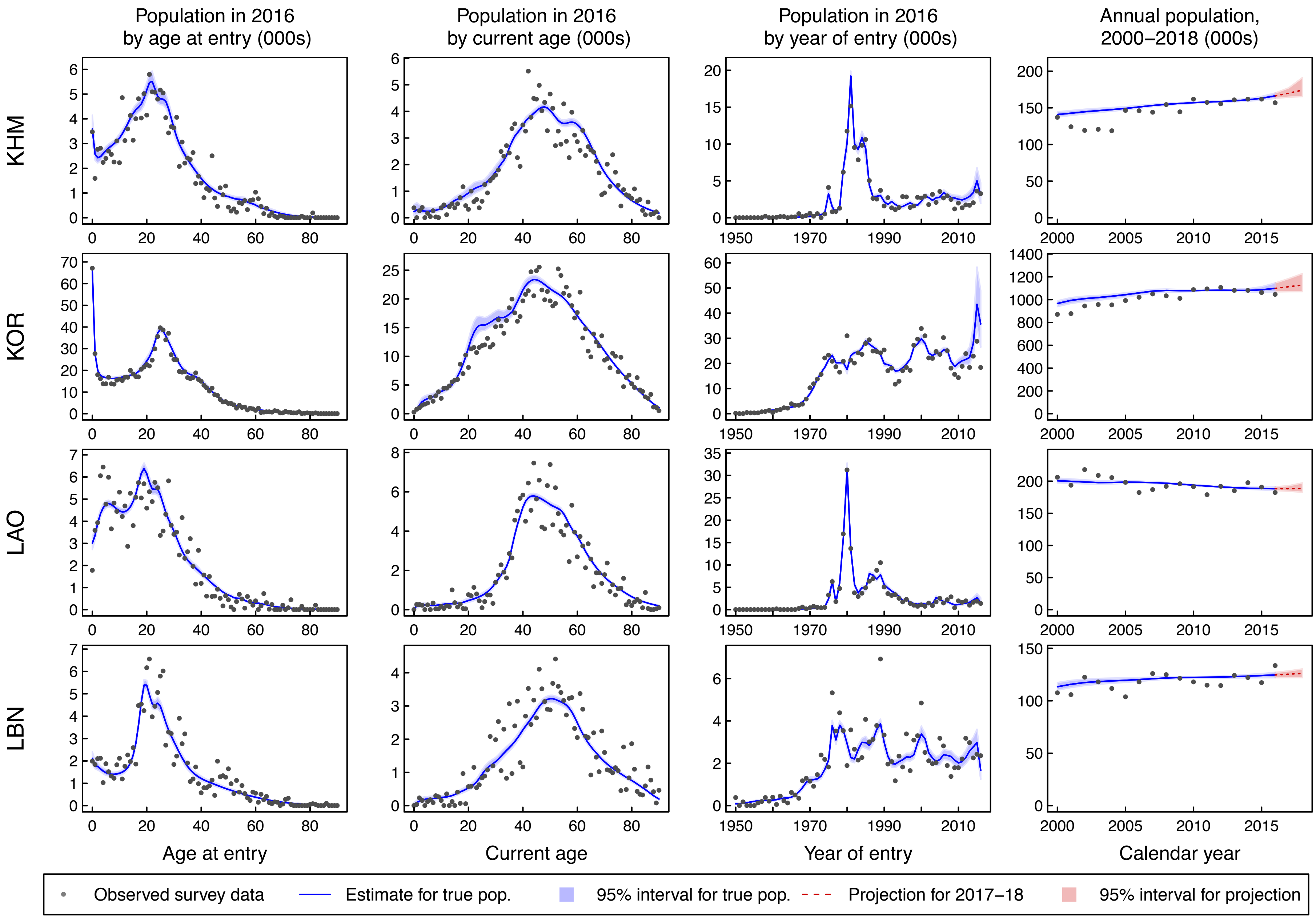

**Figure S7–18: Comparison of modelled versus raw population estimates for each country and region of origin.**

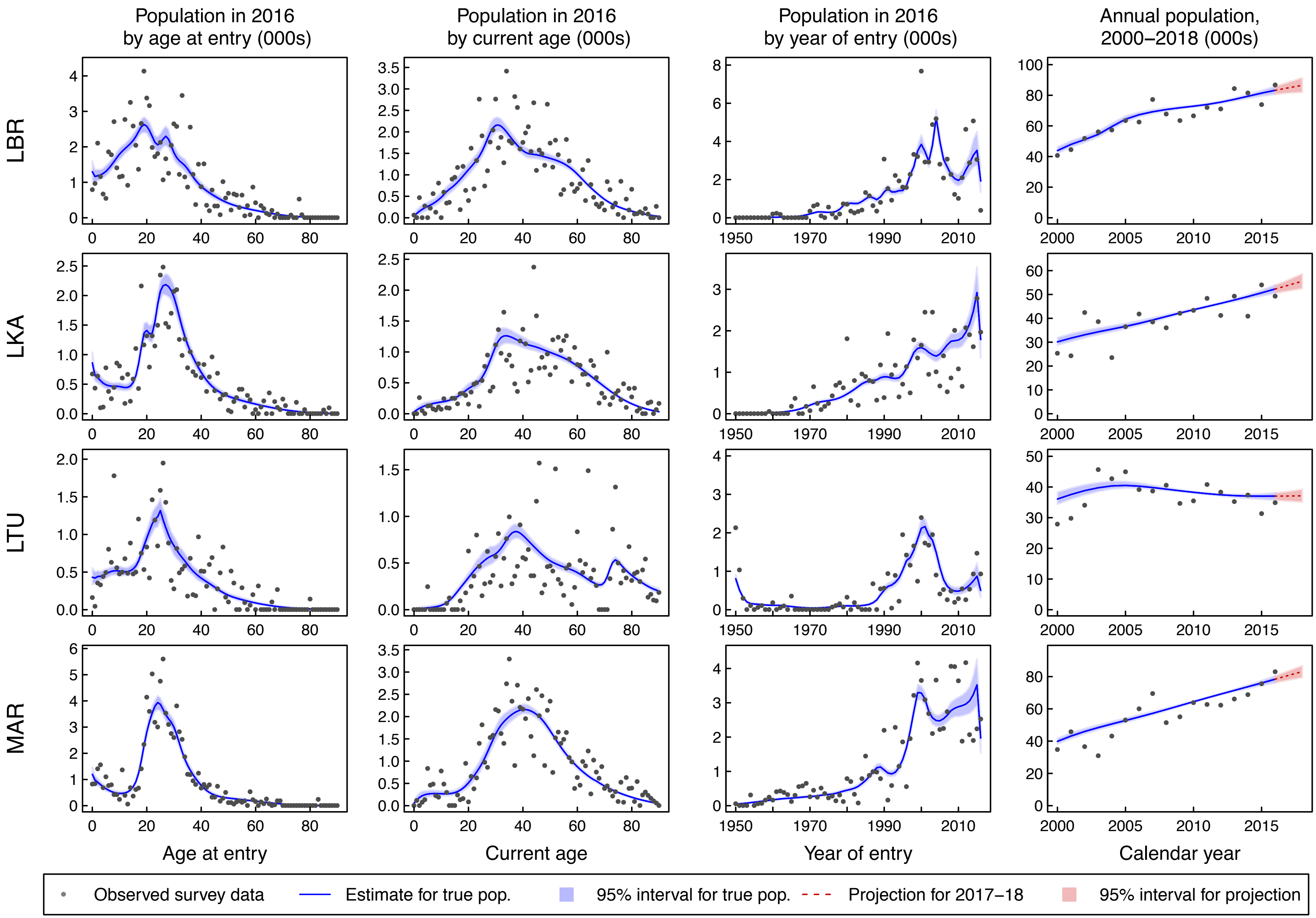

**Figure S7–19: Comparison of modelled versus raw population estimates for each country and region of origin.**

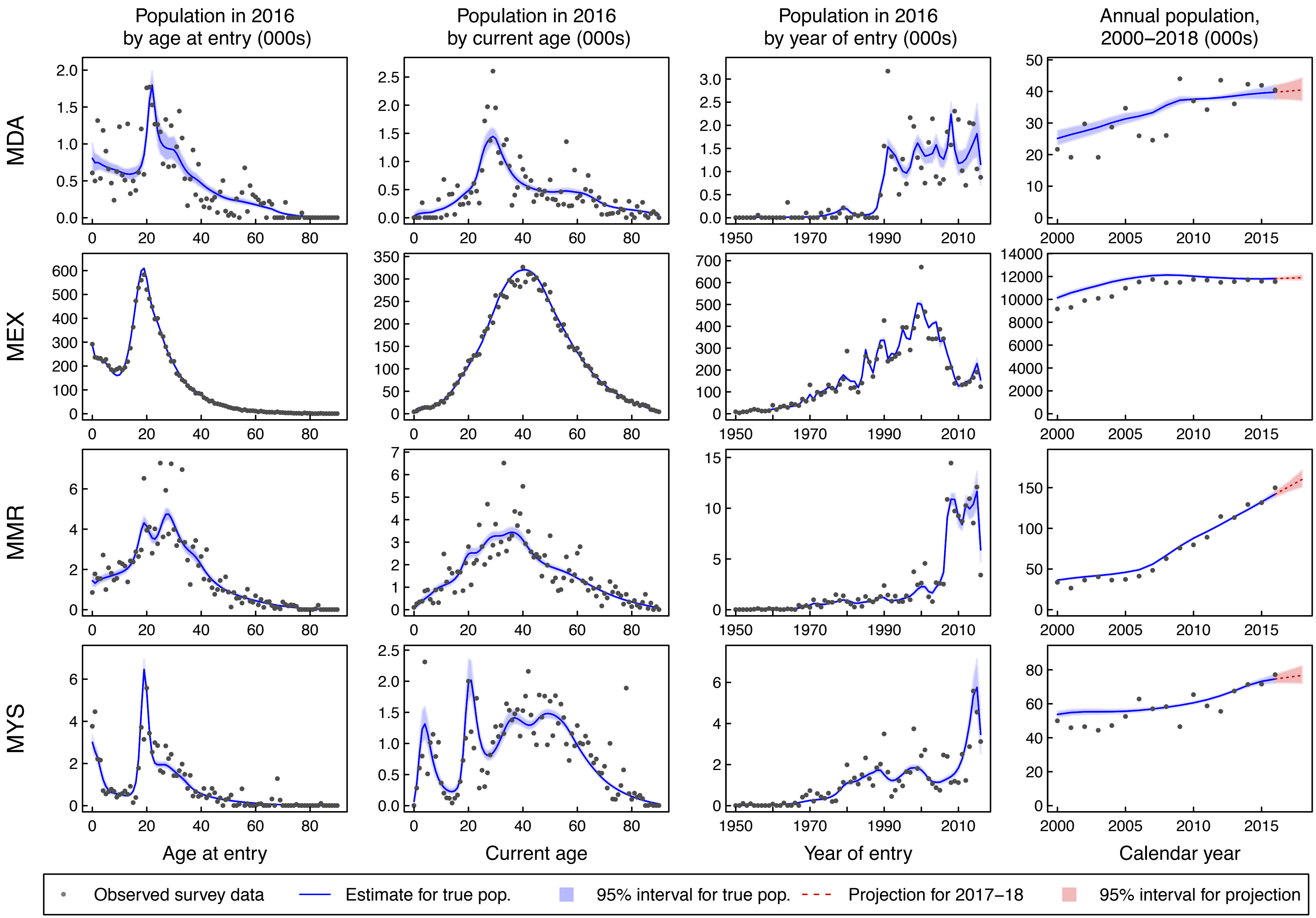

**Figure S7–20: Comparison of modelled versus raw population estimates for each country and region of origin.**

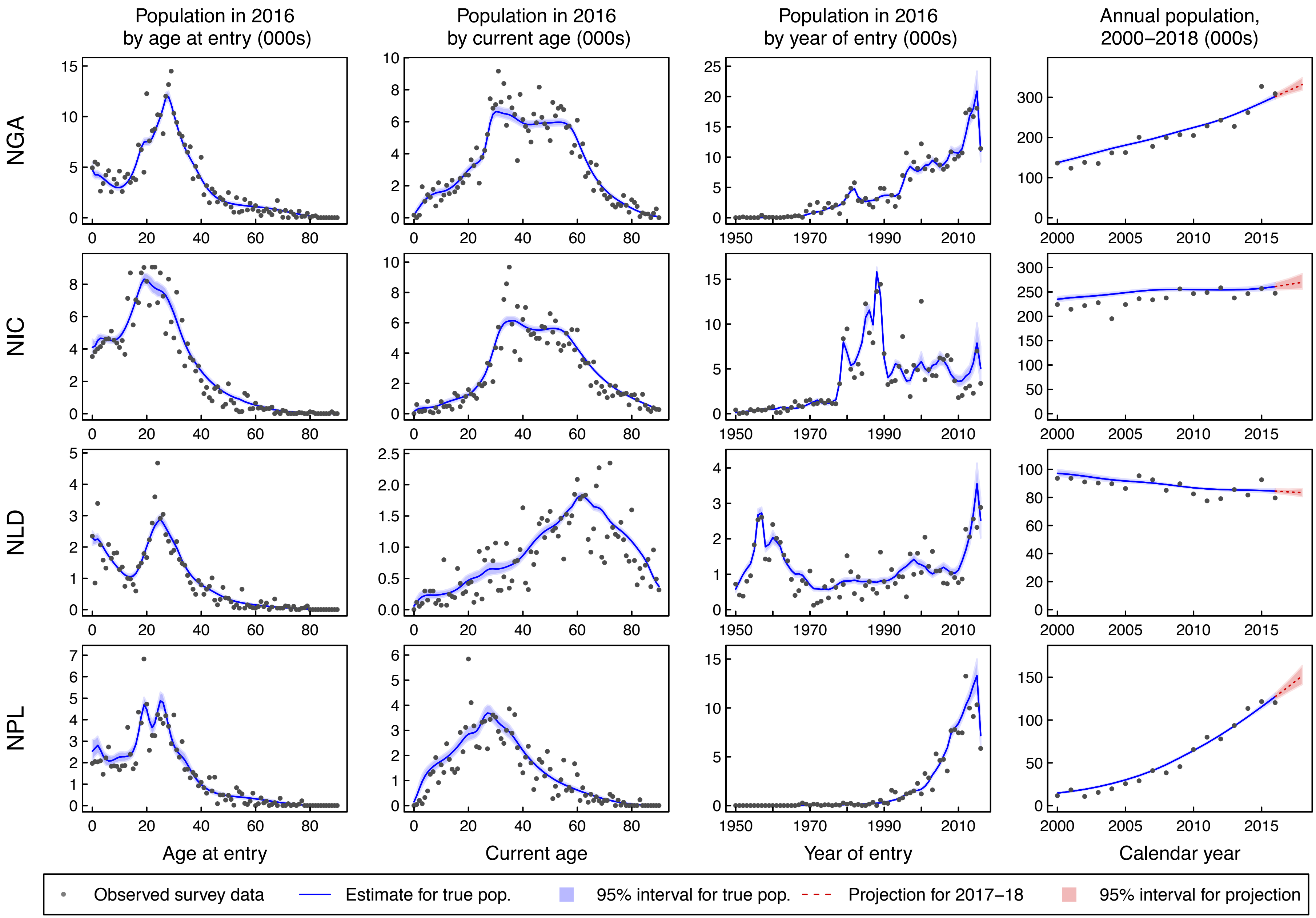

**Figure S7–21: Comparison of modelled versus raw population estimates for each country and region of origin.**

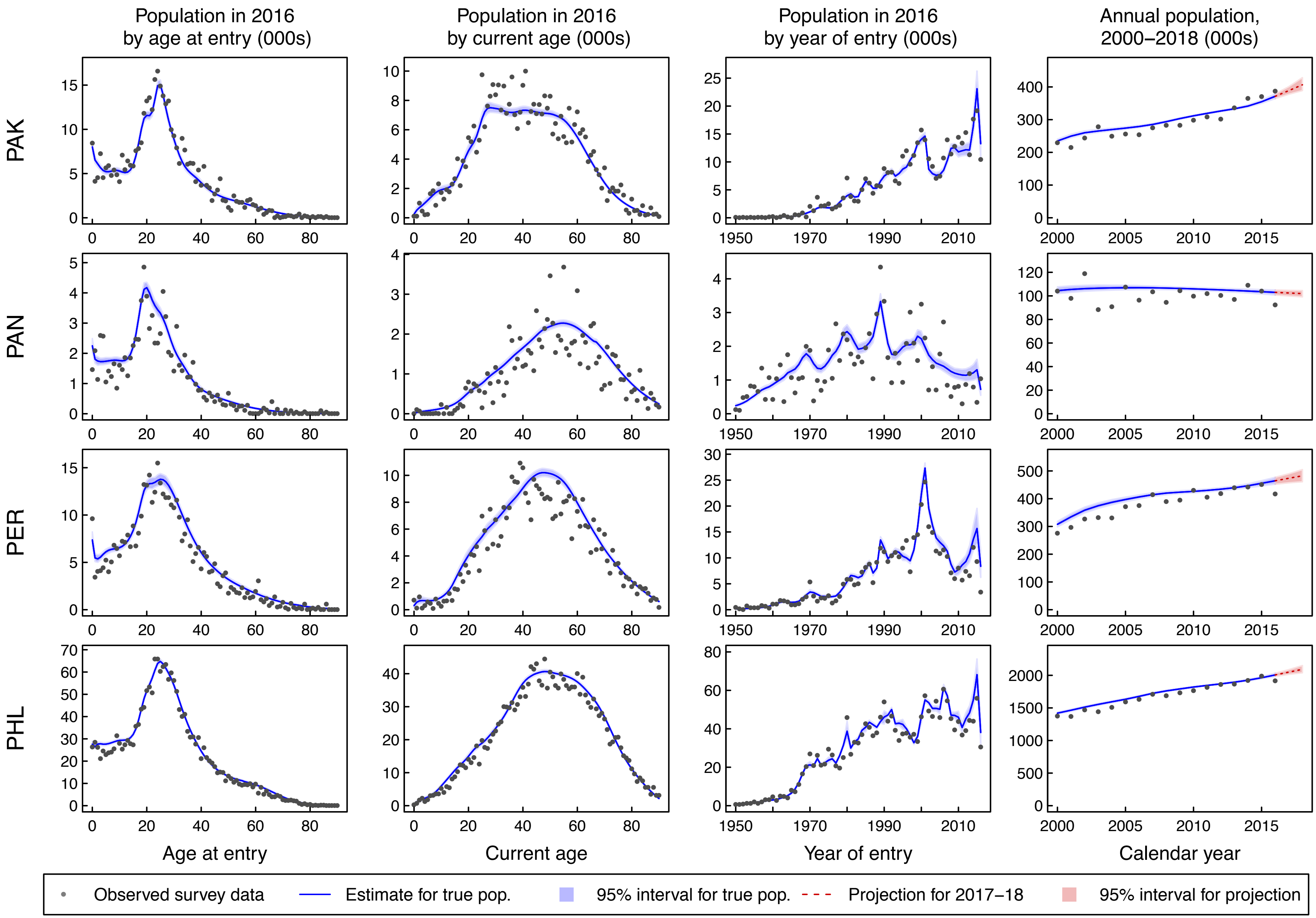

**Figure S7–22: Comparison of modelled versus raw population estimates for each country and region of origin.**

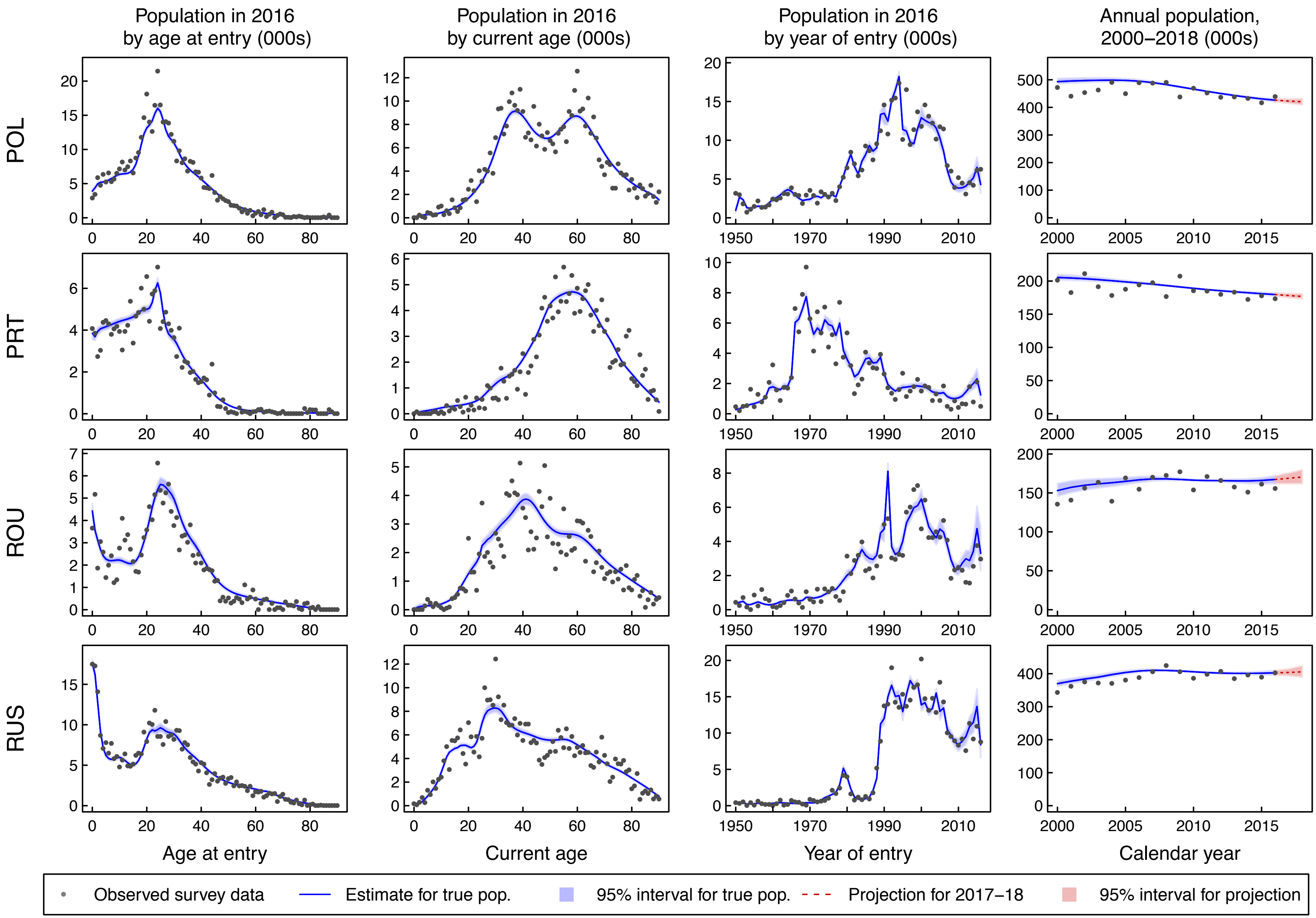

**Figure S7–23: Comparison of modelled versus raw population estimates for each country and region of origin.**

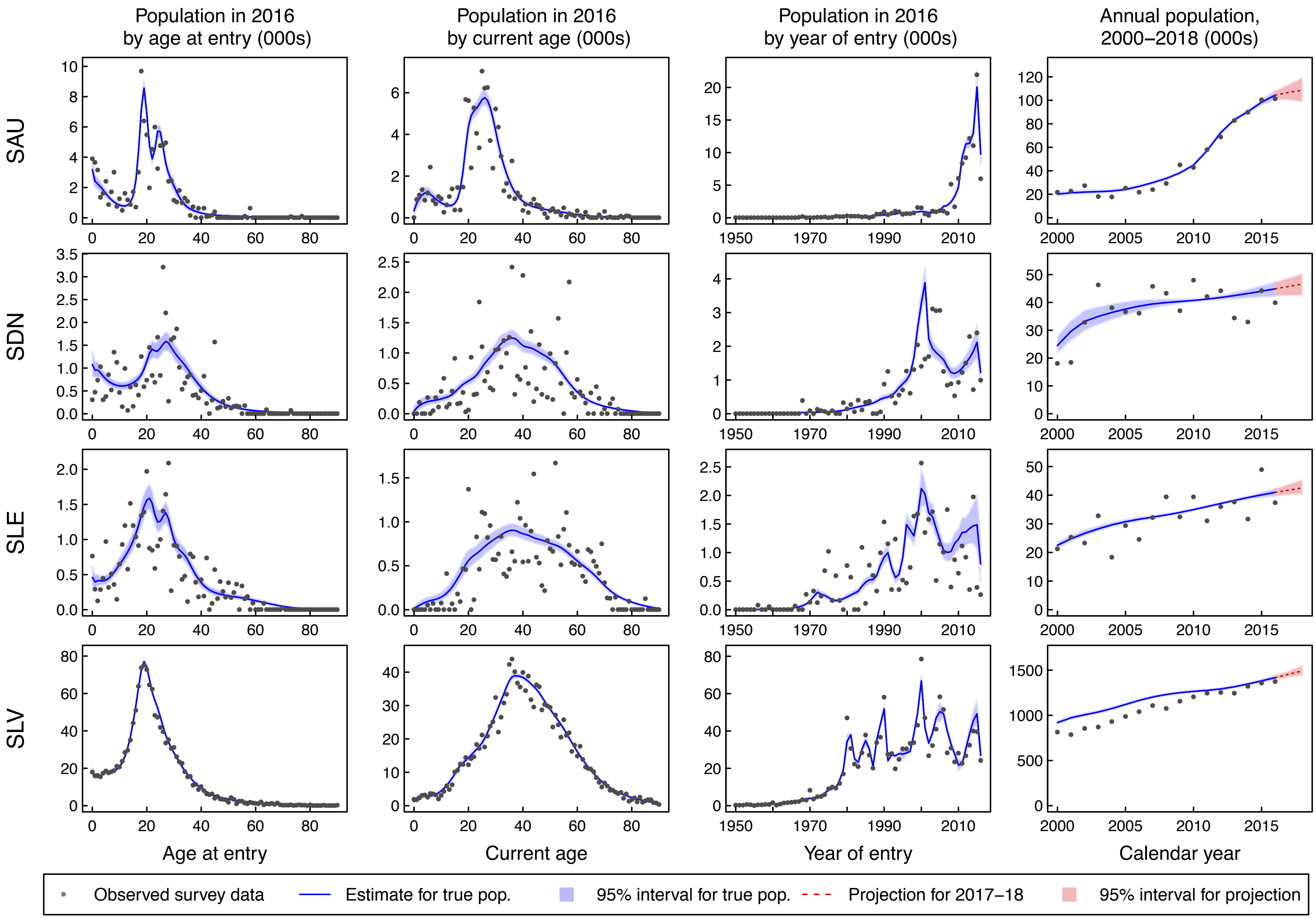

**Figure S7–24: Comparison of modelled versus raw population estimates for each country and region of origin.**

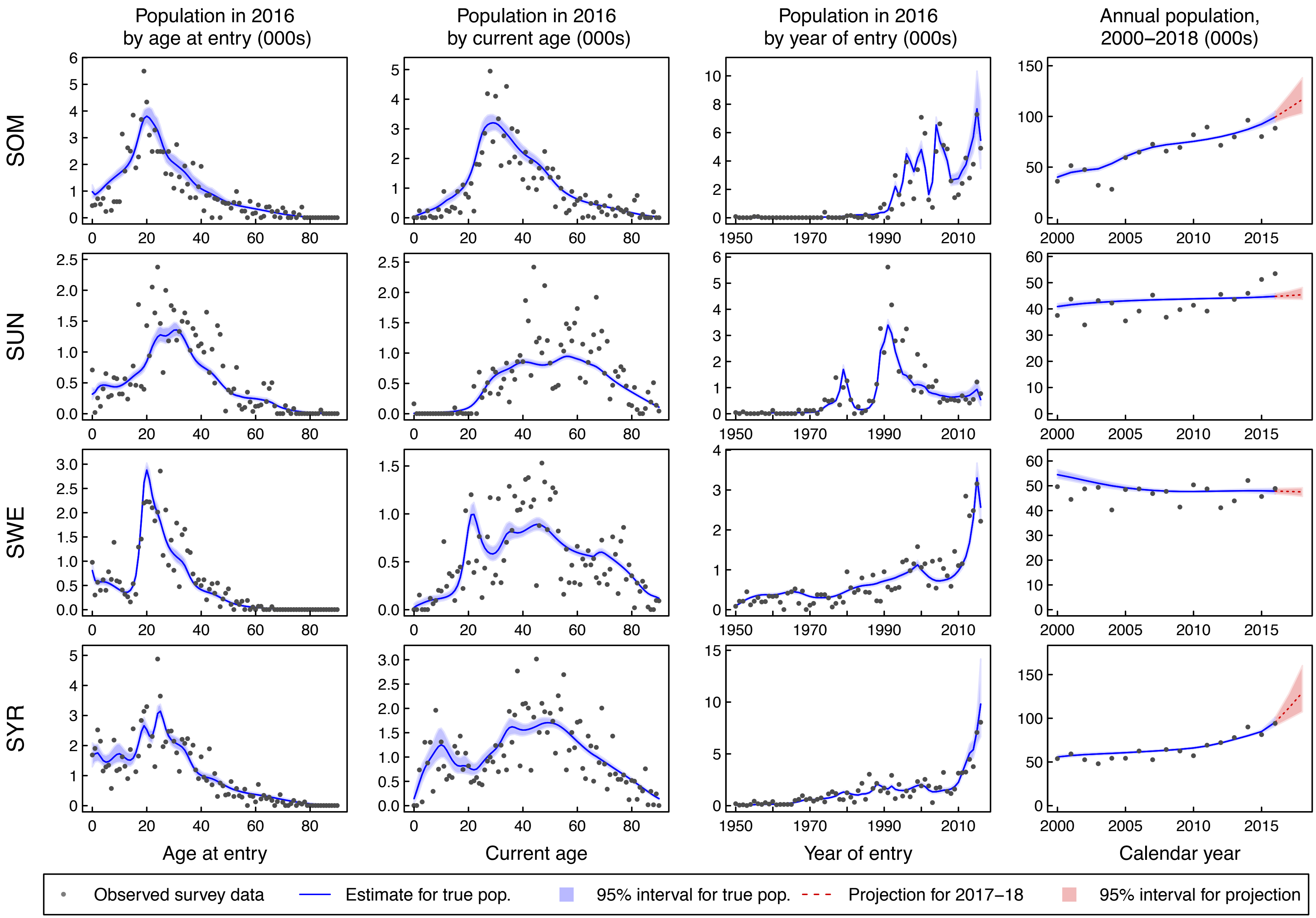

**Figure S7–25: Comparison of modelled versus raw population estimates for each country and region of origin.**

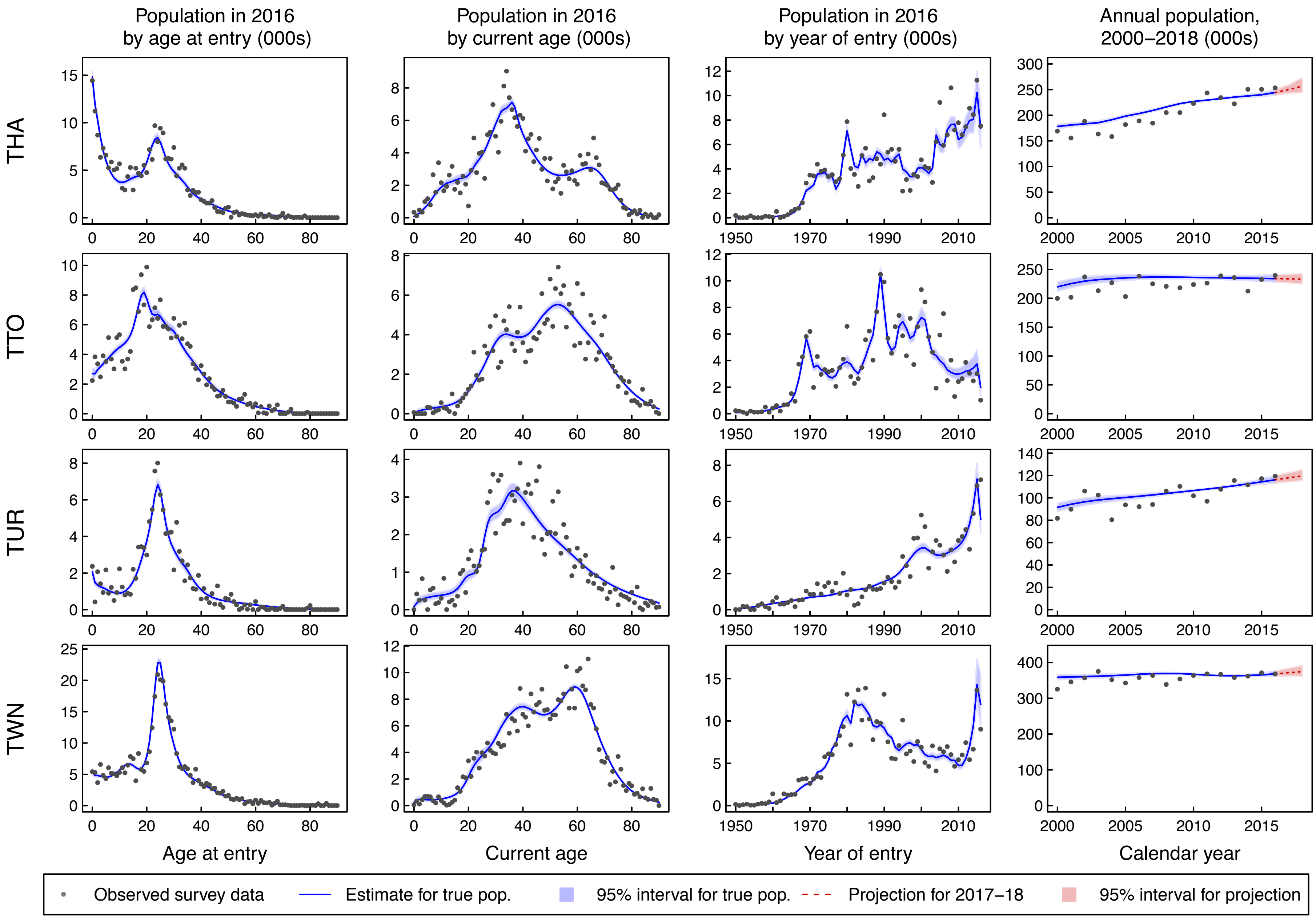

**Figure S7–26: Comparison of modelled versus raw population estimates for each country and region of origin.**

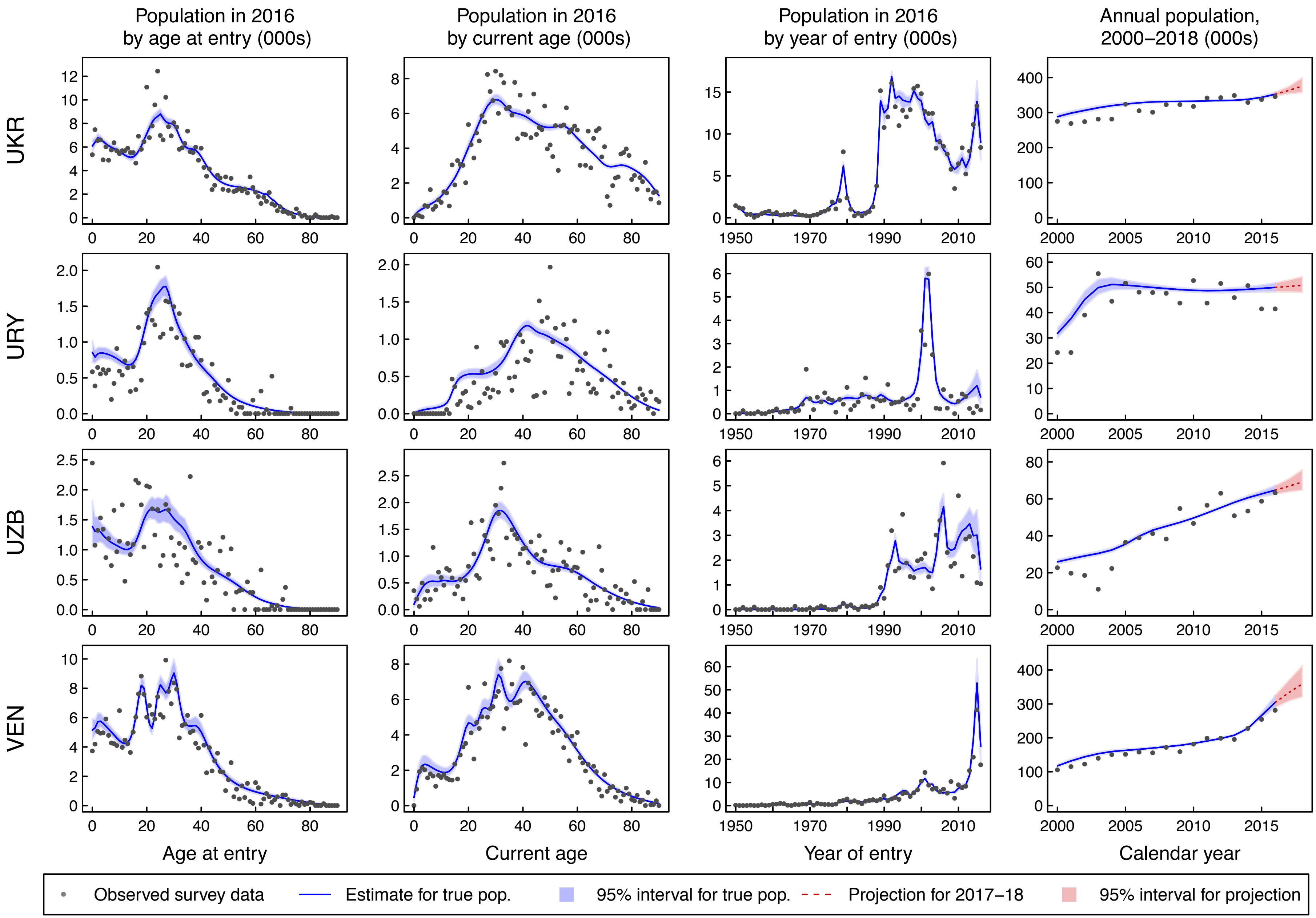

**Figure S7–27: Comparison of modelled versus raw population estimates for each country and region of origin.**

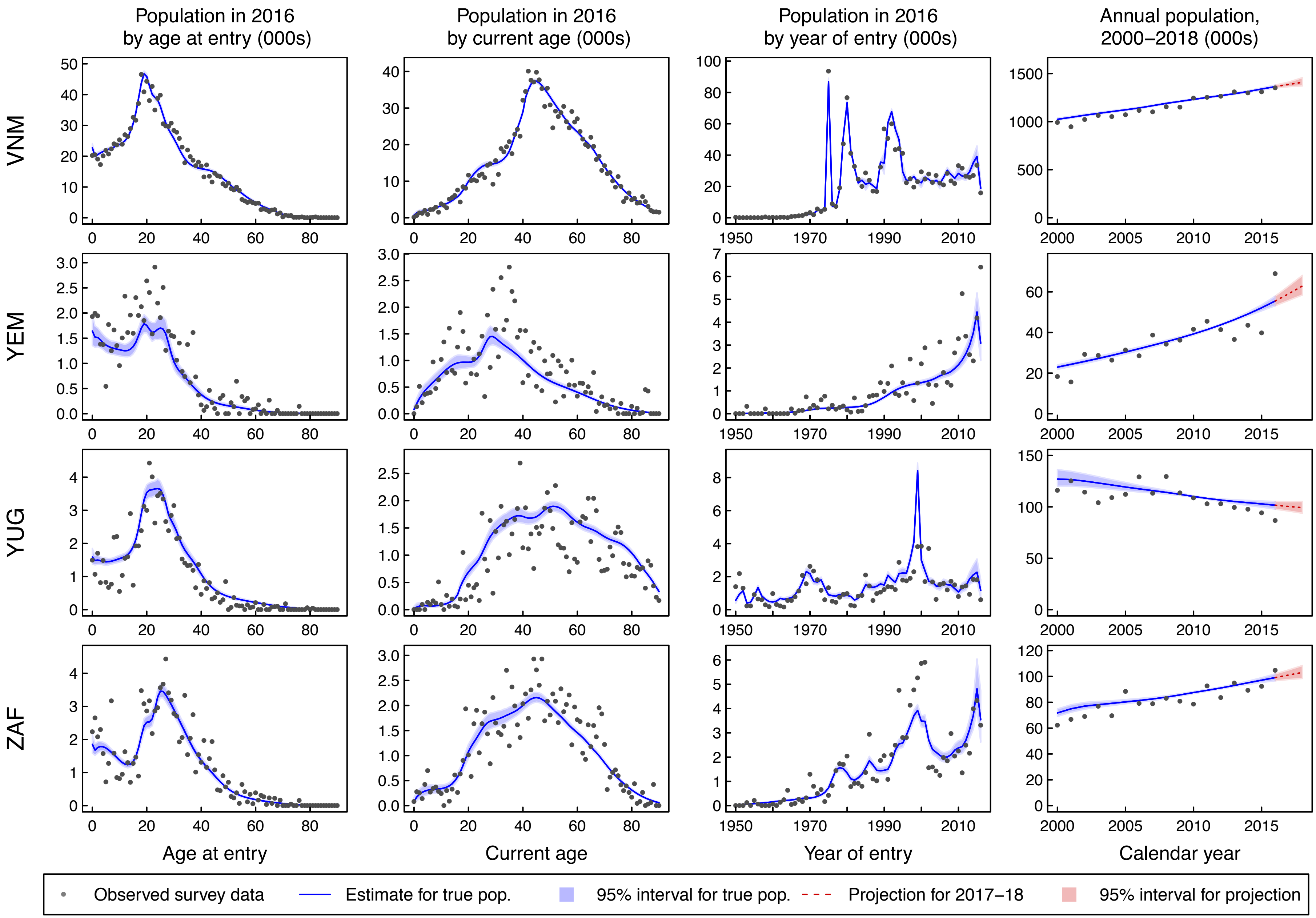